%% file: main.tex
\documentclass{article}
\usepackage{arxiv}
\usepackage{amsmath, amsfonts, amsthm}
\usepackage[utf8]{inputenc} 
\usepackage[T1]{fontenc} 
\usepackage{hyperref} 
\usepackage{url}
\usepackage{booktabs}
\usepackage{amsfonts}
\usepackage{nicefrac}
\usepackage{microtype}
\usepackage[capitalise]{cleveref}
\usepackage{graphicx}
\usepackage{doi}
\usepackage{caption, subcaption}
\usepackage{enumerate}
\usepackage{textcomp}
\usepackage{float}
\usepackage{xparse}
\usepackage{subfiles}
\usepackage{fixltx2e}
\usepackage{multicol}
\usepackage{authblk}

\newbox{\orcid}\sbox{\orcid}{\includegraphics[scale=0.06]{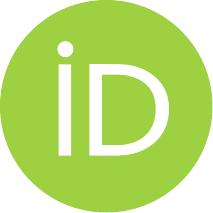}} 

\input{common/macros}

\title{Social Network Analysis and Validation of an Agent-Based Model}

\author[1]{
	\href{https://orcid.org/0000-0001-8019-9301}{\usebox{\orcid}\hspace{1mm}Karleigh~Pine\thanks{\texttt{\href{mailto:karleigh.pine@matrixresearch.com}{karleigh.pine@matrixresearch.com}}}}
}
\author[1]{
	\href{https://orcid.org/0000-0002-4135-2753}{\usebox{\orcid}\hspace{1mm}Joel~Klipfel\thanks{\texttt{\href{mailt:joel.klipfel@matrixresearch.com}{joel.klipfel@matrixresearch.com}}}}
}
\author[2,3]{
	\href{https://orcid.org/0000-0003-4718-257X}{\usebox{\orcid}\hspace{1mm}Jared~Bennett\thanks{\texttt{\href{mailto:jbennett@mobiuslogic.com}{jbennett@mobiuslogic.com}}}}
}
\author[2]{
	\href{https://orcid.org/0000-0001-8565-2427}{\usebox{\orcid}\hspace{1mm}Nathaniel Bade\thanks{\texttt{\href{mailto:nbade@mobiuslogic.com}{nbade@mobiuslogic.com}}}}
}
\author[2]{
	\href{https://orcid.org/0000-0002-2283-946X}{\usebox{\orcid}\hspace{1mm}Christian Manasseh\thanks{\texttt{\href{mailto:cmanasseh@mobiuslogic.com}{cmanasseh@mobiuslogic.com}}}}
}
\affil[1]{Matrix Research, Dayton, Ohio, USA}
\affil[2]{Mobius Logic, Tysons, Virginia, USA}
\affil[3]{UC Berkeley, Berkeley, California, USA}

\renewcommand{\shorttitle}{Social Network Analysis and Validation of an ABM}

\hypersetup{
pdftitle={Social Network Analysis and Validation of an Agent-Based Model},
pdfsubject={cs.SI, cs.MA, cs.DM. math.CO, q-bio.PE, q-bio.QM},
pdfauthor={Karleigh Pine, Joel Klipfel, Jared Bennett, Nathaniel Bade, and Christian Manasseh},
pdfkeywords={Graphs and networks, Multiagent systems, Social networking, Stochastic processes},
}

\begin{document}
\maketitle

\begin{abstract}
Agent-based models (ABMs) simulate the formation and evolution of social processes at a fundamental level by decoupling agent behavior from global observations. In the case where ABM networks evolve over time as a result of (or in conjunction with) agent states, there is a need for understanding the relationship between the dynamic processes and network structure. Social networks provide a natural set of tools for understanding the emergent relationships of these systems. This work examines the utility of a collection of network comparison methods for the purpose of tracking network changes in an ABM over time or between model parameters. Among the techniques examined is a novel graph pseudometric based on heat content asymptotics, which have been shown to distinguish many isospectral graphs which are not isomorphic. Additionally, we establish the use of observations about real-world networks from network science (\textit{e.g.} fat-tailed degree distribution, small-world property) for ABM validation in the case where empirical population data is unavailable. These methods are all demonstrated on systematic perturbations of an original model simulating the formation of friendships in a population of 20,000 agents in Cincinnati, OH.
\end{abstract}


%%==========
%% Main body
%%==========

%% Introduction
\input{sections/intro}

%% Background
\input{sections/background}

%% Experimental Setup
\input{sections/exp_setup}

%% Results
\input{sections/results}

%% Conclusion
\input{sections/conclusion}

%%=================================
%% Acknowledgement and Bibliography 
%%=================================
\section*{Acknowledgement}
Work on this research has been funded by the Air Force Research Lab (AFRL) Autonomy 
Capability Team 3 (ACT3) under contracts FA8650-20-C-1121 and FA8649-20-C-0130. 
The authors would like to thank Dr. Jared Culbertson, Dr. Scott Clouse, and 
Dr. Greg Arnold for their thoughtful feedback. Thanks also go to Dr. Razvan Veliche for 
assistance running the ABM simulations. 

\bibliographystyle{ieeetr}
\bibliography{references}

%%===========
%% Appendices
%%===========
\appendix
\setcounter{equation}{0}
\setcounter{figure}{0}
\setcounter{table}{0}
\renewcommand{\theequation}{\thesection\arabic{equation}}
\renewcommand{\thefigure}{\thesection\arabic{figure}}
\renewcommand{\thetable}{\thesection\arabic{table}}

\include{appendices/heatcontent}
\include{appendices/abmdescription}
\include{appendices/runtable}
\include{appendices/addresults}

\end{document}

%% file: common/macros.tex
\NewDocumentCommand\mainsection{o m o}{\IfNoValueTF{#3}{\section[#1]{#2}}{\section[#1]{#2}\label{sec:#3}}}
\NewDocumentCommand\mainsubsection{o m o}{\IfNoValueTF{#3}{\subsection[#1]{#2}}{\subsection[#1]{#2}\label{ssec:#3}}}
\NewDocumentCommand\mainsubsubsection{o m o}{\IfNoValueTF{#3}{\subsubsection[#1]{#2}}{\subsubsection[#1]{#2}\label{ssec:#3}}}
\NewDocumentCommand\appsection{o m o}{\IfNoValueTF{#3}{\section[#1]{#2}}{\section[#1]{#2}\label{app:#3}}}
\NewDocumentCommand\appsubsection{o m o}{\IfNoValueTF{#3}{\subsection[#1]{#2}}{\subsection[#1]{#2}\label{sapp:#3}}}
\NewDocumentCommand\appsubsubsection{o m o}{\IfNoValueTF{#3}{\subsubsection[#1]{#2}}{\subsubsection[#1]{#2}\label{ssapp:#3}}}
\NewDocumentCommand\apptable{o m o}{\IfNoValueTF{#3}{\section[#1]{#2}}{\section[#1]{#2}\label{app:#3}}}

\newcommand{\tableref}[1]{Table~\ref{table:#1}}  % ref a table
\newcommand{\secref}[1]{Section~\ref{sec:#1}}  % ref a section within a sentence
\newcommand{\ssecref}[1]{Subsection~\ref{ssec:#1}}  % ref a section within a sentence
\newcommand{\appref}[1]{Appendix~\ref{app:#1}} % ref an appendix
\theoremstyle{plain}      %[section]
\theoremstyle{plain}      \newtheorem*{thm*}{Theorem}%[section]
\theoremstyle{plain}      %[section]
\theoremstyle{plain}      %[section]
\theoremstyle{plain}      \newtheorem*{lma*}{Lemma}
\theoremstyle{plain}      %[section]
\theoremstyle{plain}      \newtheorem*{cor*}{Corollary}%[section]
\theoremstyle{plain}      %[section]
\theoremstyle{plain}      \newtheorem*{claim*}{Claim}
\theoremstyle{plain}      %[section]
\theoremstyle{plain}      \newtheorem*{prop*}{Proposition}%[section]
\theoremstyle{plain}      
\theoremstyle{plain}      
\theoremstyle{remark}     
\theoremstyle{remark}     \newtheorem*{rmk*}{Remark}
\theoremstyle{remark}     
\theoremstyle{definition} %[section]
\theoremstyle{definition} \newtheorem*{defn*}{Definition}%[section]
\theoremstyle{definition} 
\theoremstyle{definition} \newtheorem*{ex*}{Example}%[section]
\theoremstyle{plain}

%% file: sections/intro.tex
\section{Introduction}\label{sec:introduction}

At its simplest, an agent-based model (ABM) consists of a set of agents and a set of rules (deterministic or stochastic) that define how agents interact with each other and their perceived environment. As the model propagates through time, agents' states are updated to reflect these interactions and the changing environment due to other agents' actions. This bottom-up approach to simulation promotes a natural and flexible description of the system (an agent is home and the weather is nice $\rightarrow$ they go outside and ``see'' their neighbor $\rightarrow$ the agent ``waves''). These very basic descriptions, performed appropriately by agents dependent on their environment, lead to emergent phenomena that reflect behavioral changes within agents. These emergent patterns, and the changes that lead to them, are often what researchers wish to study.

ABMs fall into three categories of network formation: exogeneous formation, endogeneous formation, and co-evolutionary networks. Exogeneous networks are often static and imposed independently of the agents' states. Endogeneous networks evolve over time in response to changes in agents' states. The third type of network, a co-evolutionary model, contains a feedback loop between network structure and agents' states, where each influences the other over the course of the simulation. Our work is primarily relevant to the last two cases. Network analysis is particularly important as networks become more complicated and affected by an increasing number of variables. There is a history of studying endogeneous network formation in game-theoretic ABMs, but these models are often simplistic~\cite{namatame_agent-based_2016}. More work needs to be done to understand the relationship between dynamic models and network structure~\cite{namatame_agent-based_2016,will_combining_2020}. Our paper seeks to address this need by employing a variety of network comparison techniques for tracking changes in network structure over time and between model parameters. 

There are many different methods for network comparison, each providing 
complementary lenses to evaluate the results of an ABM. This work employs a wide variety of network
comparison techniques for evaluation of ABM social networks over time and between models, including several
methods that have not previously been considered for this purpose. Section
\ref{network-analysis} provides a detailed discussion of the network comparison 
methods we use in this paper. We also introduce in Section
\ref{network-analysis} the heat content asymptotics (HCA) pseudometric, which 
is a new method for comparing two networks based on 
\cite{mcdonald_2002, mcdonald_2003, lu_complex_2014}.\footnote{In this paper, we use 
the terms metric and pseudometric according to their definitions in mathematics. Specifically, a 
metric is positive definite, symmetric, and satisfies the triangle inequality. A pseudometric 
is positive, symmetric and satisfies the triangle inequality, but need not be positive definite.} 
The questions that we seek to answer in this paper are then ``What information does each of 
these network comparison methods provide?'' and ``Which techniques are most 
useful for characterizing an ABM's behavior?'' 

Multiple papers highlight the strong importance of model validation in ABM 
based research while simultaneously emphasizing both the lack of and need for systematic methodologies for model validation \cite{k_uuml_ppers2005, Silverman_2018, 
sabzian2019theories, Pullum2012TechniquesAI, groff_state_2019}. Indeed, Groff 
{\em et al.} note in the abstract of \cite{groff_state_2019} that ``[f]or the 
field to progress, at a minimum, standardized reporting that encourages 
transparency will be necessary.'' It is within the context seeking to develop 
ABM model validation norms we explore the two questions at the end of the 
previous paragraph. In particular, we consider a scenario in which no empirical 
data for validation of networks is available. In this case, we turn to the study 
of social networks which has identified several qualities that tend to be true 
of real-world networks ({\em e.g.} fat-tailed degree 
distribution~\cite{price_networks_1965}, small-world 
property~\cite{milgram_small_1967}, high clustering 
coefficient~\cite{watts_collective_1998}). It is natural to suppose that an ABM 
which successfully models a real-world scenario will have a network which 
satisfies these properties. Thus, 
examining these features on the networks arising from ABMs can inform validation 
of the underlying model. 
In this paper, we present a case study in which we demonstrate the use of 
network analysis as a potential means for validating an ABM.

While the case for combining network theory and agent-based modeling has been made 
before~\cite{el-sayed_social_2012,alam_networks_2012}, to the best of our knowledge, the 
use of social network characteristics for validation of an ABM has not been proposed. 
Additionally, an agent-based modeling framework can offer valuable insight into the 
processes involved in social network formation.  Unfortunately, work exploring the effect 
of particular social model parameters on graph quantities has been limited in the context of 
ABMs. In this paper, we demonstrate the value of network science for both validation and 
fine-tuning of an ABM by examining perturbations of an agent relationship formation model. 
Our particular use case is an ABM modeling the evolution of friendships in Cincinnati, OH. 
Analysis of the resulting social networks provides insight into the model and the process 
it attempts to simulate. We examine the utility of various graph quantities 
for this purpose, including the novel HCA pseudometric we introduce in \secref{background}.

The paper starts with background material on social network analysis and agent-based modeling and validation (\secref{background}). A description of the experiments performed can be found in \secref{exp_setup} along with methodology behind the creation of the ABM and definitions of the various graph measures used. Results are summarized in \secref{results} with concluding remarks in \secref{conclusion}. 

%% file: sections/background.tex
\section{Background} \label{sec:background}

\subsection{Network Analysis and Comparison}
\label{network-analysis}
A graph $G=(\mathcal{N},\mathcal{E})$ is composed of a set of nodes, $\mathcal{N}$, and connections (edges) between nodes, $\mathcal{E}$. For this paper, we work with undirected graphs. Edges, while weighted in the ABM model, are converted to unweighted edges by removing all edges weighted less than a threshold, $\alpha$.

Network comparison is a well-studied problem, either to determine the similarity of networks from different processes or the change in a single network over time. Some natural Euclidean representations for a network include the adjacency matrix, $A$, graph Laplacian, $L = D-A$, and the normalized graph Laplacian, $\mathcal L = D^{-1/2}LD^{-1/2}$, where $D$ is the diagonal matrix of node degrees. However, these matrices are only unique up to node ordering, making them ill-suited for direct network comparison. That is, different representations will arise from nodes being encoded in a different order (corresponding to permutations of rows and columns). Additionally, the large size ($|\mathcal{N}| \times |\mathcal{N}|$) of both the adjacency matrix and the graph Laplacian can be memory intensive for large graphs, although sparsity of the representation can often be exploited for storage. 

The ambiguity in representation motivates the use of graph descriptors that are invariant to node labeling for network comparison, referred to as graph invariants or graph properties. If a full graph representation is not needed, two main categories of graph invariants have proven useful for concisely representing distinguishing information: graph features and spectral methods. 

The first category includes the use of popular descriptive statistics such as the shortest path distance, degree, clustering coefficient, number and sizes of connected components, number of nodes, and number of edges. Multiple features can also be combined into hybrid descriptors such as the network portrait developed in~\cite{bagrow_information-theoretic_2019} which captures aspects of both the degree distribution and the distribution of shortest paths. Each of these features provide a different perspective of the network structure. Quantities that are computed on the node level (degree or clustering coefficient) or edge level (shortest path distance) can be summarized for a network in the form of distributions. Although this provides a method of comparing networks which is node invariant, a network is not uniquely defined by its statistics. That is, there may be many distinct graphs which can be described by the same graph features. 

However, graph summary statistics can give valuable insight into the graph structure. Real-world networks, especially social networks, are known to have graph features that satisfy certain properties. Social networks tend to have a ``fat-tailed'' degree distribution with many more low-degree nodes or high-degree nodes than are found in a random network~\cite{price_networks_1965}. Even more specifically, some researchers have found real-world networks to have degree distributions that follow a scale-free power-law distribution~\cite{barabasi_emergence_1999} although there is some debate over that fact (\cite[Section 3.2]{jackson_social_2008},~\cite{broido_scale-free_2019}). Another well-known property of real-world networks is their tendency to have a small diameter\footnote{A graph's diameter is the largest shortest path length between pairs of nodes.} and small average path lengths. This has been coined the \textit{small-world property}~\cite{milgram_small_1967}. Colloquially, one example of this phenomenon is the game ``six degrees of Kevin Bacon'' which tasks players with finding a path from any actor to Kevin Bacon by linking actors that appear in a movie together. This takes advantage of the fact that the network formed by co-starring actors is made up of many short paths between pairs of nodes. Real-world networks also tend to exhibit much higher clustering coefficients than those associated with a randomly generated network. This means that in a real-world network two nodes which share an edge are more likely to share neighbors than in a network with randomly drawn edges. This quality can be evaluated by the local clustering coefficient, a measure introduced by Watts and Strogatz~\cite{watts_collective_1998}.   

Spectral analysis, or analysis of the eigenvalues of $A$, $L$, or $\mathcal L$, can also give insight into the topology of a network and has direct implications to the modeling of physical diffusion processes on a graph. The relationship between a graph's structure and its eigenspectrum has been studied in depth~\cite{chung_spectral_1997, zhu_study_2005, belkin_laplacian_2003, bai_heat_nodate,takahashi_discriminating_2012, gera_identifying_2018}. Analyzing the spectrum of $A$, $L$, or $\mathcal L$ circumvents the issue of node ordering since all isomorphic networks have the same spectrum. Additionally, it can be shown that if two networks are similar in terms of graph edit distance, their eigenvalues will be close in Euclidean space~\cite{zhu_study_2005}. Spectral analysis also provides a principled approach for projecting nodes into a lower dimensional space (the eigenspace)~\cite{goos_inexact_2002}. However, eigenanalysis methods tend to be less interpretable than graph feature techniques and have a high computational cost. One method for mitigating the high computational cost is approximating the spectrum using random walks as is done in~\cite{lu_complex_2014}. 

A known limitation to many spectral analysis methods is the existence of 
non-isomorphic graphs which are isospectral\textemdash{}meaning 
that knowing the Laplacian spectrum for two non-isomorphic graphs may not be 
enough to distinguish them \cite{buser_1994}.
In their search for graph invariants to help 
distinguish isospectral graphs, McDonald and Meyers examine\textemdash{}among 
other graph invariants\textemdash{}the heat content asymptotics (HCA) of a graph 
\cite{mcdonald_2002, mcdonald_2003}. The heat content $q(s)$ of a graph $G$ is 
the sum of the entries of the matrix $e^{-s \mathcal L}$, and the HCA of $G$ is 
the collection of coefficients in the Maclaurin expansion in $s$ of $q$. That is, 
if we represent the Maclaurin expansion $q$ by $q(s) = \sum_{n\geq 0} q_n s^n$, 
then the HCA of $G$ is the set 
\begin{displaymath}
\text{hca}(G):= \{q_n\}_{n\geq0}.
\end{displaymath} 
McDonald and Meyers prove in their 2002 
paper that a graph's HCA determine the spectrum of the corresponding Laplacian
\cite[Theorem 1.3]{mcdonald_2002}, and, conversely, a graph's Laplacian spectrum 
plus the spectral partition of its volume determine its heat content 
asymptotics \cite[Corollary 5.3]{mcdonald_2002}. They further demonstrate in 
\cite{mcdonald_2003} that a known class of non-isomorphic graphs which are 
pairwise isospectral can be distinguished using heat content asymptotics. 
Inspired by \cite{mcdonald_2003}, Lu {\em et al}. derive in 
\cite{lu_complex_2014} a method for approximating heat content curve using 
random walks that bypasses the need to compute the spectrum of a graph 
Laplacian. We provide a formal definition of heat asymptotics along with a derivation
of the heat content curve approximation method by Lu {\em et al}.
in \appref{heatcontentapp}.

Each of these quantities described are defined on static 
graphs. For networks with a temporal component ({\em i.e.} the presence or 
strength of edges is changing over time) static graphs can be obtained by 
taking ``snapshots'' of the graph at different moments in time. Measurements can 
then be taken on each of the captured graphs. Tracking a change in quantities 
temporally can provide insight into the evolution of a network over time. 
Multiple temporal networks can be compared by taking measurements at similar 
intervals and comparing the corresponding static graphs. In this work, we compare networks both temporally and over ABM model parameters using the descriptive statistics mentioned above along with the HCA spectral-based method.

\subsection{Agent-Based Modeling}

Agent-Based Models (ABMs) are computational models for simulating interactions of agents in an environment and studying the emergent dynamics of these interactions to infer system behavior.

Introduced in the 1960s, ABMs have evolved into tools used in biology \cite{emonet_agentcell_2005}, ecology \cite{thober_agent-based_2018}, social sciences \cite{ogibayashi_model_2020}, and the stock market \cite{bonabeau_agent-based_2002}, among other diverse applications. They combine a simplified model of agent actions and interactions with a stochastic (random) aspect on priorities and action selections, allowing quick and inexpensive exploration of a wide range of complex scenarios, built from comparatively simple rules of interactions, from which general system behavior can be extracted. 

Social Interaction Models (SIMs), a realization of Cederman's third approach to social behavior modeling \cite{agent-based-modeling-political-science}, are ABMs focused on emerging interaction networks evocative of human societal interactions; synthetic societies created using agents that represent humans, with a subset of their capabilities, along with necessary environmental influences (e.g., roads/buildings/streetlights to generate traffic patterns or a simplified stock-market for agents to “play” for economic exploration), and then attempt to understand the impact of changes, in parameter values or agent capabilities, on the synthetic society.

However, one of the most pervasive issues with the application of agent-based modeling to 
study social phenomena is the problem of validation \cite{kuppers_validation_2005}. 
As noted in \cite{silverman_methodological_2018}:
\begin{quote}
Criticism of agent-based models in political science has come from a number of different 
sources, but a large portion of those criticisms focus on the difficulty of making 
sensible abstractions for social and political structures within such a model. […] 
political systems involve multiple layers of interacting components, each of which is 
understood primarily as an abstracted entity; frequently only the end results of political 
change or transition are easily observable, and even then the observer will have great 
difficulty pinpointing specific low-level effects or drives which may have influenced 
those results.
\end{quote}

Because of this, many validation frameworks have been suggested, including methods to compare data against real-world data sources, mathematical models, or other ABMs, and how to verify model functionality matches high level design \cite{sabzian_theories_2019,pullum_techniques_2012}. These may be done on the entire model or performed on subsystems to give evidence for the whole. Towards this end, we demonstrate the use of network analysis methods for validating the social interaction subsystem of an ABM.

\subsection{Network Analysis for ABMs}

The symbiotic relationship between ABMs and social network analysis has been reviewed in~\cite{namatame_agent-based_2016,will_combining_2020} and~\cite{alam_networks_2012}. There has been incorporation of social networks into ABMs in fields such as epidemiology~\cite{el-sayed_social_2012}, marketing~\cite{beretta_cultural_2018,pearce_feed-tariffs_2018}, and social modeling~\cite{zhuge_agent-based_2018}. In the case where networks evolve endogeneously or co-evolutionarily with ABM agent states there is a need for deeper understanding of the effect of dynamic social processes on network structure~\cite{namatame_agent-based_2016,will_combining_2020}.    

Even though the value of incorporating social networks into ABMs has been established, several network comparison techniques have not been considered for this purpose ({\em e.g.} network portrait, HCA), nor has there been a systematic review of comparison methods. Additionally, there has been relatively little work connecting social network properties to validation of a model. Particularly, we are interested in which graph features are useful for validation and how various ABM parameters can influence the graph topology. A few studies which address the latter point include Jin {\em et al}.~\cite{jin_structure_2001} and Pujol {\em et al}.~\cite{pujol_how_2005}. Jin {\em et al}. find that networks exhibiting features of real world networks can be produced using only a few simple interaction rules. Pujol {\em et al}. also examine the types of complex networks that emerge from variations of a social exchange model. In~\cite{zhuge_agent-based_2018} ABM networks are forced to have realistic qualities (in terms of degree distribution and clustering) by fitting these distributions as networks develop. Our work differs from~\cite{zhuge_agent-based_2018} in that we do not enforce these characteristics, instead evaluating them for validation after the model evolves. 

Several tools exist for analyzing ABM networks using network science techniques. The Repast Simphony ABM software~\cite{north_complex_2013} provides internal tools for simulating networks but then recommends external tools (such as Pajek~\cite{batagelj_pajek_2004}, JUNG~\cite{Madadhain2005}) for analysis. The ReSoNetA library~\cite{holzhauer_developing_2010} has also been developed for incorporation of social network analysis with the Repast ABMs, although it tends to focus on node level quantities such as centrality and prestige as opposed to analysis of the global topology. 

In addition to the utility ABMs have for simulating the formation of social networks, social network analysis also offers a framework for understanding the processes of an ABM. Understanding what information is provided by different graph analysis techniques is important because different metrics may prove more valuable based on the application of the ABM. For instance, when analyzing two variations of an ABM simulating disease propagation, the random seed variations alone may result in different people being infected. As researchers, we may not be concerned on exactly \textit{who} is infected as long as the aggregate behavior is the same (measured in number of infections over time, for instance). Thus, looking at metrics such as graph edit distance would not make sense in this problem space, and could grow computationally untractable. Improving our understanding for what information can be gained from various network quantities is a goal of this work.

%% file: sections/exp_setup.tex
\section{Experimental Setup} \label{sec:exp_setup}

\subsection{Model Definition}
We designed an ABM to represent a medium-sized city familiar to us, Cincinnati Ohio, using the Repast simulation framework \cite{north_complex_2013}. Using demographic and socioeconomic data, including the US Decennial Census \cite{u_g_census_decennial_2020}, the yearly ACS from the Census Bureau \cite{noauthor_american_2020}, OpenStreetMap \cite{openstreetmap_contributors_open_2022}, and the Longitudinal Employer-Household Dynamics \cite{noauthor_longitudinal_2020}, we generated a reduced population (20K agents) with statistically representative characteristics of the true Cincinnati population \cite{mobius_2022} (\appref{abmapp}).

As our focus is on the evolving social interaction network, we kept the simulations short (6 months) and did not implement any life changes ({\em e.g.} moving, changing jobs, school graduations, etc.). Instead, we concentrated on agents' friends - how they make friends, how they spend time with friends, and eventually how they lose friends. We define friends as non-family members that an agent regularly spends time with, either through random daily interactions ({\em e.g.} travel to work, eating at the same restaurants) or through planned activities ({\em e.g.} going to the mall, an evening at a bar). Further explanation of daily activities can be found in the second part of \appref{abmapp}.

\begin{figure*}[h!]
    \centering 
    \def\scale{0.8}
    \begin{subfigure}[b]{0.32\textwidth}
        \centering
        \includegraphics[width=\scale\textwidth]{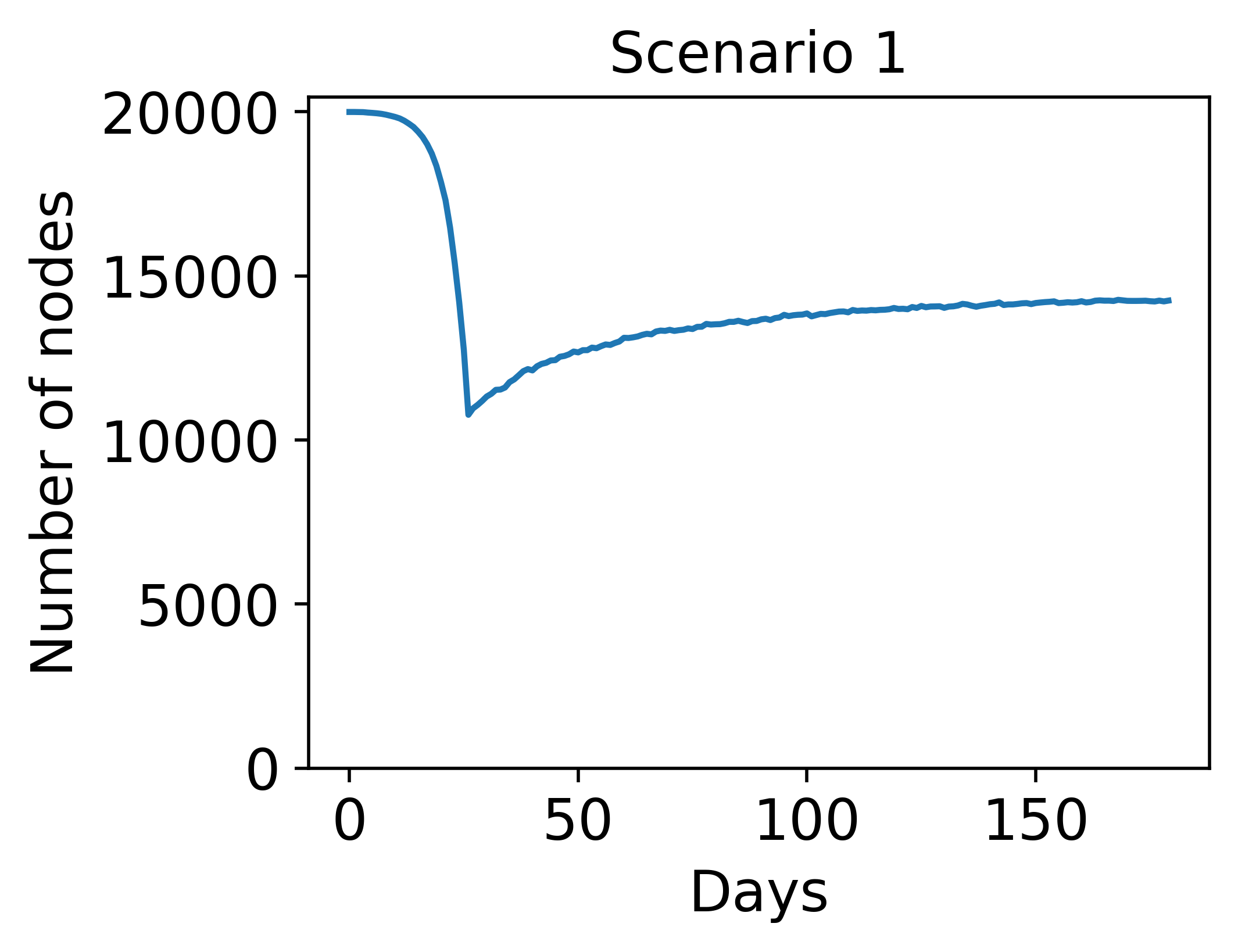}
        \caption{}
        \label{node1}
    \end{subfigure}
    \hfill
    \begin{subfigure}[b]{0.32\textwidth}
        \centering
        \includegraphics[width=\scale\textwidth]{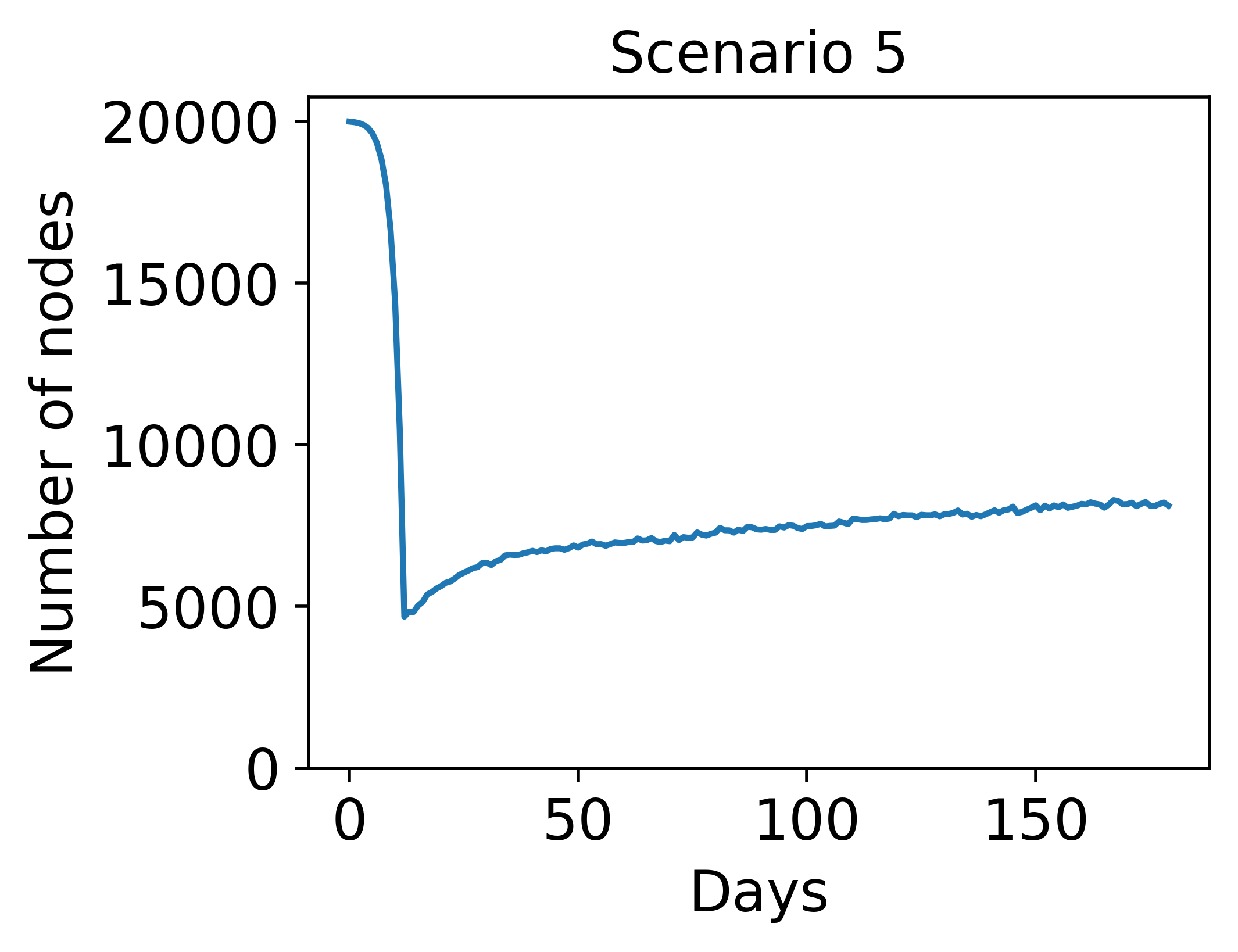}
        \caption{}
        \label{node5}
    \end{subfigure}
    \hfill
    \begin{subfigure}[b]{0.32\textwidth}
        \centering
        \includegraphics[width=\scale\textwidth]{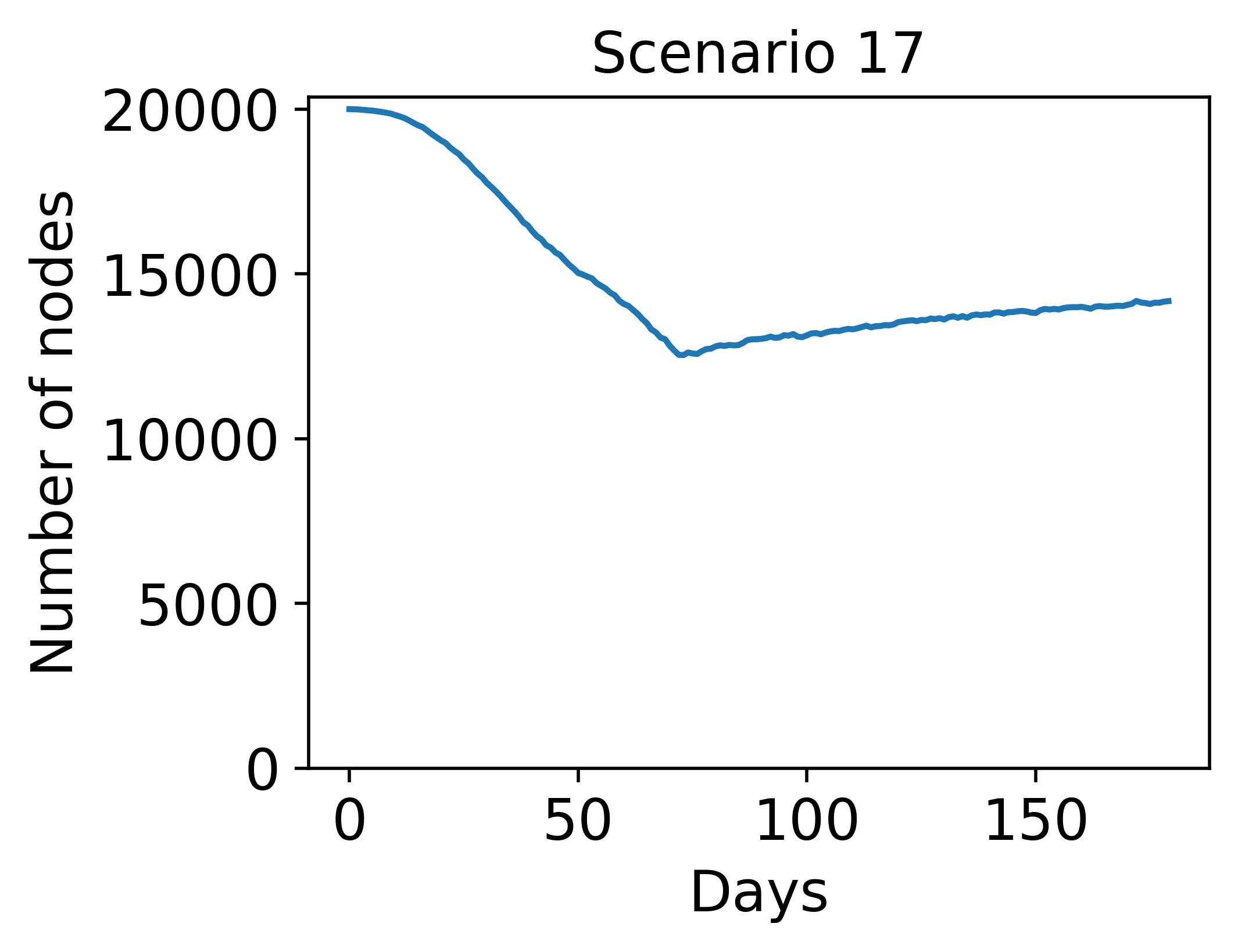}
        \caption{}
        \label{node17}
    \end{subfigure}
    \begin{subfigure}[b]{0.32\textwidth}
        \centering
        \includegraphics[width=\scale\textwidth]{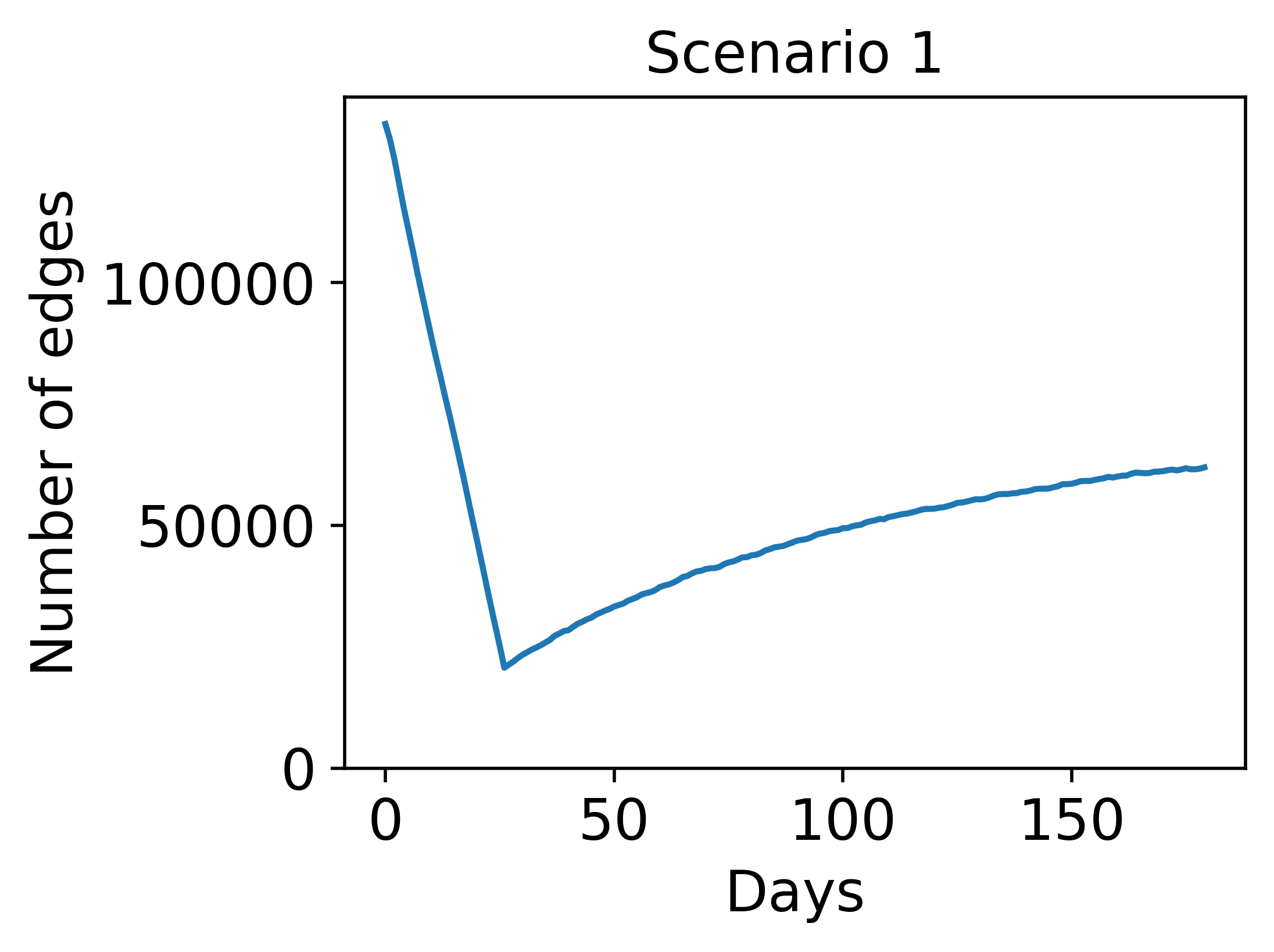}
        \caption{}
        \label{edge1}
    \end{subfigure} 
    \hfill
    \begin{subfigure}[b]{0.32\textwidth}
        \centering
        \includegraphics[width=\scale\textwidth]{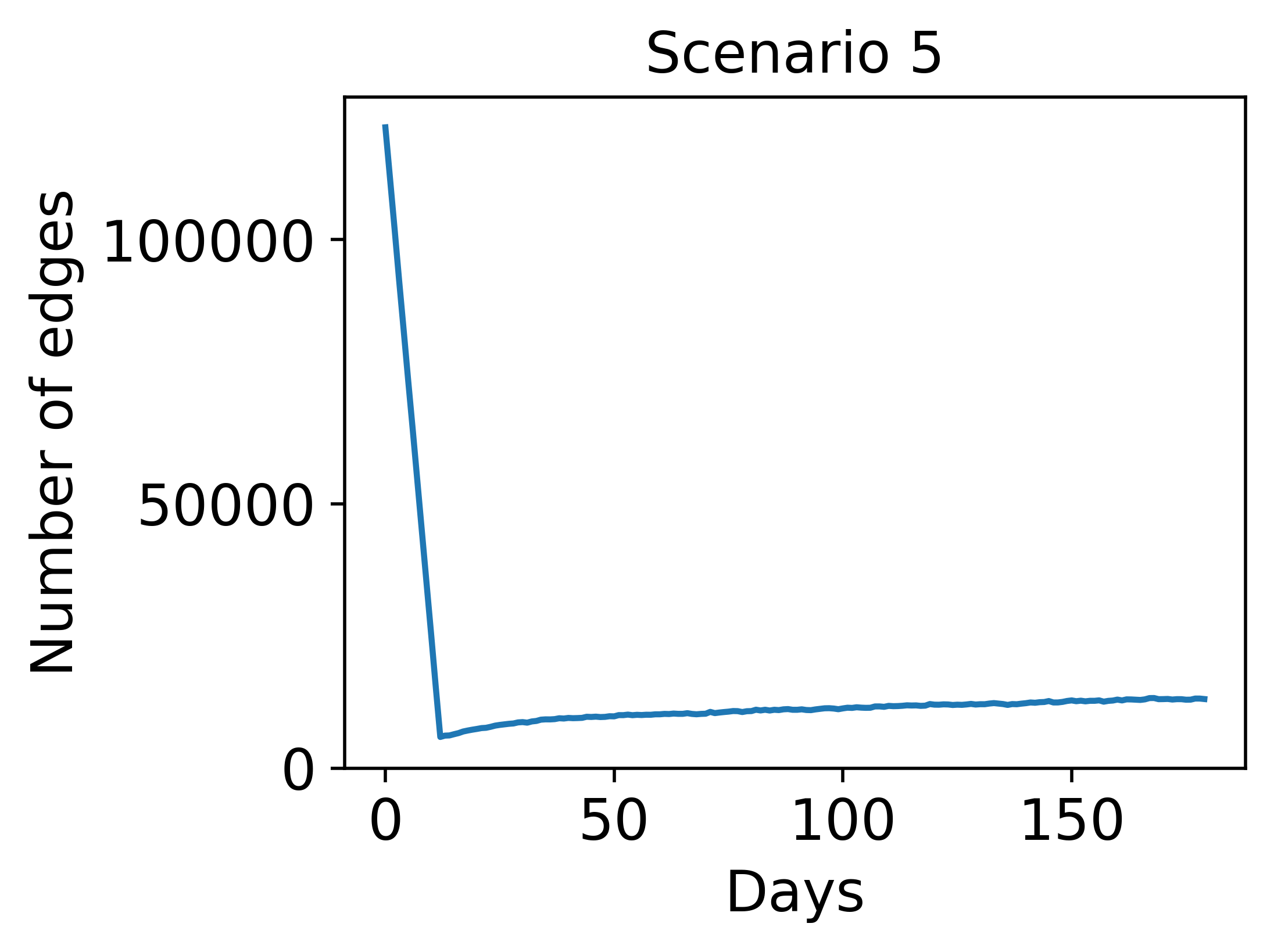}
        \caption{}
        \label{edge5}
    \end{subfigure} 
    \hfill
    \begin{subfigure}[b]{0.32\textwidth}
        \centering
        \includegraphics[width=\scale\textwidth]{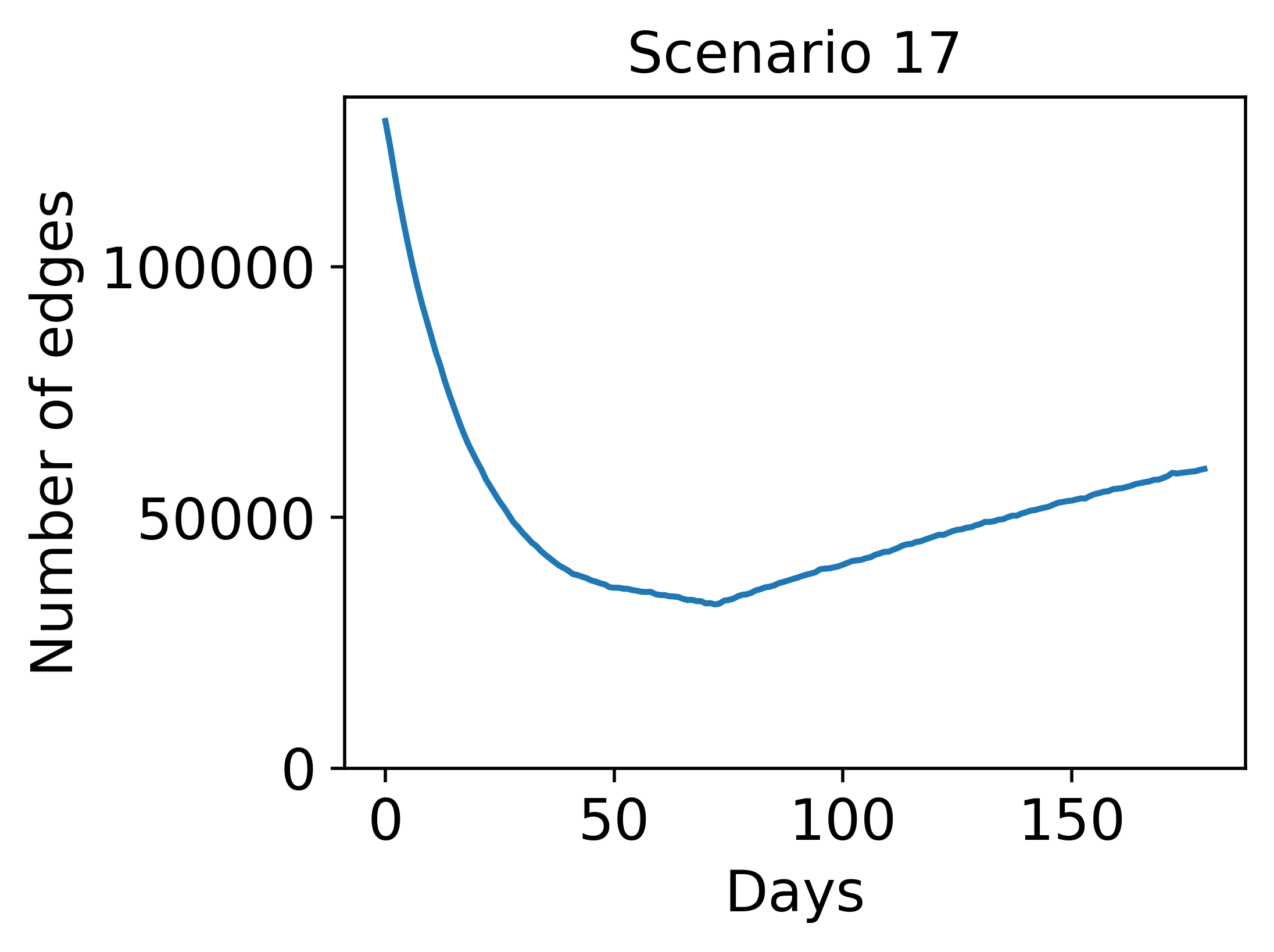}
        \caption{}
        \label{edge17}
    \end{subfigure} 
    \begin{subfigure}[b]{0.32\textwidth}
        \centering
        \includegraphics[width=\scale\textwidth]{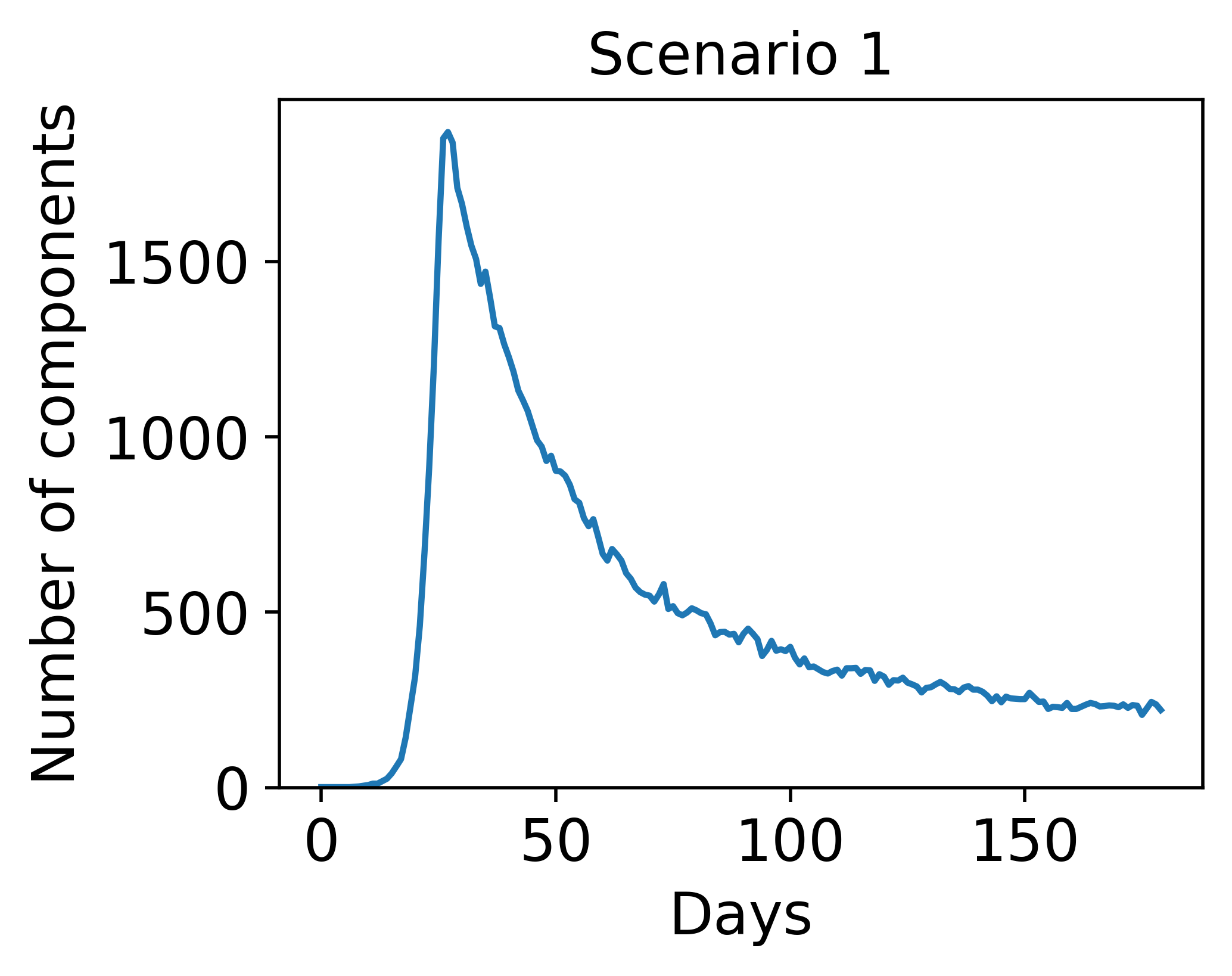}
        \caption{}
        \label{comp1}
    \end{subfigure}
    \hfill
    \begin{subfigure}[b]{0.32\textwidth}
        \centering
        \includegraphics[width=\scale\textwidth]{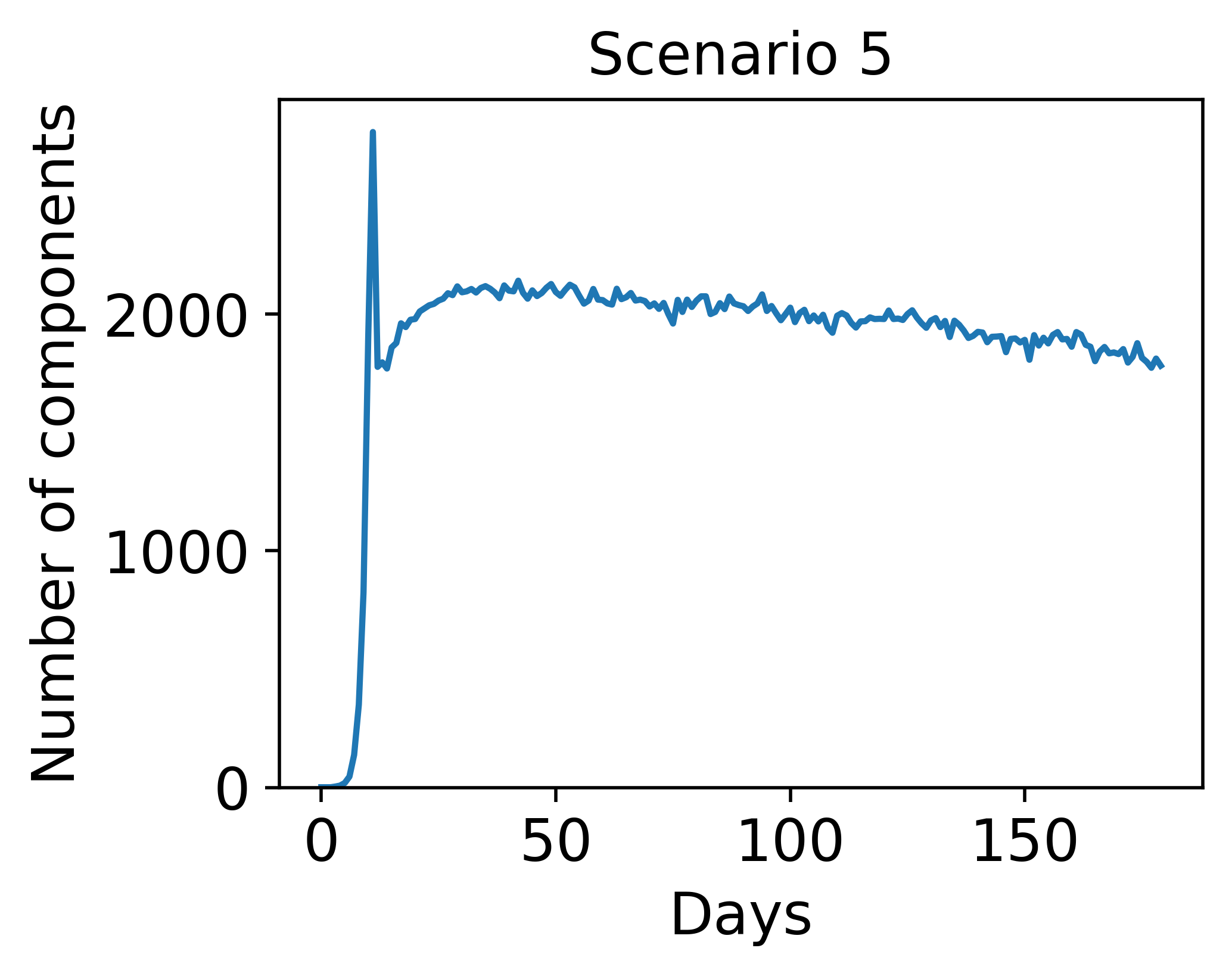}
        \caption{}
        \label{comp5}
    \end{subfigure}
    \hfill
    \begin{subfigure}[b]{0.32\textwidth}
        \centering
        \includegraphics[width=\scale\textwidth]{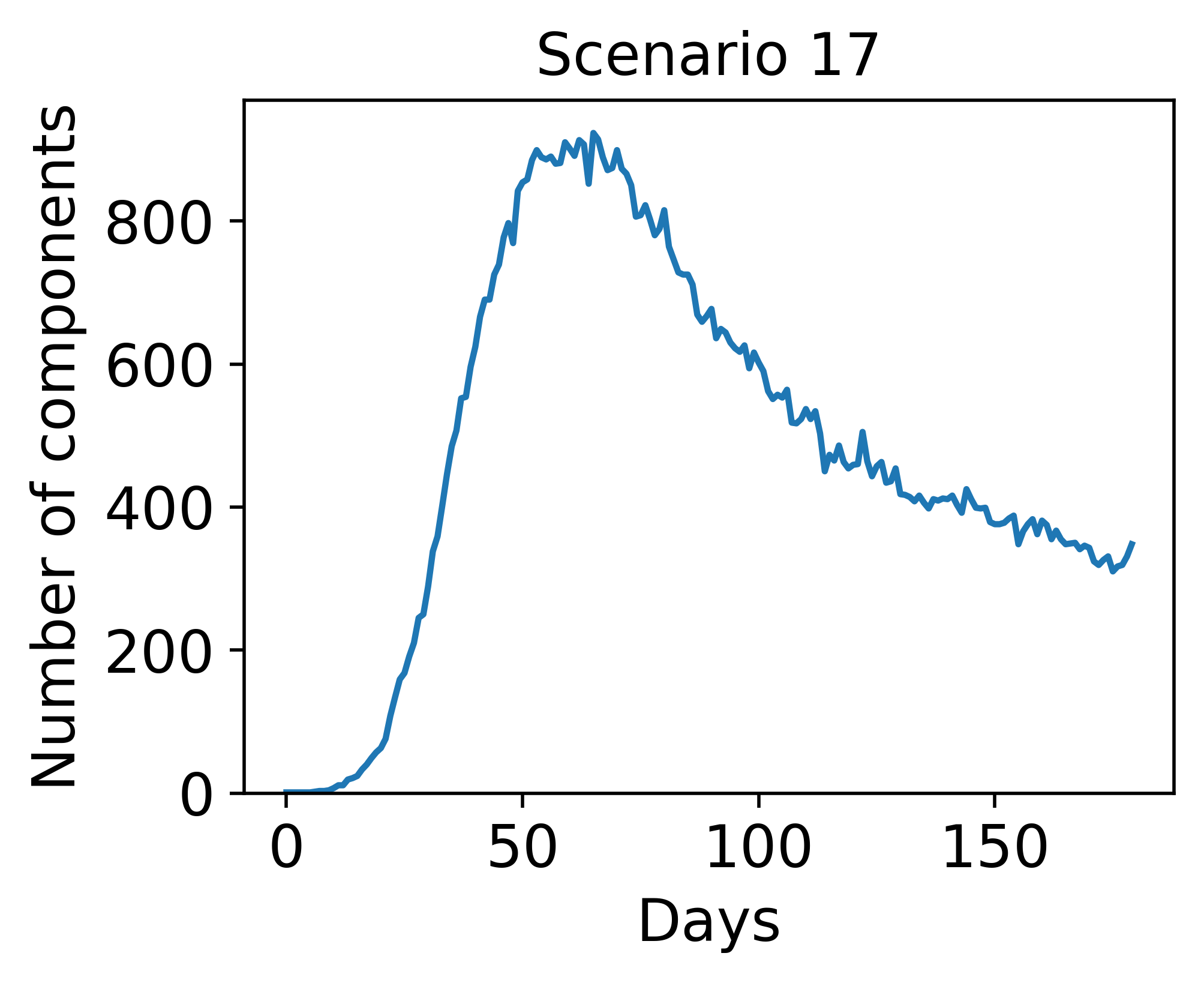}
        \caption{}
        \label{comp17}
    \end{subfigure}
    \caption{\textbf{Number of nodes, edges and components for various scenarios.} }
    \label{fig:features}
\end{figure*}

Since the synthetic population has no history, all agents are initialized with ``friends'' chosen randomly. This does not generally reflect friendship networks in real life, as friends are made through continued, personal interactions between people, but it coincidentally highlights the impact of agent actions on friendship networks. We assume that friendships require effort (represented as interactions here, formalized in Table \ref{table:parameters}) to maintain, and agents must spend time with friends in order to maintain that friendship. This allows agents to strengthen friendships by spending some part of a day with their friends, increasing the strength of that friendship by the amount of time spent together (see~\appref{abmapp}
for a deeper description). Thus, the initial, random connections do not receive the interactions between agents that are required, leading to a steep drop in the number of edges in the graph (see Figs. \ref{edge1}, \ref{edge5}, \ref{edge17}). The exact timing of this drop depends on the rate of friendship decay (Table \ref{table:parameters}, ``Decline Rate'') and the threshold chosen to binarize the social graph. The remaining behavior demonstrates long-term impact of agent actions on the social fabric of our synthetic society. 

To understand how agent actions impact network topology, we explore several parameters related to making, keeping, and losing friends (see Table \ref{table:parameters}). To make new friends, we assume that an agent needs to randomly interact with another agent on a repeated basis (Table \ref{table:parameters}: ``New Friend Interactions''). Once a friend is made, it is necessary to nurture that friendship, otherwise it eventually dissolves. Since we have no reference for how this is done in reality, we define two possible methods - linear daily decline in friendship strength, or a multiplicative decline that is inversely related to friendship strength (Table \ref{table:parameters}: ``Friendship Decline Function''). Additionally, we vary the rate that friendships decline to explore the impact on our social networks (Table \ref{table:parameters}: ``Decline Rate'' and ``Minimum Decline Rate''). In total, this combination of parameters creates 36 unique scenarios (see \tableref{runs} in \appref{runtable}). For robustness, we replicate all simulations 5 times (using different random seeds) and check for consistency between seed repetitions. As results were comparable, we focus on a single realization here for clarity.

To create a social graph at each time step, we threshold the friendship strength at a value of 0.3 to remove low-value relationships. More precisely, the values of the associated adjacency matrix are defined
\begin{displaymath}
A_{i,j} = 
\begin{cases}
1 & \text{if } \alpha_{ij} \geq 0.3, \\
0 & \text{if } \alpha_{ij} < 0.3   \\ 
\end{cases},
\end{displaymath}
where $\alpha_{ij}$ is the strength of the friendship between person $i$ and person $j$. We note that this research could be extended to evaluate the effect of different thresholds or to use weighted graph edges, but for simplicity we leave this for future work. 

\begin{table*} 
    \centering
    \begin{tabular}{|c|c|}
        \hline 
        Parameter & Values \\
        \hline\hline
        New Friend Interactions & 1, 3, 5, 8 \\
        \hline 
        Friendship Decline Function &  \(\displaystyle f_1(x)= x - dr\) or \(\displaystyle f_2(x)=x - dr \cdot max(mr, 1.0-x)\) \\
        \hline 
        Decline Rate ($dr$) & 0.025, 0.05, 0.075 \\
        \hline 
        Minimum Decline Rate ($mr$) & 0.05, 0.01 \\ 
        \hline 
        Seed & 5 seeds for each set of parameters \\
        \hline 
    \end{tabular}
    \caption{\textbf{Parameters Impacting the Friendship Network.} Friendships are built up over time, according to how often one agent interacts with another. The number of interactions required to make a new friend was tested from 1 (very friendly, a single interaction) to 8 (two agents must interact at least 8 times for a chance to become friends). Once two agents are friends, they must continue to interact or the strength of that friendship \textit{x} will begin to decrease (``Friendship Decline Function''). The friendship declined at one of three rates (``Decline Rate''), calculated daily, with a minimum rate (``Minimum Decline Rate'') specified for function $f_2()$.}
    \label{table:parameters}
\end{table*}

\subsection{Evaluating Networks} \label{ssec:network-comparison}

Each ABM run resulted in a dynamic network evolving over time. The friendship networks were saved at the end of each simulation day for analysis. For our experiments, each ABM run was made up of 180 simulation days and thus 180 individual social interaction graphs. These static networks were evaluated using the set of properties defined later in this section (\ssecref{graph-measures}: {\em Graph Properties}). These measurements, viewed in series, offered insight into the corresponding ABM run. To compare multiple ABM runs, we performed a pairwise comparison of graph measurements from the same time step (\ssecref{graph-metrics}: {\em Graph Comparisons}) then aggregated over all time steps (\ssecref{CompABM}: {\em Comparing ABM Runs}).

% Graph metrics for evaluation 
\subsubsection{Graph Properties}\label{ssec:graph-measures}
Networks formed by friendships within representative ABM scenarios were examined 
across timesteps using complementary graph properties. While we primarily use 
graph features for their interpretability (7 of 8 analyses), we include the 
spectral-based quantity, HCA, for its novel, orthogonal approach. The graph 
properties we opted to explore include: 

\noindent\textbf{Number of nodes} - Number of nodes in the network with at least one connection (isolated nodes are not included in this count). 

\noindent\textbf{Number of edges} - This is the total number of connections in the network ($|\mathcal{E}|$). 

\noindent\textbf{Number of components} -  A subgraph $S$ of $G$ is connected if for any two vertices $u$ and $v$ a path can be traversed from $u$ to $v$ along edges in $S$. A subgraph $S$ is considered a connected component if it is a maximal connected subgraph, {\em i.e.} if it is not contained in any other connected subgraph of $G$. 

\noindent\textbf{Clustering coefficient} - The local clustering coefficient, introduced in~\cite{watts_collective_1998}, is a measure of connectedness of a node to its neighbors. It is measured as the number of edges in a node's neighborhood (made up of a node and its immediate neighbors) divided by the number of possible edges in the neighborhood.

\noindent\textbf{Degree distribution} - The distribution of node degrees, $d$, over all nodes in $\mathcal{N}$. 

\noindent\textbf{Shortest path distribution} - The shortest path between two nodes is the smallest number of edges needed to traverse the graph from one node to another. The shortest path distribution is the distribution of shortest path lengths over all pairs of nodes. 

\noindent\textbf{Network portrait} - Defined in~\cite{bagrow_information-theoretic_2019}, the network portrait is a matrix $B$ whose $(l,k)^\text{th}$ entry is given by the number of nodes with $k$ nodes at distance $l$, where distance is defined as the number of edges traversed. This matrix has a number of interesting properties. For instance, the zeroth row of the network portrait gives the number of nodes in the network, the first row gives the degree distribution of the network, the second row gives the distribution of next nearest neighbors, and so forth. 

\noindent\textbf{Heat content asymptotics} - 
    Maclaurin series coefficients for a graph's heat content curve, which is the sum 
    of the entries in the matrix $e^{-s \mathcal L}$.

\subsubsection{Graph Comparisons}\label{ssec:graph-metrics}
Graph differences were quantified using distances or divergences between their 
respective measurements. Since our goal was to maximize the amount of 
information gained from changes in graph composition, we applied distances 
well-suited to the properties being considered. These comparison tools, and the graph 
quantities they were applied to, include:

\noindent\textbf{Wasserstein distance} - Often referred to as the ``earth mover's distance'' in computer science, the Wasserstein distance was used for comparisons involving the degree distribution or the shortest path distribution. Given two distributions $u$ and $v$, the Wasserstein distance can be interpreted as the amount of ``work'' needed to move distribution $u$ to match distribution $v$. Mathematically, this is defined 
\begin{displaymath}
    W_1(u,v) = \displaystyle\inf_{\pi \in \Gamma(u,v)} \int_{\mathbb{R} \times \mathbb{R}} |x-y| \; d\pi(x,y),
\end{displaymath}
where $\Gamma(u,v)$ is the set of probability distributions with marginal distributions $u$ and $v$ respectively. 

\noindent\textbf{Network portrait divergence} -  For comparisons of network portraits between networks, we use the network portrait divergence proposed by Bagrow and Bollt~\cite{bagrow_information-theoretic_2019}. This is a variation on the Jenson-Shannon divergence and is thus denoted $D_{JS}$ following the notation in~\cite{bagrow_information-theoretic_2019}. We direct the reader to the original paper for the full derivation.  

\noindent\textbf{Heat content asymptotics pseudometric} - In light of the work 
by McDonald and Meyers in which they demonstrated that two isospectral graphs 
may be distinguished from their respective heat content asymptotics, we 
introduce the HCA pseudometric $d_{\text{hca}}$ as the $\ell^2$ 
difference between two graphs' HCA.\footnote{$d_{\text{hca}}$ is not positive 
definite and therefore a pseudometric rather than a metric.} That is, for graphs 
$G_i$, $G_j$ with HCA 
$\text{hca}(G_i) = \mathop{\left\{ q_n^{(i)} \right\}}_{n\geq0}$ and 
$\text{hca}(G_j) = \left\{ q_n^{(j)} \right\}_{n\geq0}$, we define
\begin{displaymath}
    d_{\text{hca}} \big(G_i, \, G_j\big) 
        := \sqrt{ \sum_{n\geq 0} \left( q_n^{(i)} - q_n^{(j)}\right)^2 }.
\end{displaymath}
In practice, we use estimates of the first $N$ terms from each graph's heat 
content asymptotics to approximate their $d_{\text{hca}}$ difference, where for 
the results given in \secref{results}, we set $N = 20$. As we soon discuss,
we base our estimates of a graph's HCA on its corresponding heat content curve, 
which we estimate numerically using the method introduced by Lu {\em et al.} in 
\cite{lu_complex_2014}.

Extending Lu's method to estimate the heat content curve of a graph $G$, we 
treat $G$ as a graph domain embedded in a larger graph\footnote{See \appref{heatcontentapp}, 
or \cite{mcdonald_2003} for a definition of a graph 
domain.} in such a way that the boundary of $G$ consists of the 
$\texttt{num\_bndry}$ number of nodes of lowest degree, where 
$\texttt{num\_bndry}$ represents the maximum of either 1 or 2\% of the total 
number of nodes (rounded to the nearest integer). In the case where $G$ is not 
connected, we analyze each component independently and sum the resulting curves. 
The resulting heat content curve $\tilde q(s)$ for the entire graph is 
normalized by the total number of nodes in the graph's interior (vertex set 
minus boundary nodes).\footnote{Note that we use the variable $s$ in 
$\tilde q(s)$ to denote time propagation with regard to heat dissipation on a 
graph. We do so to avoid confusion in Subsection \ref{ssec:CompABM} and later 
where we use the variable $t$ to represent time when considering time evolving 
graphs.}

We then use $\tilde q(s)$ to estimate a graph's HCA by
computing an orthogonal polynomial representation 
for the graph's heat content curve on the interval $[0, 5]$, 
expanding this representation, and then extracting the monomial coefficients. 
The orthogonal polynomials $\{ p_n \}_{n=1}^N$ we use are generated by applying 
Gram-Schmidt to the monomial list $\{s^n\}_{n=1}^N$, where $N$ is set to the 
desired number of HCA coefficients. 
For our experiments, we set $N=20$ as noted previously. In numerically computing the 
$L^2$ inner product between $\tilde q(s)$ and a orthogonal polynomial $p_n(s)$, 
we uniformly sample both functions twenty times over the interval 
$[0, 5]$ and apply Simpson's rule to 
integrate the product of $\tilde q(s)$ with $p_n(s)$.
% \end{itemize}

\begin{figure*}[h!]
    \centering 
    \def\scale{0.6}
    \begin{subfigure}[t]{0.48\textwidth}
        \centering
        \includegraphics[width=\scale\textwidth]{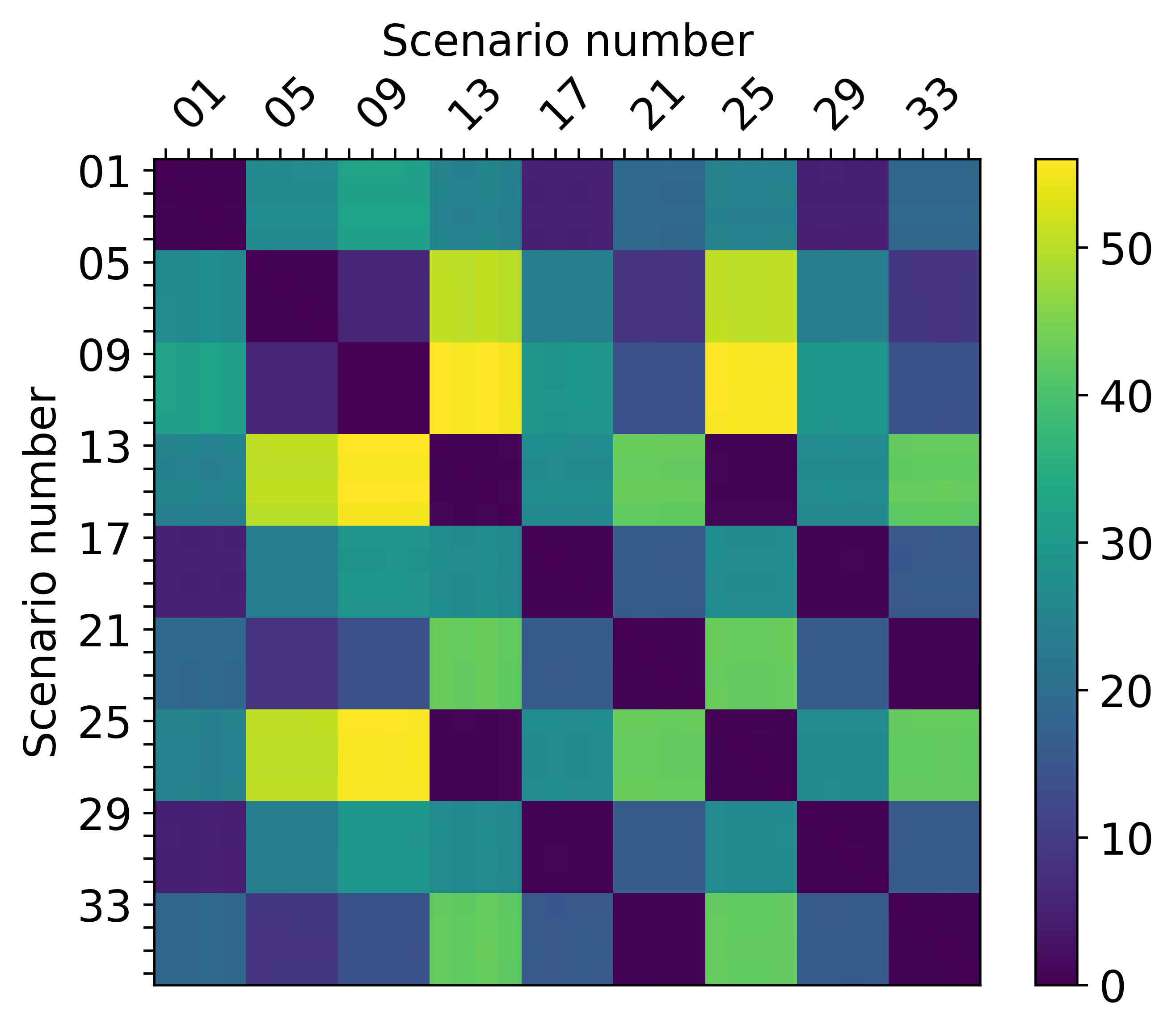}
        \caption{Comparison of degree distribution between runs.}
    \end{subfigure}
    \hfill
    \begin{subfigure}[t]{0.48\textwidth}
        \centering
        \includegraphics[width=\scale\textwidth]{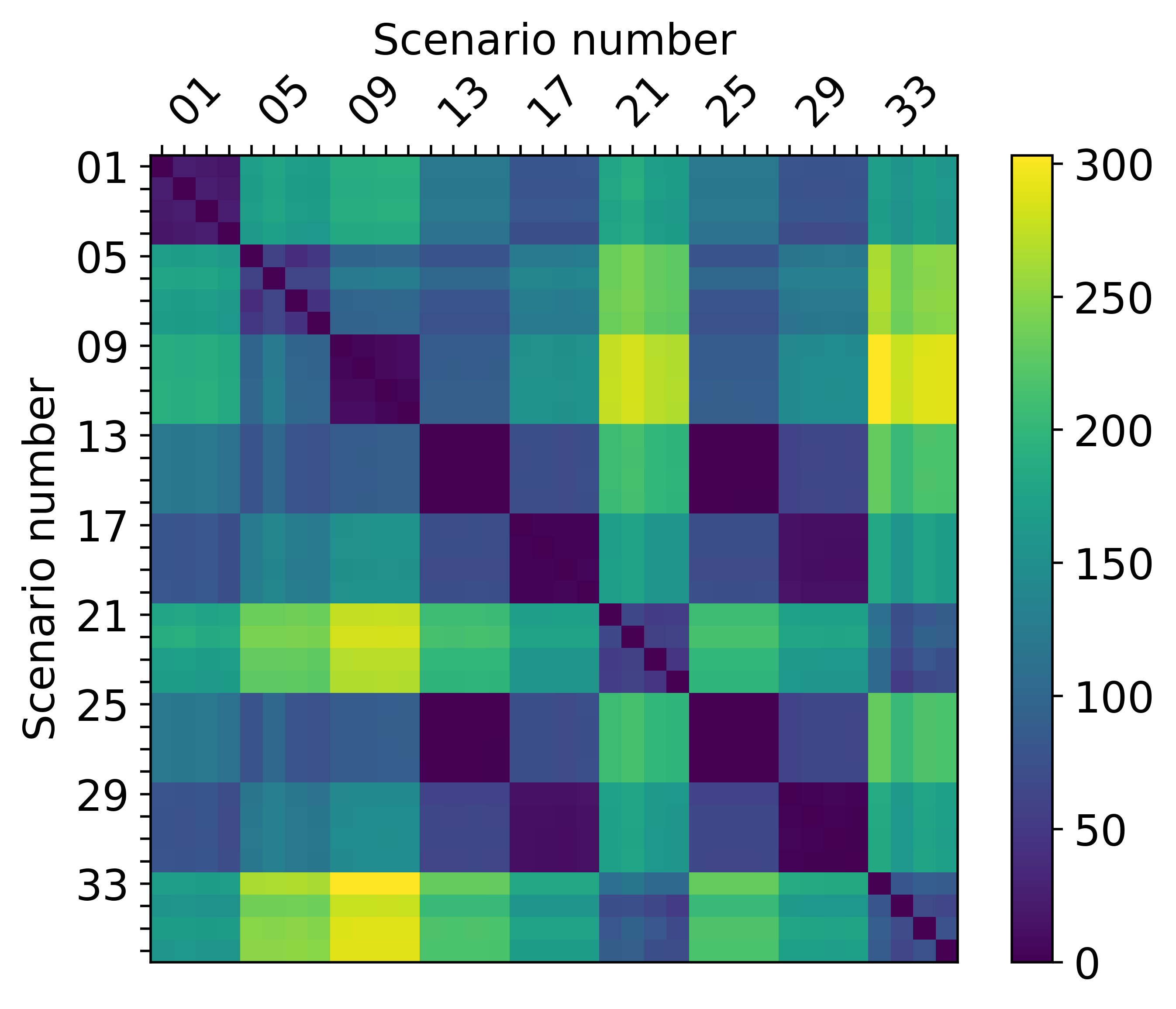}
        \caption{Comparison of shortest path distribution between runs.}
    \end{subfigure} \\

    \begin{subfigure}[t]{0.48\textwidth}
        \centering
        \includegraphics[width=\scale\textwidth]{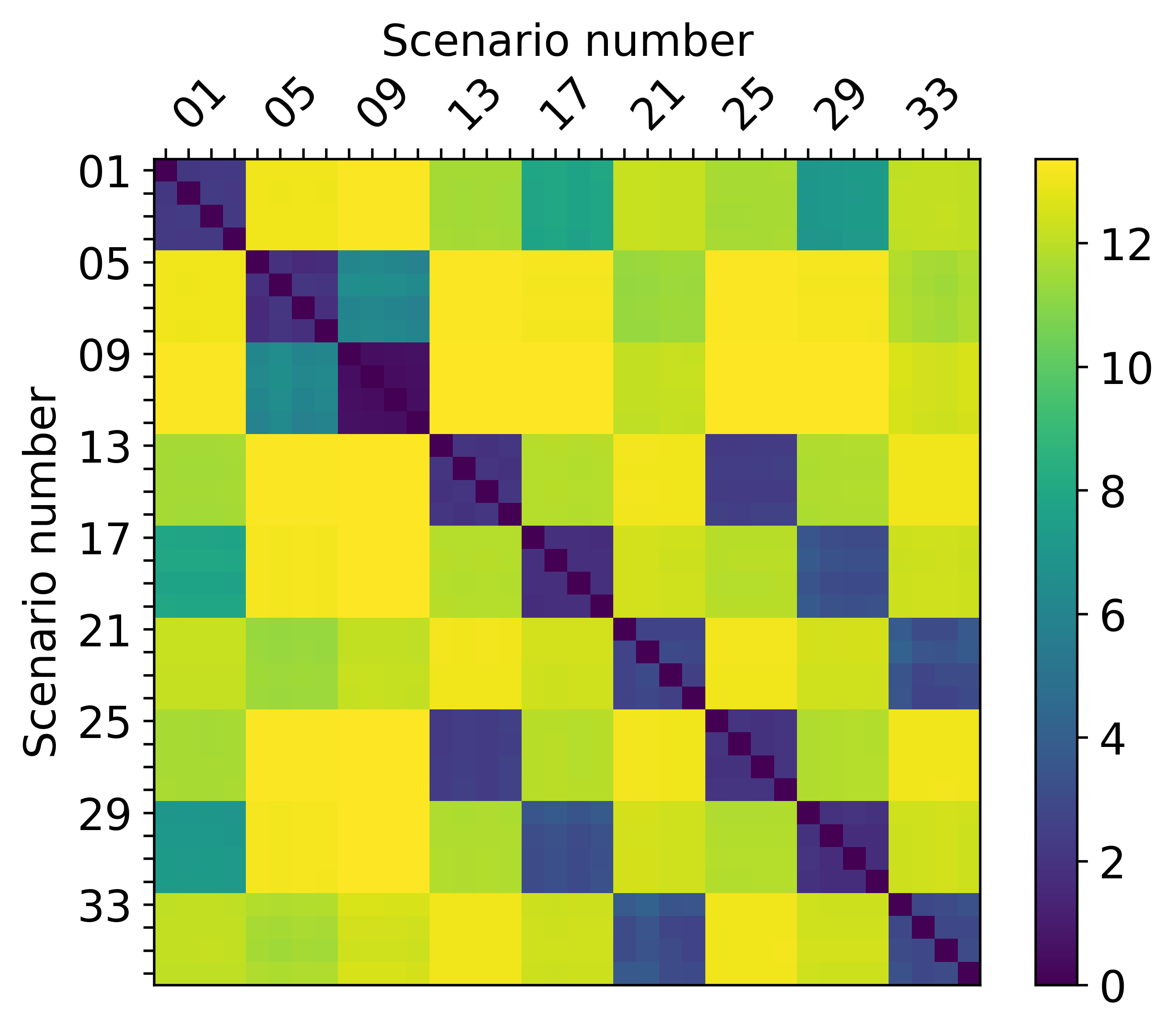}
        \caption{Network portrait divergence between runs.}
    \end{subfigure}
    \hfill
    \begin{subfigure}[t]{0.48\textwidth}
        \centering
        \includegraphics[width=\scale\textwidth]{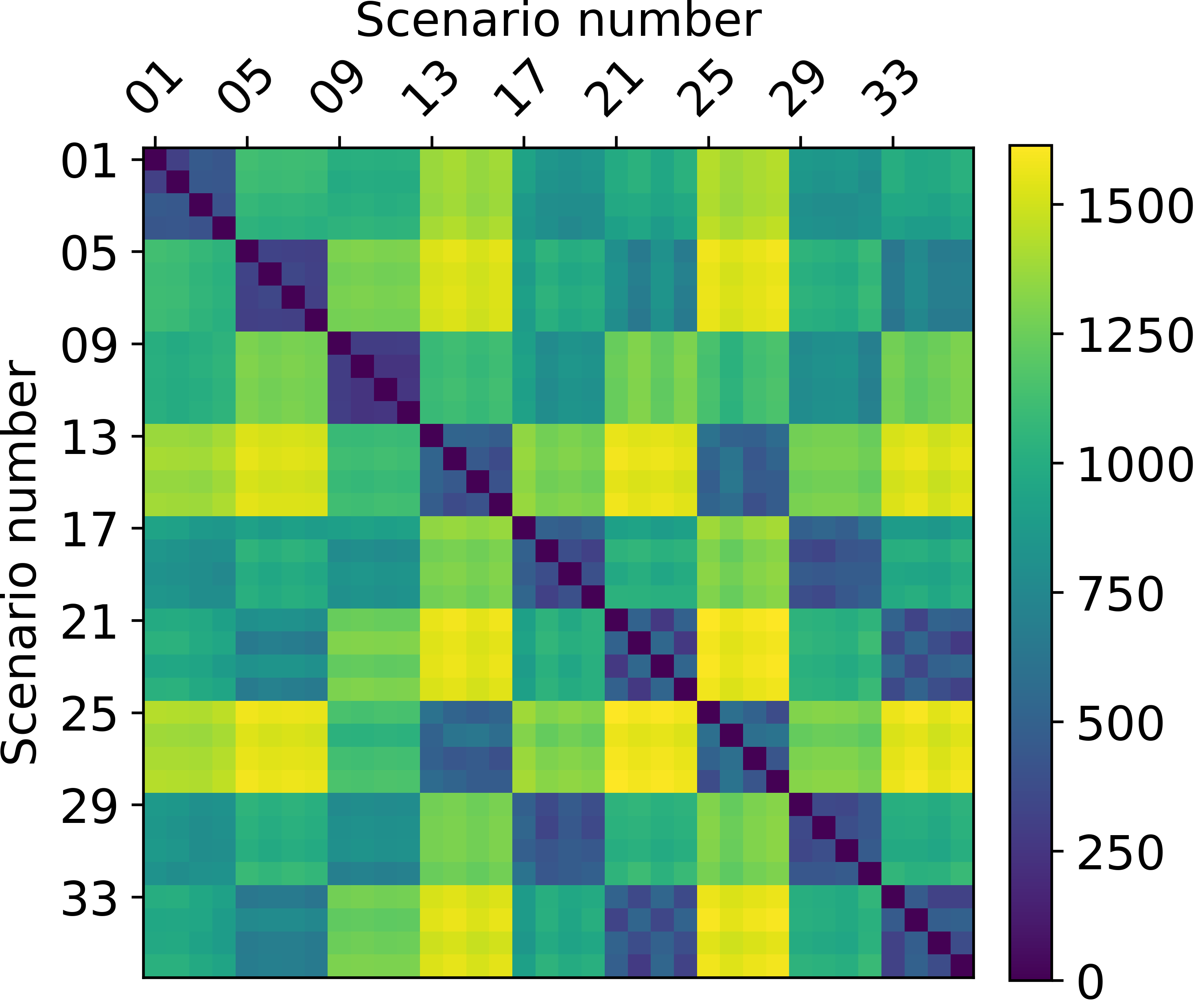}
        \caption{Heat content asymptotic difference between runs.}
        \label{sfig:hcbetweenpop}
    \end{subfigure}
    \caption{\textbf{Between-run comparison.} Confusion matrices comparing 
        degree distribution, shortest path distribution, and network portrait 
        between runs from different ABM scenarios. Comparisons are computed 
        based on \ssecref{graph-measures}. Each ``Scenario number'' corresponds 
        to the parameters defined in 
        Table~\ref{table:runs}. 
        The block structure 
        corresponds to ``New Friend Interactions'', demonstrating minimal dependence 
        on this parameter.}
    \label{fig:betweenpop}
\end{figure*}

\subsubsection{Comparing ABM Runs}\label{ssec:CompABM}
The network portrait divergence, Wasserstein distance, and HCA comparisons were 
used to compute the difference between the features of two 
\textit{individual networks}. To compare two ABM \textit{runs} we looked at the 
$2$-norm of the pairwise comparison of the graphs with corresponding time steps. 
Accordingly, we let $G_i(t)$ denote the social network associated with run $i$ 
at time step $t$, and respectively let $D_{i}(t)$, $S_i(t)$, and $N_i(t)$ be the 
degree distribution, shortest path distribution, and network portrait divergence
corresponding to $G_i(t)$. We then used the following comparisons as a measure 
of difference between two runs
\begin{align*}
    \hat{D}_{i,j} 
        &:= \left\| W_1\big(D_{i}(\cdot), \, D_{j}(\cdot)\big)\right\|_2, 
        \\[0.5 em]
    \hat{S}_{i,j} 
        &:= \left\| W_1\big(S_i(\cdot), \, S_j(\cdot) \big) \right\|_2, 
        \\[0.5 em]
    \hat{N}_{i,j} 
        &:= \left\| D_{JS} \big(N_i(\cdot), \, N_j(\cdot) \big) \right\|_2, 
        \\[0.5 em]
    \left(\hat{D}_{\text{hca}}\right)_{i,j}
        &:= \left\| d_{\text{hca}} \big(G_i(\cdot), \, G_j(\cdot)\big)  \right\|_2,
\end{align*}
where $W_1$ is the Wasserstein distance, $D_{JS}$ is the network portrait 
divergence, $d_\text{hca}$ is the HCA pseudometric, and the above 2-norms
are taken over the time step parameter $t$ for the discrete values 
$t = 0, \ldots, 179$ in accordance with ABM experiment runs.

%% file: sections/results.tex
\section{Results} \label{sec:results}

We find that temporal network changes are more easily observed with simplistic 
properties, such as number of nodes, edges and components. Other graph 
properties which return more complex information offer useful information when 
compared in aggregate (\textit{i.e.} degree distribution, shortest path 
distribution, network portrait, heat content asymptotics). We summarize the 
results of both types of comparisons for the purpose of evaluating sensitivity 
of parameters in Section~\ref{ssec:network-analysis}. Properties which are 
characteristic of social networks (\textit{i.e.} clustering coefficient and 
degree distribution) are evaluated for model validation in 
Section~\ref{ssec:model-validation}. Finally, in Section~\ref{ssec:utility} we
briefly present some insights gained from working with the four methods
introduced in Section~\ref{ssec:CompABM} for comparing time dependent graphs.

\subsection{Model Parameter Sensitivity}\label{ssec:network-analysis}

A preliminary step to model validation and parameter sensitivity analysis
was to ensure the simulation was consistent 
across seeds. Ideally, changing the seed should have a very small effect on the 
global network structure. For each scenario, the runs associated with each seed 
were compared using the techniques outlined above. When compared to the 
magnitude of the between-scenario comparisons it was found that the seed had 
little effect on the network structure in terms of degree distribution, shortest 
path distribution, or network portrait. While HCA comparison picks up more 
differences in network structure between different seeds in a scenario, those 
differences are still small compared to the cross scenario 
differences.\footnote{See \appref{addresults}
for an HCA comparison of different seeds for selected scenarios.} 

Since seed was found to have little effect on the model, sensitivity analysis 
on the other model parameters (listed in Table~\ref{table:parameters}) was 
performed by comparing a single seed from each scenario. The results are shown 
in Fig.~\ref{fig:betweenpop}. In each of the plots (based on degree distribution,
shortest path distribution, network portrait, and HCA) there is a clear block 
structure. This shows that the network structure is affected very little by the 
number of interactions required before ``befriending'' another agent 
(Table~\ref{table:runs}: ``New Friend Interactions''). 
Additionally, 
scenarios 13-24 show strong similarity to scenarios 25-36. Based on 
Table~\ref{table:runs}, 
this behavior indicates very little change in network 
structure due to the minimal decline rate for multiplicative friendship decline. 

To investigate the effect of the remaining two parameters, the rate of 
friendship decline and the functional form of that decline, network quantities 
were compared for the remaining non-redundant scenarios: 1, 5, 9, 13, 17, and 21. 
Representative plots using scenarios 1 and 5 are presented in the first two 
columns of Fig.~\ref{fig:features}. These scenarios are associated with a 
decline of 0.025 and 0.05. It can be seen from comparing these plots that a 
higher relationship decline rate results in a social graph with fewer edges and 
more connected components. This information can help fine-tune the friendship 
parameters. Notice that with a decline of 0.05 only about 8,000 of the 20,000 
people in the population end up having at least one connection. This indicates 
that the researcher may want to use a lower decline rate or a different 
functional form altogether to mimic a more realistic scenario.  

\begin{figure*}
    \centering 
    \def\scale{0.5}
    \begin{subfigure}[b]{0.32\textwidth}
        \centering 
        \includegraphics[width=\textwidth]{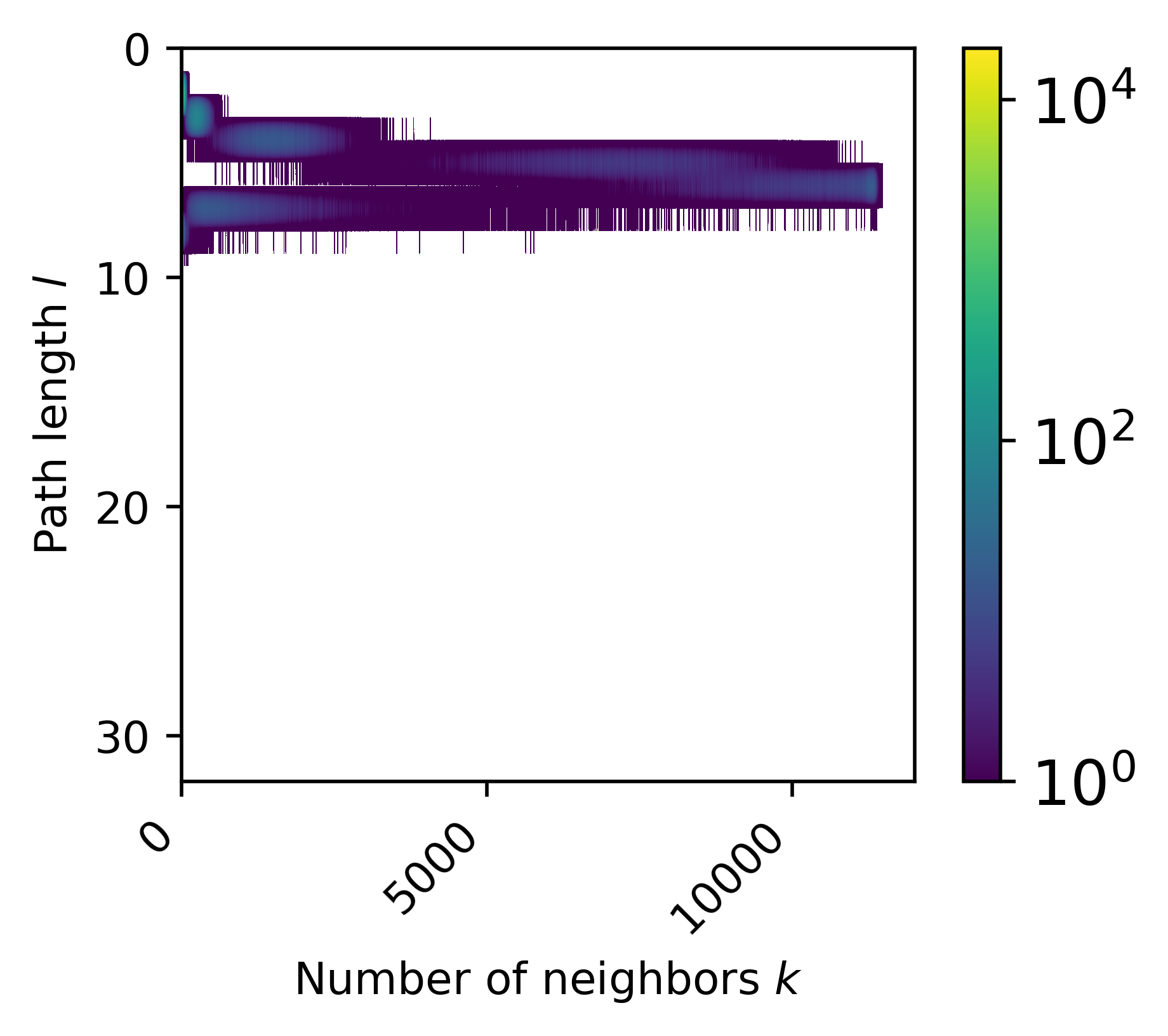}
        \caption{Network portrait at time 0.}
    \end{subfigure}
    \hfill
    \begin{subfigure}[b]{0.32\textwidth}
        \centering 
        \includegraphics[width=\textwidth]{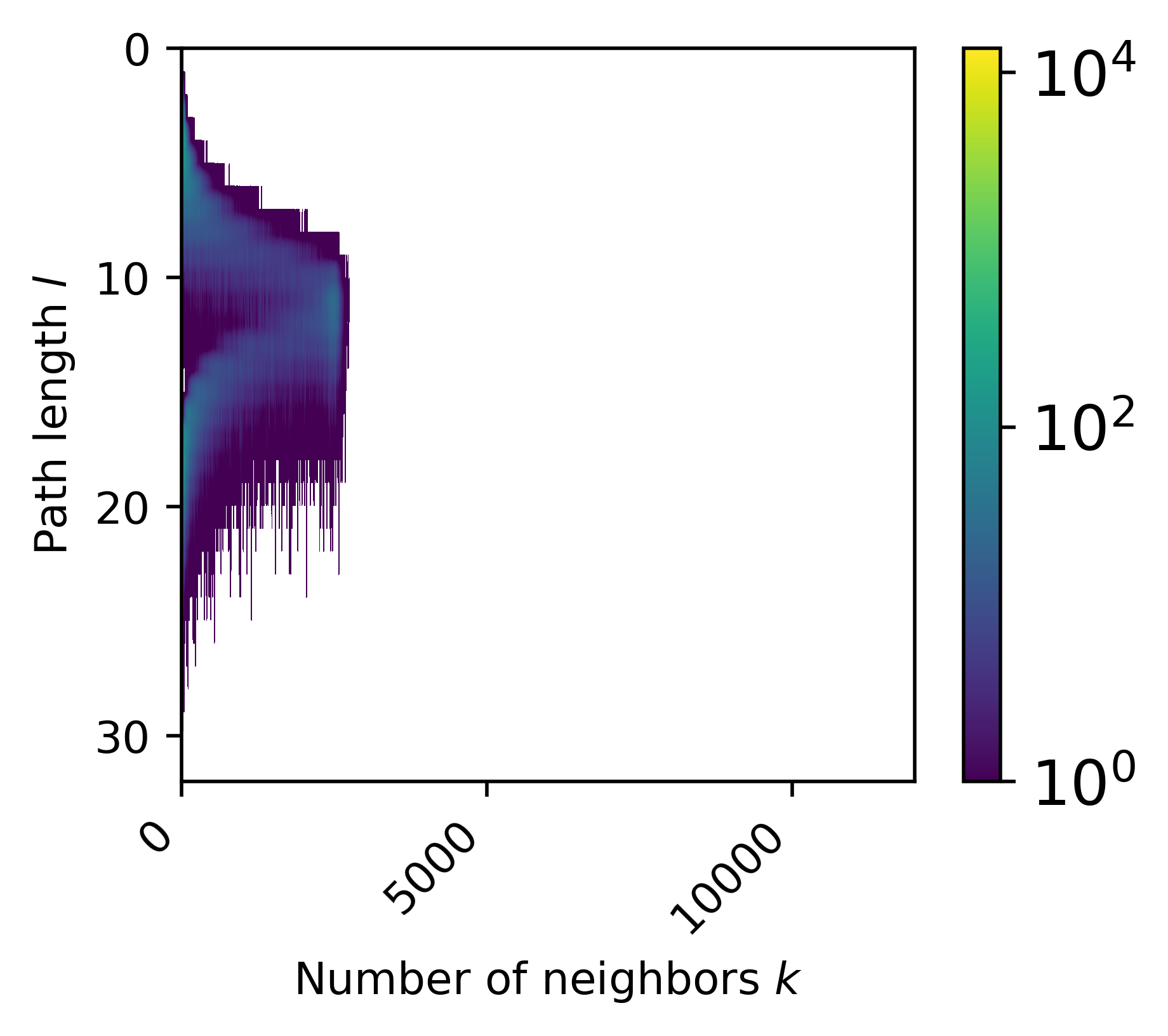}
        \caption{Network portrait after 90 days.}
    \end{subfigure} 
    \hfill
    \begin{subfigure}[b]{0.32\textwidth}
        \centering 
        \includegraphics[width=\textwidth]{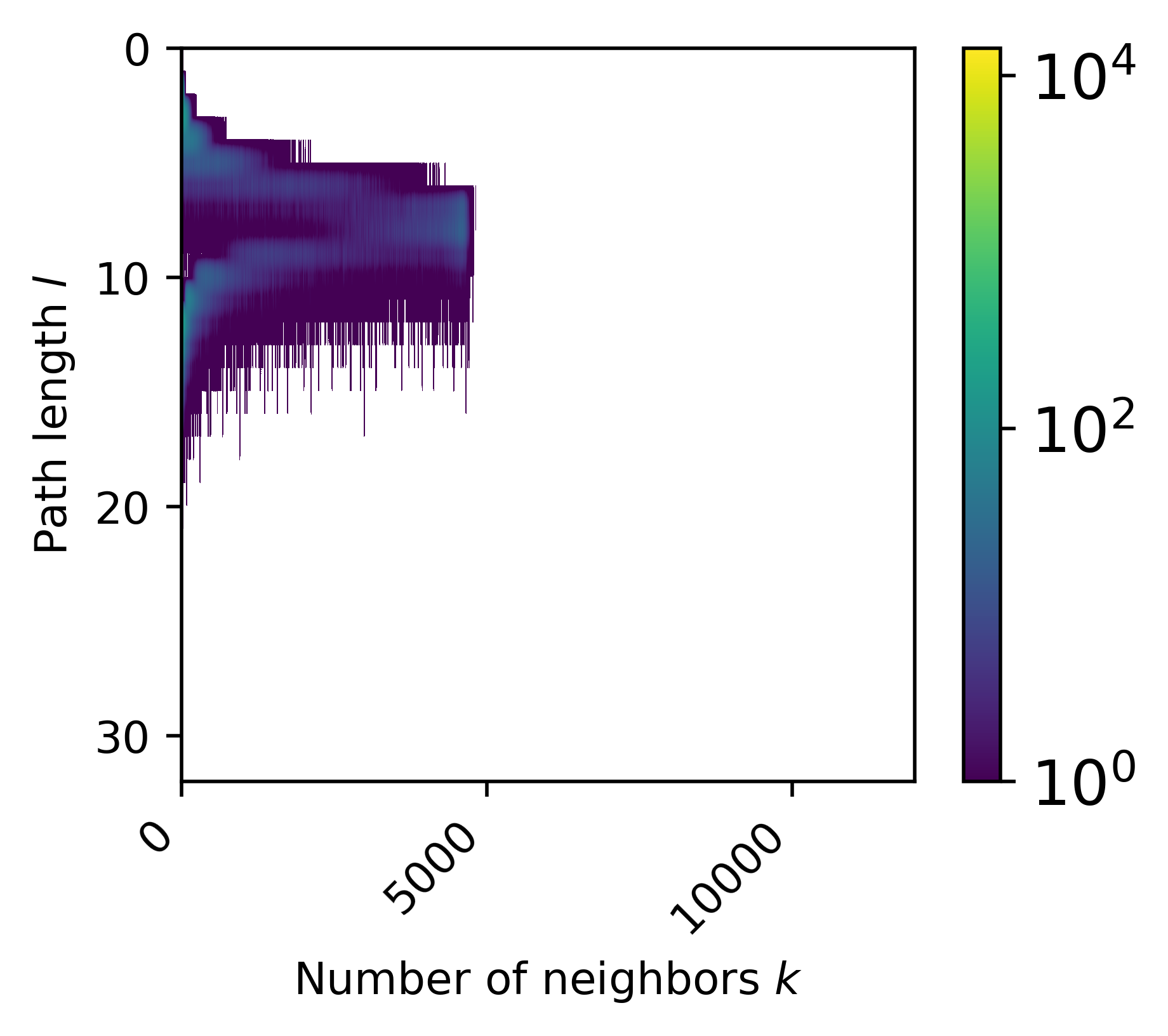}
        \caption{Network portrait after 179 days.}
    \end{subfigure}
    \caption{\textbf{Network portrait (Scenario 1).} The $(l,k)^\text{th}$ entry of the network portrait of a graph gives the number of nodes that have $k$ neighbors at distance $l$.}
    \label{fig:scenario1-portrait}
\end{figure*}

Viewing the graph quantities temporally as in Fig.~\ref{fig:features} also reveals 
a sharp decrease in the number of nodes and edges around day 25. The exact 
timing is an artifact of our parameterization\textemdash{}the friendship decay 
rate (Table~\ref{table:parameters}, ``Decline Rate'') and the threshold for 
binarizing the social network graphs\textemdash{}but the behavior demonstrates 
correct specification of the underlying social network mechanism in the 
simulation. Additionally, this is accompanied with a sharp increase in connected 
components. After this critical point, the nodes and edges increase and the 
number of connected components decreases. This pattern can be explained as the 
decaying and eventual removal of edges associated with the initial random social 
network imposed in the ABM. These edges are then replaced by new edges formed by 
the evolution of the ABM. This pattern can also be seen in the network portraits 
(shown for scenario 1 in Fig.~\ref{fig:scenario1-portrait}). At the beginning of 
the run ($\text{time} = 0$) most nodes were connected by relatively short paths 
(the $y$-axis has a small support). As the edges decay, nodes get farther apart 
in terms of path length. Toward the end of the run ($\text{time} = 179$) the 
path lengths are shorter once again.

The effect of using the multiplicative form of friendship decline can be observed in the last column of 
Fig.~\ref{fig:features} where results are included for scenario 17. We find that using the multiplicative decline function 
results in a dip in edges that is more gradual than with the linear decline 
function. In general, the multiplicative form seemed to be much less 
sensitive to the decline rate than the linear form. 

\begin{figure*} 
    \centering 
    \def\scale{0.9}
    \begin{subfigure}[t]{0.32\textwidth}
        \centering 
        \includegraphics[width=\scale\textwidth]{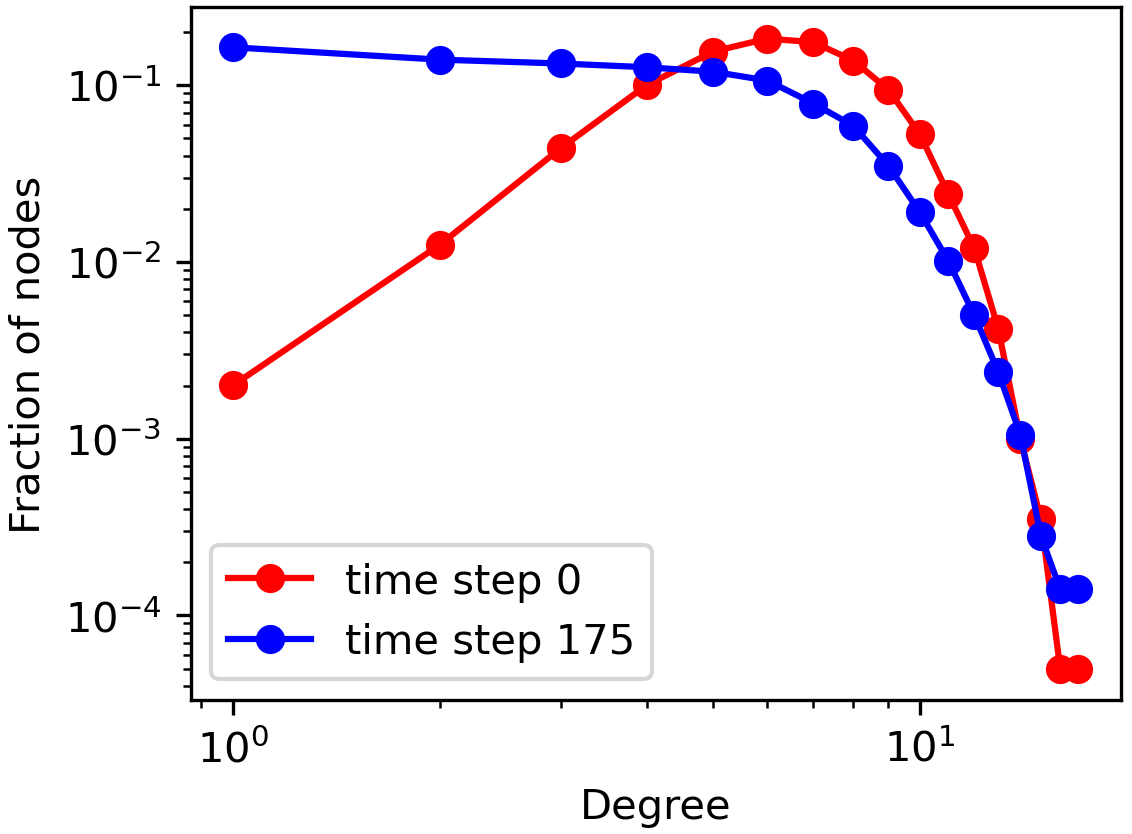}
        \subcaption{Log-log degree distribution}
        \label{fig:scenario1-degreedist}
    \end{subfigure}
    \hfill
    \begin{subfigure}[t]{0.32\textwidth}
        \centering 
        \includegraphics[width=\scale\textwidth]{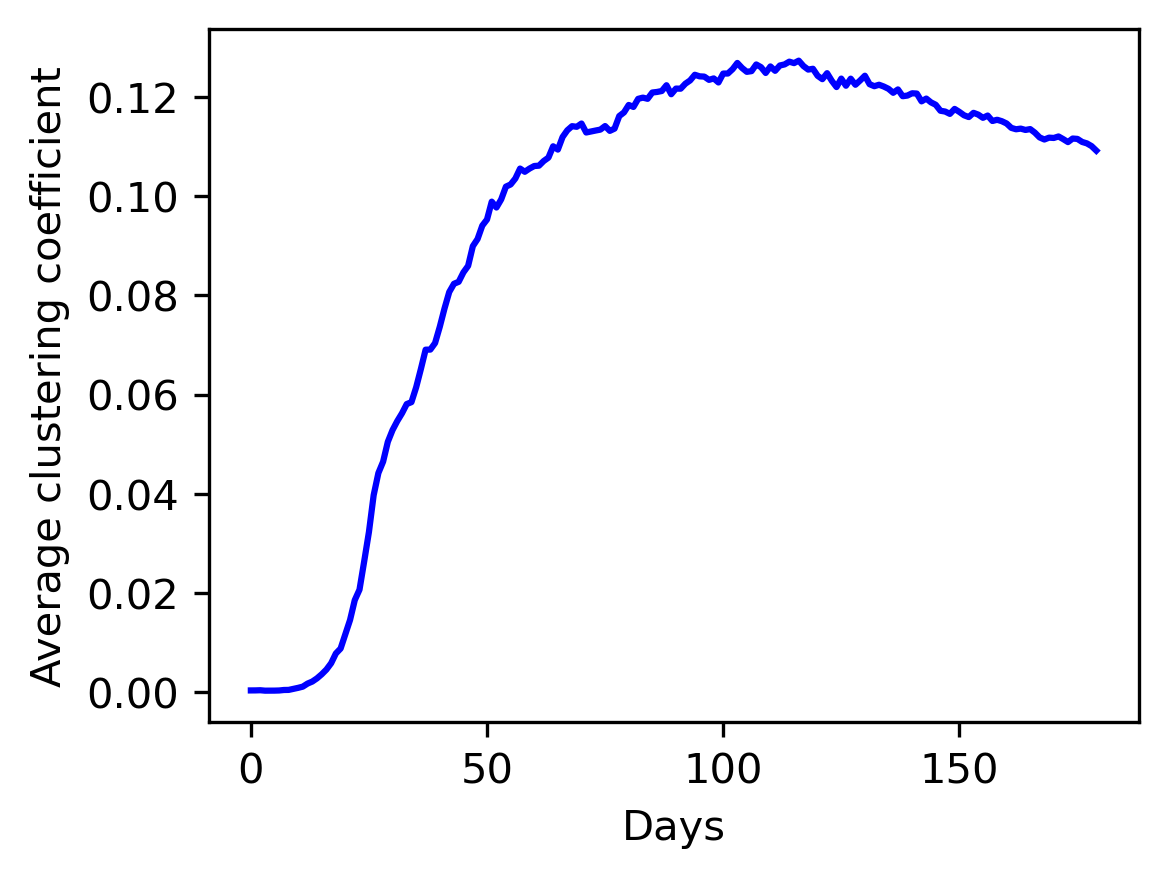}
        \subcaption{Clustering coefficients over time}
        \label{sfig:scenario1-clusteringcoeffs}
    \end{subfigure}
    \hfill
    \begin{subfigure}[t]{0.32\textwidth}
        \centering 
        \includegraphics[width=\scale\textwidth]{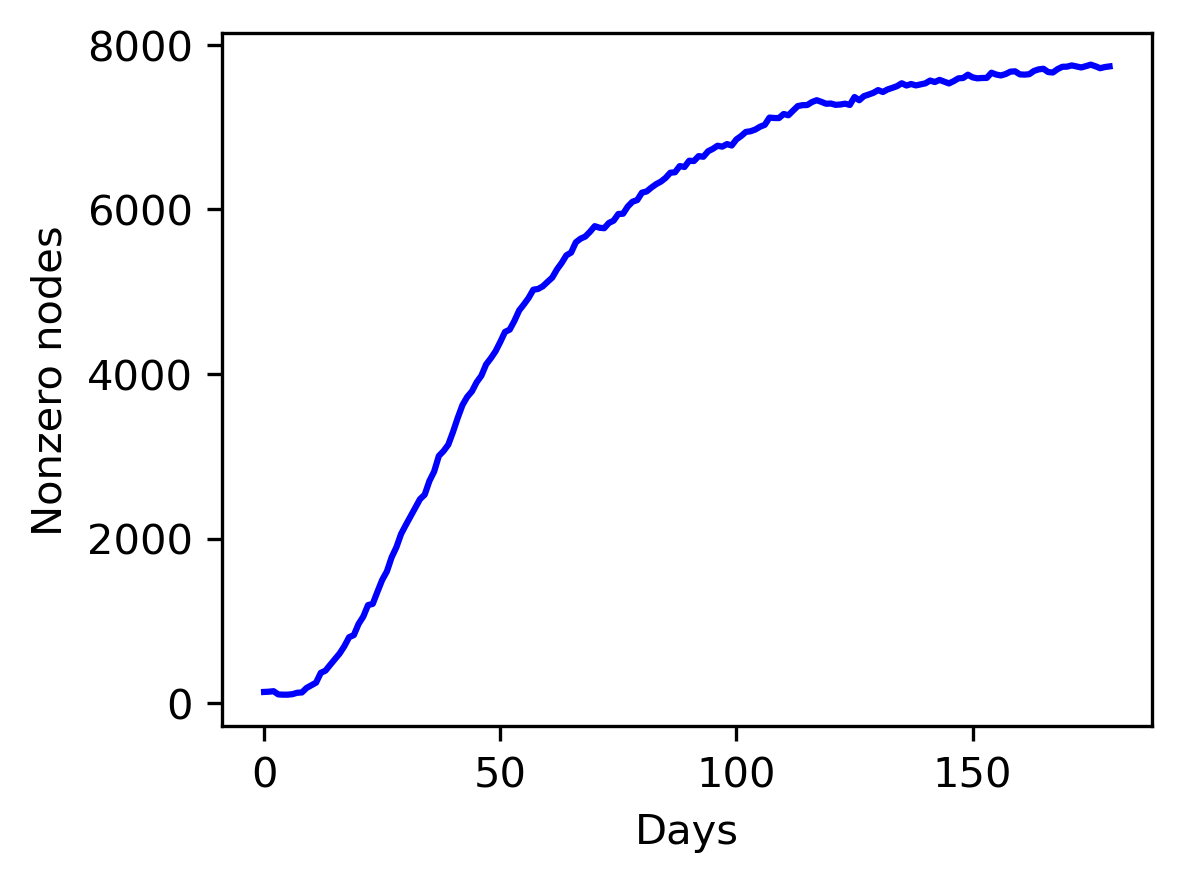}
        \subcaption{Number of nodes with nonzero clustering coefficients}
        \label{sfig:scenario1-clusteringnodes}
    \end{subfigure} 
    \caption{\textbf{Log-log degree distribution and clustering (Scenario 1).} 
    In (\protect\subref{fig:scenario1-degreedist}) we see the degree distribution moving toward a fat-tailed distribution characteristic of real-world networks between time step 0
    and time step 175. In 
    (\protect\subref{sfig:scenario1-clusteringcoeffs}) and (\protect\subref{sfig:scenario1-clusteringnodes})
    we see the average clustering coefficient of the ABM social graph nodes 
    increases over time, as do the number of nodes with nonzero clustering coefficient.
    The number of nodes with nonzero clustering coefficients is equivalent to the number of
    nodes with more than one connection.}
    \label{fig:scenario1-cluster}
\end{figure*}

\subsection{Model Validation}\label{ssec:model-validation}
The next step in model validation is to determine to what extent the formation 
of this new network mimics a real-world network. To do this, we examine the 
degree distribution and clustering coefficient of the networks over the course 
of the run. These plots are shown for scenario 1 in 
Fig.~\ref{fig:scenario1-cluster}, although the other scenarios show a similar 
pattern. While these networks do not satisfy a power-law distribution, we find 
that the degree distribution moves toward a more ``fat-tailed'' degree 
distribution over time (Fig.~\ref{fig:scenario1-degreedist}), which is 
characteristic of real-world networks~\cite{price_networks_1965}. Clustering 
also increases over time with both an increase in the average clustering 
coefficient (Fig.~\ref{sfig:scenario1-clusteringcoeffs}), and the number of 
nodes with nonzero clustering coefficient 
(Fig.~\ref{sfig:scenario1-clusteringnodes}). This provides validation that 
although these qualities were not explicitly coded into the ABM, the system 
evolved to portray characteristics of a real-world network over time. These 
observations are also important to the researcher as they can inform the need to 
run the model for a ``burn-in'' period to allow the connections to settle to a more realistic network before 
starting to make inferences.   

\begin{figure*}[h!]
    \centering 
    \def\scale{0.8}
    \begin{subfigure}[b]{0.32\textwidth}
        \centering
        \includegraphics[width=\scale\textwidth]{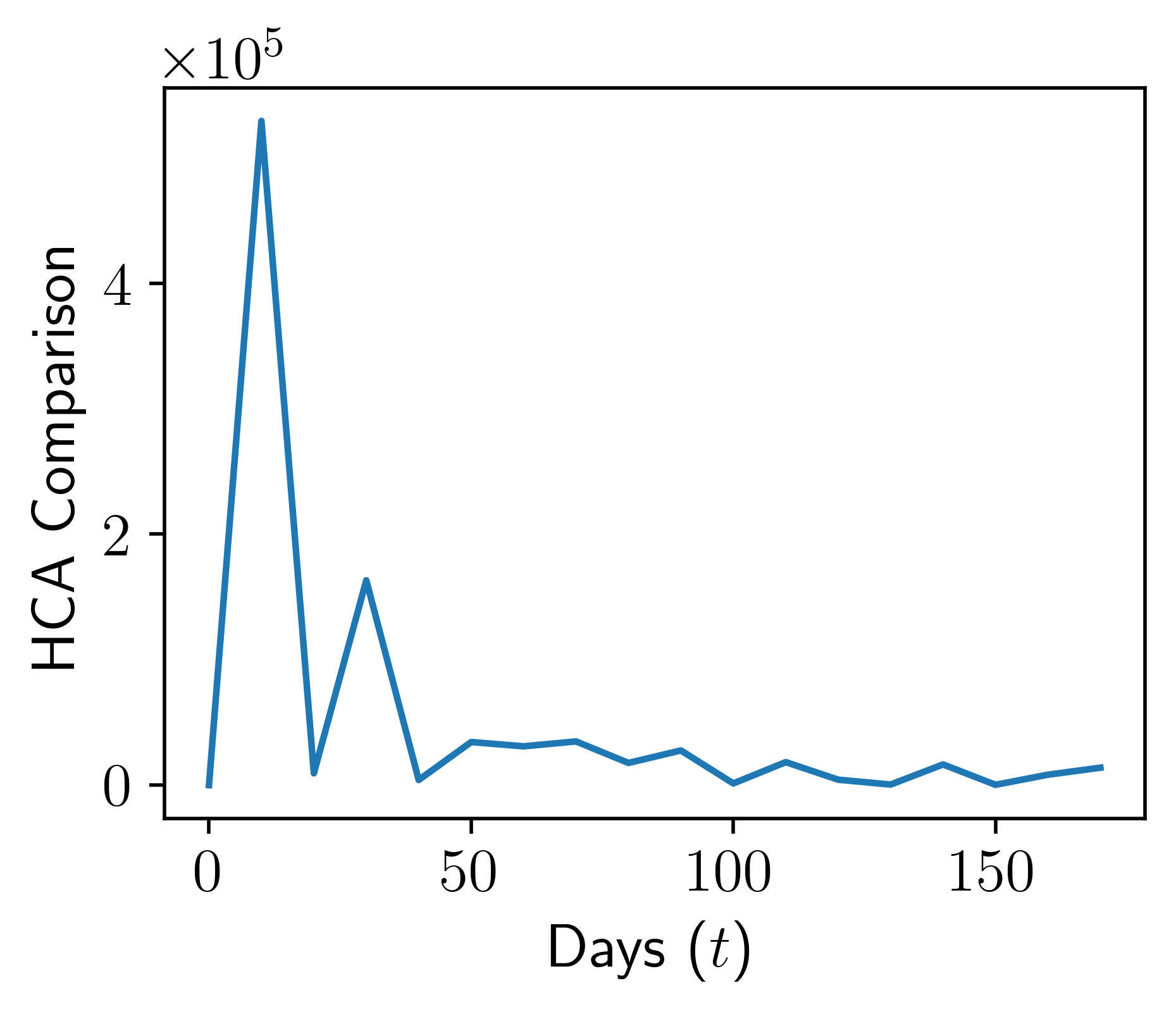}
        \caption{Scenario 1}
        \label{sfig:hca_pop01}
    \end{subfigure}
    \hfill
    \begin{subfigure}[b]{0.32\textwidth}
        \centering
        \includegraphics[width=\scale\textwidth]{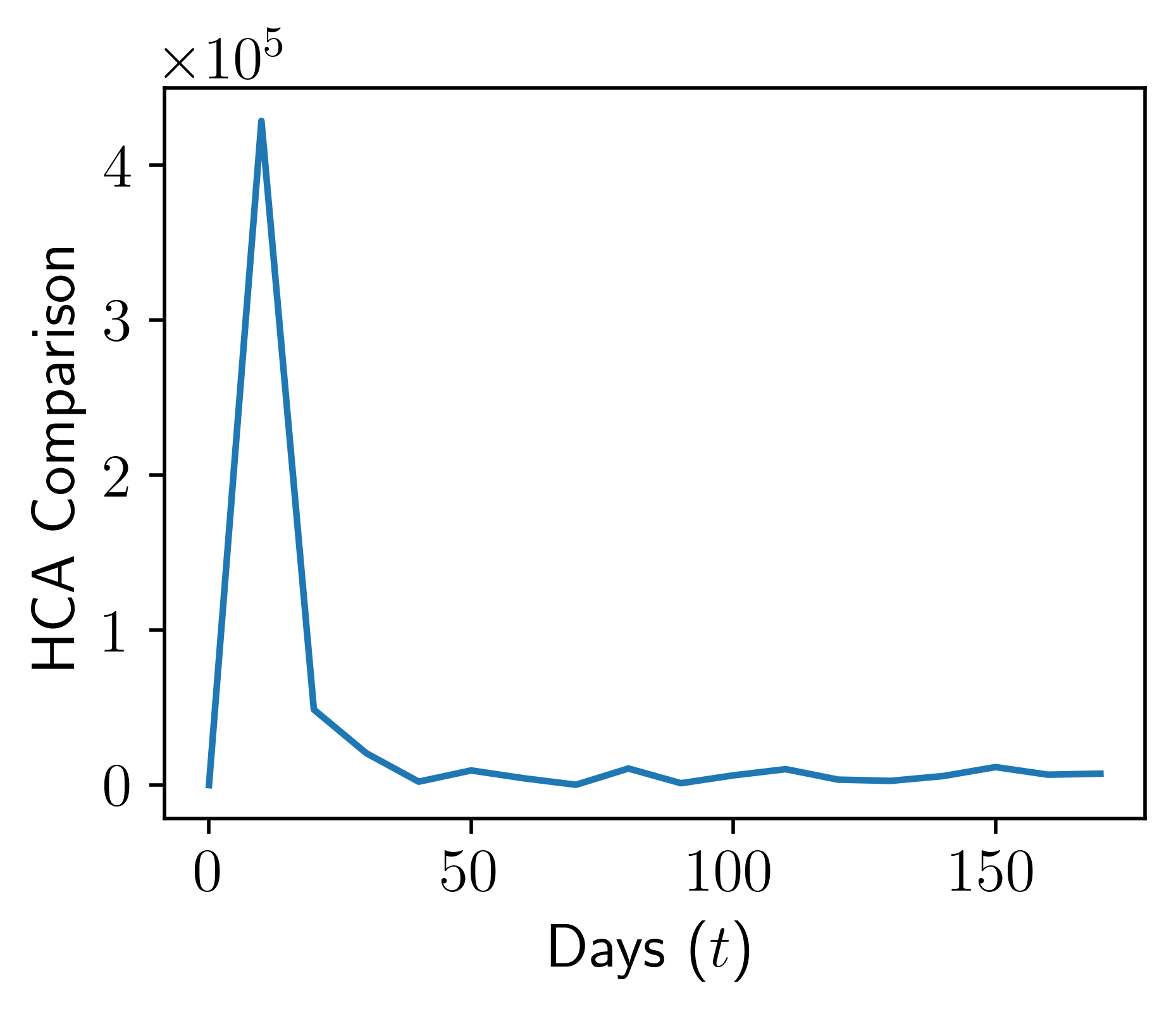}
        \caption{Scenario 5}
        \label{sfig:hca_pop05}
    \end{subfigure}
    \hfill
    \begin{subfigure}[b]{0.32\textwidth}
        \centering
        \includegraphics[width=\scale\textwidth]{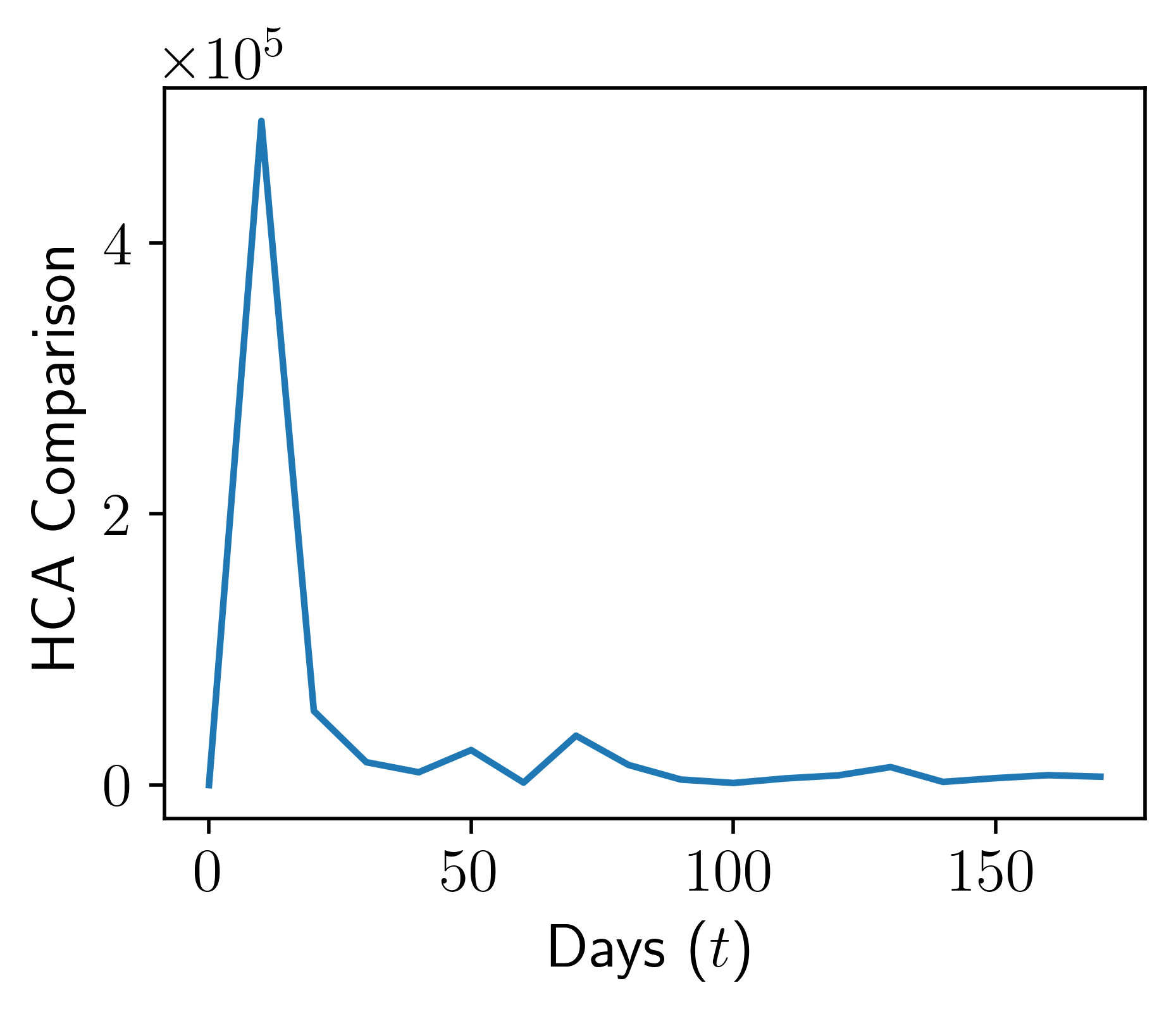}
        \caption{Scenario 17}
        \label{sfig:hca_pop17}
    \end{subfigure}
    \caption{\textbf{HCA difference between ABM friendship network at timestep $t$ and 
    $t-1$}. That is, if $G(t)$ represents a given scenario's friendship network 
    at timestep $t$, then the above are plots of the curve 
    $d_{\text{hca}}\big( G(t-1), \, G(t)\big)$. }
    \label{fig:hca_time_comp}
\end{figure*}

\subsection{Utility of Network Science Analytic Techniques}\label{ssec:utility}
In addition to their relevance for parameter sensitivity analysis, the plots in 
Fig.~\ref{fig:betweenpop} also provide insights into the usefulness of the four featured
graph comparison methods. The plots in 
Fig.~\ref{fig:betweenpop} corresponding 
to the shortest path distribution and HCA pseudometric show significantly more 
variation within each block when compared to the plots of the other two graph 
comparison methods, which indicates that the shortest path distribution and HCA 
pseudometric are more sensitive to network differences than either the degree 
distribution or network portrait. Additionally, both the network portrait and 
HCA plots in Fig.~\ref{fig:betweenpop} show distinct zero diagonals with 
strictly non-zero off-diagonal values, which is a requirement for the positive 
definite property of metrics.\footnote{Recall that for a graph comparison method 
to be positive definite, it must be zero when comparing two isomorphic graphs 
and non-zero when comparing two non-isomorphic graphs. As such, a graph 
comparison method which is positive definite can be safely relied upon to 
distinguish two non-isomorphic graphs. While neither the network portrait nor 
the HCA pseudometric are guaranteed to be positive definite, that their 
corresponding plots in Fig.~\ref{fig:betweenpop} each have a clear, distinct 
zero diagonal indicates they may be more useful in distinguishing similar, but 
distinct graphs than the other methods used in this paper.} Since the HCA-based 
graph comparison exhibits both of these qualities, in this scenario it is the 
best network comparison method out of the four used in this paper for 
distinguishing differences between networks. Where the HCA comparison is at a 
disadvantage to other the methods is that its results are less intuitive to 
interpret. Each of the other methods are directly based on observable graph 
features, whereas a graph's HCA is the product of its overall structure in a way 
that is less obvious.

Each of the graph features used for the analysis provided insight into different
aspects of the structure. For comparison of entire runs, the degree 
distribution, shortest path distribution, network portrait divergence, and HCA had 
some consistencies between them (block structure, similarity of scenarios 13-24 
and scenarios 25-36) although other pairwise comparisons were not consistent 
between the quantities. Viewing summary statistics about individual runs 
temporally (number of nodes, edges, and components) was helpful in understanding 
broadly how the structure changes over time. The network portrait and HCA 
comparison, while offering interesting information, were more convoluted to 
understand than the simpler quantities. Both methods also required significantly 
more computation time. Quantities such as the degree distribution, clustering, and 
shortest path distribution were helpful in relating the network structure to 
characteristics of real-world networks. The HCA pseudometric appears very 
sensitive to even slight changes in a network's structure, and, as such, is
ideally suited for detecting potentially important changes to an ABM's underlying 
network. The HCA's ability to detect structural differences was observed not 
only in the comparison of entire runs (as in Fig.~\ref{fig:betweenpop}), but 
also when comparing multiple time steps for the same scenario. Such HCA 
comparisons appear to reveal additional network structural changes not 
elucidated by other methods. For a comparison with Fig. \ref{fig:features}, 
we show in Fig. \ref{fig:hca_time_comp} plots of the HCA difference between
a scenario's friendship network at timesteps $t$ and $t-1$. In \appref{addresults},
we also provide specific plots of the HCA 
difference between a scenario's graph at time $t$ and its initial graph.

%% file: sections/conclusion.tex
\section{Conclusion} \label{sec:conclusion}

Network analysis of ABM-based social graphs provided key insights into the underlying model. Examining which scenarios produced significant changes in network trajectories allowed for elimination and fine-tuning of the various parameters based on their sensitivity. In practice, these analyses can help to narrow the number of scenarios run to only include those with substantially different behavior, resulting in computational savings. 

Analysis of the network structure over time also allowed for the discovery of a ``burn-in'' stage in which the initial random relationships decayed and more meaningful connections were established. While this process makes sense in theory, it was a revelation that was only discovered through the social network analysis. Verification that these new networks satisfied qualities common to real-world networks (i.e. fat-tailed degree distribution, clustering, and the small-world property) provided validation of the behavior and relationship formation models. 

Network analysis offered a collection of tools to evaluate various aspects of the graph structure. Even the most basic network statistics (such as number of nodes, edges, and components) gave insight into the change in structure over time. More sophisticated quantities, such as the network portrait or the heat content, while useful for certain applications, were less intuitive in their interpretation from a modeling perspective. 

There are many variations of social models and network quantities that can be evaluated in a similar manner. Additionally, the effect of including network edges weights or varying their threshold could be explored. Consequently, we do not intend for this paper to be comprehensive, but rather to establish a starting point for using network science to understand the relationship between agent-based simulations and networks they produce, and demonstrate the utility of employing ABM network analysis as a part of model validation.

%% file: appendices/heatcontent.tex
\section{Heat Content} \label{app:heatcontentapp}

In this section we provide an overview of the relevant mathematical 
background behind graph heat content curves, as presented in 
\cite{mcdonald_2003} and \cite{lu_complex_2014}. A detailed discussion 
of the Laplacian operator on graphs is available in \cite{lim_2020}. 
For the context of this paper, the graph (vertex) Laplacian is best thought of
as an operator on functions from a graph vertex set to an associated ring
(such as the real line) which is carefully defined to be a discrete equivalent
to the differential Laplace operator. 

In the context of following discussion, we use $G=(\mathcal N, \mathcal E)$ to 
represent a graph on vertex set $\mathcal N$ and edge set $\mathcal E$ without vertex or edge 
weights, denote by $C^0(G)$ to be the space of all real valued functions on $\mathcal N$, 
and define the graph (vertex) Laplacian $L: C^0(G) \to C^0(G)$ to be the 
operator given by
\begin{equation} \label{eq:LaplaceDefn}
    L f(v) = \sum_{w \in \mathcal N_G(v)} \Big[ f(w) - f(v) \Big]
\end{equation}
for $f \in C^0(G)$, where
$\mathcal N_G(v) := \big\{ w \in \mathcal N ~:~ (v, w) \in \mathcal E, \text{ or } (w, v) \in \mathcal E \big\}$
is the set of nearest neighbors of $v$ in $G$. 
This formulation of the vertex Laplacian is equivalent
to defining $L$ as $L:= \text{Deg}(G) - A_G$, where $\text{Deg}(G)$ and 
$A_G$ respectively denote the degree and adjacency matrices corresponding to $G$. 
Throughout this section we continue to use $\mathcal L$ to denote the normalized 
Laplacian\textemdash{}{\em i.e.} 
$\mathcal L := \text{Deg}(G)^{-1/2} \, L \, \text{Deg}(G)^{1/2}$.

%%===============================
%% Subsection: Heat Content Curve
%%===============================
\subsection{Graph Heat Content Curve}

The concept of a graph heat content curve as introduced in \cite{mcdonald_2003} 
and \cite{lu_complex_2014} is based solving the heat equation initial
value problem (IVP) with Dirichlet boundary conditions (DBC) on a graph. However, as
graphs lack an intrinsic notion of a boundary, a special class of subgraph, 
called a {\em domain}, which has a well defined interior and boundary
must be introduced. 

\begin{defn*}[{\bf Graph domain, interior and boundary nodes}]
    Suppose $D=(\mathcal N', \mathcal E')$ is a subgraph of a graph 
    $G=(\mathcal N, \mathcal E)$ ($\mathcal N' \subseteq \mathcal N$, 
    $\mathcal E' \subseteq \mathcal E$). A node (vertex) $u \in \mathcal N'$
    of $D$ is called an {\em interior node} (or vertex) of $D$ if every node
    $v \in \mathcal N$ adjacent to $u$ in $G$ is also adjacent to $u$ 
    in the subgraph $D$. That is, if
    \begin{displaymath}
        (u,v) \in \mathcal E \implies (u,v) \in \mathcal E'
    \end{displaymath}
    for every $v \in \mathcal N'$, then $u$ is an interior node of $D$. 
    
    The the interior $iD$ of $D$ is collection of all interior nodes of $D$, 
    and the boundary $\partial D$ of $D$ is the collection of all 
    non-interior nodes of $D$ ({\em i.e.} $\partial D := \mathcal N' \setminus iD$).
    Elements of $\partial D$ are referred to as boundary nodes of $D$.

    A subgraph $D$ of $G$ is called a {\em domain} of $G$ provided
    $iD \ne \emptyset$.
\end{defn*}

As discussed in \cite{mcdonald_2003}, there is a natural way to define
the graph (vertex) Laplacian $L_D$ corresponding to a domain 
$D = (\mathcal N', \mathcal E')$ of a graph $G=(\mathcal N, \mathcal E)$ based 
on an induced inclusion map. 
More precisely, if we denote $C^0(G)$ to be the space of all real valued 
functions on $\mathcal N$ and define
\begin{displaymath}
    C^0(G, D) := \{ f \in C^0(G) \, : \, f|_{\mathcal N\setminus \mathcal N'} = 0 \},
\end{displaymath}
then there are natural isomorphisms 
$J_D\, :\, C^0(D) \to C^0(G, D)$
which induce an inclusion map 
\begin{equation}
    I_D\, : \, C^0(D) \to C^0(G)
\end{equation}
and a projection map
\begin{displaymath}
    P_D\, : \, C^0(G) \to C^0(D).
\end{displaymath}
In this context, $L_D$ is simply defined to be
\begin{displaymath}
    L_D := L I_D
\end{displaymath}
Further, under the basis $\{\delta_v\}$, where $\delta_v$ denotes
the indicator function corresponding to a vertex $v \in V'$, 
the domain Laplacian $L_D$ has the block 
form representation 
\begin{equation}\label{eq:laplaceblock}
    L_D = 
        \begin{pmatrix}
            L_{iD, iD} & L_{\partial D, iD} \\
            L_{iD, \partial D} & L_{\partial D, \partial D} \\
        \end{pmatrix}
\end{equation}
where an operator of the form $L_{*, *'}$ is determined by
\begin{equation}\label{eq:iDLaplacian}
    L_{*, *'} = P_{*'} L_{D} I_*.
\end{equation}
Given the importance of the operator $L_{iD, iD}$ in the rest of this 
section, we follow the example in \cite{mcdonald_2003} and introduce the
shorthand $L_{iD} := L_{iD, iD}$. Further, since \eqref{eq:iDLaplacian} in 
conjunction with \eqref{eq:LaplaceDefn} imply $L_{iD}$ satisfies
\begin{equation*}
    L_{iD}f(v) = \sum_{w \in \mathcal N_{iD}(v)} \Big[ f(w) - f(v) \Big]
\end{equation*}
for $f \in C^0(iD)$, the operator $L_{iD}$ is the vertex Laplacian 
corresponding to the interior $iD$ of the domain $D$.

With the introduction of graph domain and corresponding Laplace operators, 
the heat equation IVP with DBC on a domain $D$ of a graph $G$ can be stated as
\begin{align}\label{eq:heativp}
    \left\{
        \begin{aligned}
            \mathcal L h 
                &= - \frac{\partial h}{\partial s} &&\text{ on } iD \times (0, \infty) \\
            h(v, 0) 
                &= 1 &&\text{ on } iD \\
            h(v, s) 
                &= 0 &&\text{ on } \partial D
        \end{aligned}
    \right. 
\end{align}
A solution $h$ to \eqref{eq:heativp} is 
\begin{align} \label{eq:heatsoln}
    h(u, s) 
        &= e^{-s \mathcal L_{iD}} 1_{iD}(u) \nonumber \\
        &= \sum_{n=0}^\infty \frac{s^n}{n!}\mathcal L_{iD}^n 1_{iD}(u),
\end{align}
where $1_{iD}$ denotes the indicator function on the interior $iD$ of the domain $D$.

Since a normalized vertex Laplacian is known to be diagonalizable, assuming 
$|iD| = n$, there exist eigenvalues $\lambda_1 \leq \ldots \leq \lambda_n$
and corresponding eigenvectors $\phi_1, \ldots, \phi_n$ of $\mathcal L_{iD}$ so that
$\mathcal L_{iD}$ can be decomposed as 
\begin{equation}\label{eq:LiDdecomp}
    \mathcal L_{iD} = \Phi \Lambda \Phi^{-1}
\end{equation}
where
\begin{align*}
    \Phi &= \begin{pmatrix} \phi_1~\ldots~\phi_n \end{pmatrix} \\
    \Lambda &= \text{diag} \begin{pmatrix} \mathop{\{\lambda_i\}}_{i=1}^n \end{pmatrix}.
\end{align*}
Denoting the rows of $\Phi^{-1}$ as $\mathop{\{\pi_i\}}_{i=1}^n$
allows us to write the matrix $e^{-t\mathcal L_{iD}}$ as 
\begin{align*}
    e^{-s\mathcal L_{iD}} 
        &= \Phi e^{-s \Lambda} \Phi^{-1} \\
        &= 
            \begin{pmatrix} 
                \displaystyle
                \sum_{i=1}^n e^{-\lambda_i s}\phi_i(v) \pi_i(w) 
            \end{pmatrix}_{v, w \in iD}
\end{align*}
As such, \eqref{eq:heatsoln} becomes 
\begin{equation} \label{eq:hclosedform}
    h(u, s) = \sum_{v\in iD} \sum_{i=1}^n e^{-\lambda_i s}\phi_i(u) \pi_i(v).
\end{equation}

\begin{defn*}[{\bf Heat content asymptotics}]
    Let $G$ be a graph and $D$ be a domain of $G$ with non-empty boundary. 
    Let $h(u, s)$ be as in \eqref{eq:heatsoln}, then
    the heat content $q(s)$ of $D$ as a function of time is defined 
    to be the sum of $h(u, s)$ over all vertices in the interior $iD$ of the 
    domain $D$. That is
    \begin{equation}\label{eq:hc}
        q(s) 
            := \sum_{u \in iD} h(u, s) 
            = \sum_{u, v \in iD} \sum_{i=1}^{|iD|} e^{-\lambda_i s}\phi_i(u) \pi_i(v)
    \end{equation}
    Alternatively, one can define $q(s)$ can be the sum of the entries of the matrix
    representation for the operator $e^{-s \mathcal L_{iD}}$. As noted in 
    \cite{mcdonald_2002} and \cite{mcdonald_2003}, $q(s)$ has the representation
    \begin{equation*}
        q(s) = \sum_{n\geq 0} q_n s^n.
    \end{equation*}
    McDonald and Meyers refer to the coefficients $\{q_n\}_{n\geq 0}$ as a graph domain's 
    {\bf heat content asymptotics}, and use the notation
    $\text{hca}(G)$ to denote the set of heat content asymptotics corresponding to the 
    graph $G$ \cite{mcdonald_2002, mcdonald_2003}.
\end{defn*}

%%======================================================================
%% Subsection: Approximating the Heat Content Curve Through Random Walks
%%======================================================================
\subsection{Approximating the Heat Content Curve Through Random Walks}

Noting the computational cost of computing the spectrum for a large matrix, 
Lu {\em et al.} introduced in \cite{lu_complex_2014} a method for 
approximating \eqref{eq:hc} based random walks on the graph which does not
require computing the eigendecomposition of the matrix $\mathcal L_{iD}$. In what 
follows, we present a derivation of Lu's method.

Let $M := \text{Deg}(iD)^{-1} A_{iD}$ denote a transition matrix on 
the interior $iD$ of a graph domain $D$, where $\text{Deg}(iD)$ and
$A_{iD}$ respectively denote the degree and adjacency matrices 
corresponding to $iD$. Observe that since 
$L_{iD} =  \text{Deg}(iD) - A_{iD}$, the interior domain random walk 
Laplacian $L_r := \text{Deg}(iD)^{-1} L_{iD}$ can be written as
\begin{equation*}
    L_r 
        = \text{Deg}(iD)^{-1}\big[\text{Deg}(iD) - A_{iD}\big] 
        = I_{|iD|} - M,
\end{equation*}
where $I_{|iD|}$ represents the $|iD| \times |iD|$ identity matrix.
Further, the relation
\begin{equation*}
    L_r = \text{Deg}(iD)^{-1/2} \mathcal L_{iD} \text{Deg}(iD)^{1/2} 
\end{equation*}
allows us to additionally conclude 
\begin{equation}\label{eq:Lrdecomp}
    L_r 
        = 
            \left[ \text{Deg}(iD)^{-1/2} \Phi \right]
            \Lambda 
            \left[ \text{Deg}(iD)^{-1/2} \Phi \right]^{-1},
\end{equation}
where $\Phi$ and $\Lambda$ are as in \eqref{eq:LiDdecomp}.

Define
\begin{displaymath}
    M_L := (1 - \delta) I_{|iD|} + \delta M = I - \delta L_r
\end{displaymath}
to be the lazy random walk transition matrix with transition probability
$0<\delta\leq1$. The 
probability distribution $P_s$ of a walker at time $s$ in terms of its initial 
distribution $P_0$ is given by $P_s = \left(M_L\right)^k P_0$. Parameterizing time $s$ by 
$s = k \delta$ we find
\begin{equation*}
    \left(M_L\right)^k = \left( I_{|iD|} - \frac{s}{k} L_r \right)^k \to e^{-s L_r},
    \qquad \text{as } k \to \infty.
\end{equation*}
Thus, in light of \eqref{eq:Lrdecomp} the element $\left(M_L\right)^k(v, w)$ of 
the matrix $\left(M_L\right)^k$ corresponding to nodes $v, w\in iD$ satisfies
the limit
\begin{equation*}
    \lim_{k\to\infty} \left(M_L\right)^k(v, w) 
        = 
            \sum_{i=1}^{|iD|} 
                e^{-\lambda_i s} \phi_i(v) \pi_i(w) 
                \sqrt{\frac{d_w}{d_v}},
\end{equation*}
where $d_v$ and $d_w$ are used to represent the respective degrees of nodes
$v, w \in iD$. That is, for large values of $k$, the matrix element 
$\left(M_L\right)^k(v, w) \sqrt{ d_v / d_w }$ approximates the corresponding
element in the matrix representation for $e^{-t \mathcal L_{iD}}$. Recalling that
the heat content $q(s)$ of a domain $D$ is sum of entries in the matrix corresponding 
to the operator $e^{-s \mathcal L_{iD}}$, we obtain the following the approximation:
\begin{equation}\label{eq:qapprox}
    q(s) \approx \tilde q(s) 
        := \sum_{v, w \in iD} \left(M_L\right)^k(v, w) \sqrt{\frac{d_v}{d_w}},
\end{equation}
for $k \gg 1$.

%% file: appendices/abmdescription.tex
\section{ABM Setup and Simulation} \label{app:abmapp}

This section provides a more detailed discussion on the population setup and 
simulation procedure for the agent-based model. The first section covers the synthetic 
population generation that accurately replicates an actual population in terms of demographic 
data (including age, sex, race, family size, and income) as well as building 
location and use (houses, offices, malls, groceries, etc.). The second section 
explains the daily activity of individual agents and where interactions that may 
lead to new friendships arise.

%%===============================
%% Subsection: Synthetic Population Generation
%%===============================
\subsection{Synthetic Population Generation}

Initialization of a SIM begins with population setup. The simulated population needs 
to reflect the population of interest for the social or behavioral analysis to 
be valid. Such population characteristic include:
\begin{itemize}
    \item Demographic characteristics: The basic information about the structure of the population in that region, for example age, gender, and/or race distribution. Provided by the US Decennial Census \cite{u_g_census_decennial_2020} or yearly Census ACS \cite{noauthor_american_2020}. 
    \item Social Organization: This includes information such as household composition: adults, children, race and gender of head of household, and family structure. Provided by the US Decennial Census \cite{u_g_census_decennial_2020} or yearly Census ACS \cite{noauthor_american_2020}. 
    \item Socio-economic characteristics: The employment status, individual and/or family income, type of job or industry sector. Block level estimates can be obtained from ACS \cite{noauthor_american_2020}. 
    \item Buildings' type and placement: Homes, businesses, hospitals, schools and other education institutions, malls and shopping centers, and social buildings like gyms, restaurants, pubs, and places of worship. Some of these can be extracted from OSM \cite{openstreetmap_contributors_open_2022} shapefiles for the region of interest, while others must be researched or confirmed with local authorities / experts. 
    \item Work / study characteristics: The Longitudinal Employer-Household Dynamics \cite{noauthor_longitudinal_2020} information on commuting to and from jobs (LODES), Job-to-Job flows (J2J), and Post-Secondary Employment Outcomes (PSEO). These allow the use of statistical methods to assign employment or (ongoing) education level pursued by any agent in the simulation. Note that schools and upper education institutions enrollment is a necessary information here. 
\end{itemize}

Combining these sources in a consistent manner to create a synthetic population 
with desired characteristics is not trivial. The problem is essentially that of 
creating joint distributions at a given granularity level (city, neighborhood, 
census block group, census block).  

\begin{figure}[h]
    \centering 
    \def\scale{1.0}
    \includegraphics[width=\linewidth]{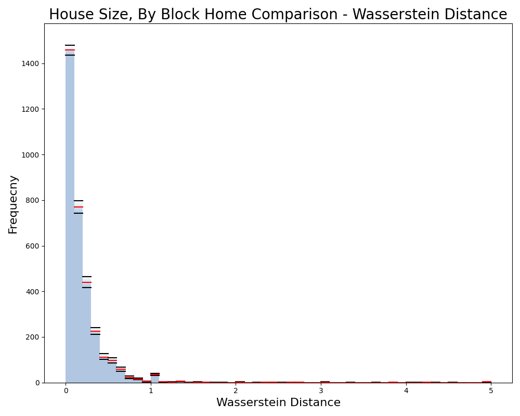}
    \caption{\textbf{Household Size Distribution.} After matching the gender, age, and race of each synthetic agent to the marginal distributions of a real population at the block level, we use the Wasserstein Distance to measure the discrepancy of our household sizes against the Decennial Census (kept out during the population generation). Orange is the median and black are the 95\% quantiles.}
    \label{fig:cinciwass}
\end{figure}

We developed a method that builds a population exactly fitting the population 
demographics - any joint demographic information is first exhausted and then 
further marginals are treated as independent. We then construct households of 
the exact type supplied, and then sort the population into households in such a 
way as to be maximally consistent with the known characteristics of the household. 
Using generally accurate data sources like the US Census \cite{u_g_census_decennial_2020} 
leads to tighter margins on the match between the generated population's characteristics 
and the public ones. However, we can also target different population demographics 
(e.g., the number of humans in an area) with custom distributions. Jobs are assigned 
to agents based on demographics and location using the LODES datasets. Schools 
can be assigned either based on neighborhood and age range, or school capacity 
for schools drawing from an entire region. Populations are validated against 
additional data held back from the creation process. A sample of the population 
used in this study is shown in Figure~\ref{fig:cinciwass}. Further discussion 
can be found in \cite{mobius_2022}.

Code for the population generation, in Python, is available upon reasonable request.

%%======================================================================
%% Subsection: Approximating the Heat Content Curve Through Random Walks
%%======================================================================
\subsection{Simulation Execution}

After generating population statistics, the simulation uses the particular demographic 
profile of each agent to setup a simulation specific to them. This setup involves 
logging the home, office, and/or school that an agent will use for the 
duration of the simulation. Homes and offices or schools (as appropriate) are 
provided by the population generation, but malls are chosen to be 
close to the home. Bars and other entertainment buildings are picked randomly during 
runtime, to simulate groups making decisions to go out for an evening. 

The simulation begins at midnight on the first day, and all agents begin at home 
(even second-shift workers). Agents with full-time day jobs or in school full-time 
get up get up between 0600 and 0800. On the first day, they calculate routes from 
home to work/school and back, calculating such that they arrive at school/work 
about 0900, and spend approximately 8hrs there (Normally distributed). Workers 
following a night shift follow the same distribution, but 12hrs later. Agents 
travel to work/school by car, bike, or on foot based on the population distribution 
of car ownership and distance from home. At a rate of once per 
week, adult agents attempt to go to an entertainment building with members of their 
office or class. This action is dependent on travel time to/from the building and any 
prior plans that have been made, prioritizing family time if the agent is a 
parent. 

To simulate interactions between agents, the simulation checks agents that are with a small distance 
of each other. Agents within cars or on bikes cannot interact with other agents. 
When an agent arrives at a building, they are logged as being within that building, 
and all agents within a building are able to interact. Family members or being at 
home are not counted as part of an agents' interactions. When two agents interact, 
this interaction is logged in a hash for each agent. 

At the end of each day, the simulation updates the friend networks. First, it checks 
if an agent interacted with a friend at all that day. If they did interact, then 
the weight of that friendship is increased or decreased, based on a chance of positive 
interaction, by the amount of time the agents spent together, up to a maximum weight 
of 1. If two friends did not interact, their friendship weight decays by 
one of the two methods shown in Table~\ref{table:parameters}.
If an agent can have 
more friends, the simulation checks the interaction hash to see if any agent in 
that hash has a meaningful interaction. A meaningful interaction may occur after a 
specified number of interactions (``New Friend Interactions,'' 
Table~\ref{table:parameters})
and a random draw between 0 and 1. This then adds 
that agent to the friend network, with a random friendship weight between 0 and 1. 
This does not add the second agent to the first agent's set of friends, allowing 
for unidirectional friendship to occur. Any agent that was previously interacted 
with, but was not interacted with that day, is cleared from the interaction hash. 
After this, the state of the simulation is written out, and it continues with the 
next day, repeating this pattern until the end of the simulation. 

Code for the simulation, in Java, is available upon reasonable request.

\begin{figure}[h]
    \centering 
    \def\scale{1.0}
    \includegraphics[width=\linewidth]{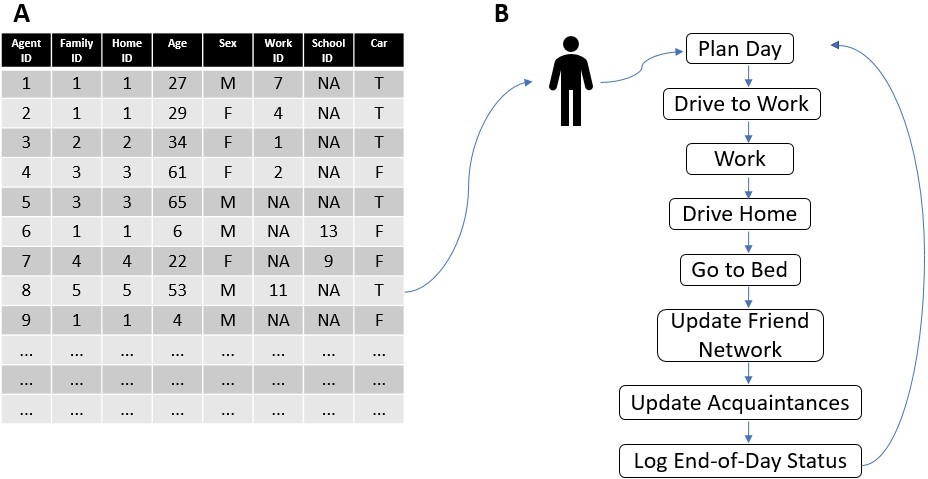}
    \caption{\textbf{Representative Population and Daily Activity.} This infographic 
    shows a population setup table and an outline of the daily activites. 
    \textbf{A.} Example population from the population generator - agents with 
    family members, homes, jobs, school, age, gender, etc. 
    \textbf{B.} Using information from \textbf{A.}, the simulation begins each day 
    by planning activities for each agent. Once all of the schedules are planned 
    and aligned, the agents' day begins, with agents getting up, going to work or 
    school, possibly going out in the evenings or heading home. At the end of each 
    day, the simulation calculates all statistics for each agent, updates those stats, 
    saves the state and any logging information for analysis later, and then everthing 
    repeats the next day. }
    \label{fig:sched}
\end{figure}

%% file: appendices/runtable.tex
\section{ABM Run Parameters} \label{app:runtable} 

\noindent

\vfill
\vspace{-0.5\baselineskip}
\begin{minipage}{\textwidth}
    \centering
    \captionof{table}[Parameters for Each ABM Scenario]{Parameters for Each ABM 
        Scenario\footnote{Column headings correspond to the parameters in 
        Table~\ref{table:parameters}. Each scenario was run with 5 different 
        seed values. The minimum decline rate is only used with the second 
        friendship decline function ($f_2$).}} 
    \label{table:runs}
    \begin{tabular}{|c||c|c|c|c|}
        \hline 
        Scenario & New Friend & Friendship Decline & Decline Rate & Minimum Decline \\
        & Interactions & Function & & Rate \\
        \hline\hline
        1 & 1 & $f_1$ (linear) & 0.025 & N/A \\
        \hline 
        2 & 3 & $f_1$ (linear) & 0.025 & N/A \\
        \hline 
        3 & 5 & $f_1$ (linear) & 0.025 & N/A \\
        \hline 
        4 & 8 & $f_1$ (linear) & 0.025 & N/A \\
        \hline 
        5 & 1 & $f_1$ (linear) & 0.05 & N/A \\
        \hline 
        6 & 3 & $f_1$ (linear) & 0.05 & N/A \\
        \hline 
        7 & 5 & $f_1$ (linear) & 0.05 & N/A \\
        \hline 
        8 & 8 & $f_1$ (linear) & 0.05 & N/A \\
        \hline 
        9 & 1 & $f_1$ (linear) & 0.075 & N/A \\
        \hline 
        10 & 3 & $f_1$ (linear) & 0.075 & N/A \\
        \hline 
        11 & 5 & $f_1$ (linear) & 0.075 & N/A \\
        \hline 
        12 & 8 & $f_1$ (linear) & 0.075 & N/A \\
        \hline 
        13 & 1 & $f_2$ (multiplicative) & 0.025 & 0.05 \\
        \hline 
        14 & 3 & $f_2$ (multiplicative) & 0.025 & 0.05 \\
        \hline 
        15 & 5 & $f_2$ (multiplicative) & 0.025 & 0.05 \\
        \hline 
        16 & 8 & $f_2$ (multiplicative) & 0.025 & 0.05 \\
        \hline 
        17 & 1 & $f_2$ (multiplicative) & 0.05 & 0.05 \\
        \hline 
        18 & 3 & $f_2$ (multiplicative) & 0.05 & 0.05 \\
        \hline 
        19 & 5 & $f_2$ (multiplicative) & 0.05 & 0.05 \\
        \hline 
        20 & 8 & $f_2$ (multiplicative) & 0.05 & 0.05 \\
        \hline 
        21 & 1 & $f_2$ (multiplicative) & 0.075 & 0.05 \\
        \hline 
        22 & 3 & $f_2$ (multiplicative) & 0.075 & 0.05 \\
        \hline 
        23 & 5 & $f_2$ (multiplicative) & 0.075 & 0.05 \\
        \hline 
        24 & 8 & $f_2$ (multiplicative) & 0.075 & 0.05 \\
        \hline 
        25 & 1 & $f_2$ (multiplicative) & 0.025 & 0.01 \\
        \hline 
        26 & 3 & $f_2$ (multiplicative) & 0.025 & 0.01 \\
        \hline 
        27 & 5 & $f_2$ (multiplicative) & 0.025 & 0.01 \\
        \hline 
        28 & 8 & $f_2$ (multiplicative) & 0.025 & 0.01 \\
        \hline 
        29 & 1 & $f_2$ (multiplicative) & 0.05 & 0.01 \\
        \hline 
        30 & 3 & $f_2$ (multiplicative) & 0.05 & 0.01 \\
        \hline 
        31 & 5 & $f_2$ (multiplicative) & 0.05 & 0.01 \\
        \hline 
        32 & 8 & $f_2$ (multiplicative) & 0.05 & 0.01 \\
        \hline 
        33 & 1 & $f_2$ (multiplicative) & 0.075 & 0.01 \\
        \hline 
        34 & 3 & $f_2$ (multiplicative) & 0.075 & 0.01 \\
        \hline 
        35 & 5 & $f_2$ (multiplicative) & 0.075 & 0.01 \\
        \hline 
        36 & 8 & $f_2$ (multiplicative) & 0.075 & 0.01 \\
        \hline 
    \end{tabular}
\end{minipage}
\vfill
\clearpage

%% file: appendices/addresults.tex
\section{Additional Results} \label{app:addresults}

\noindent
\begin{minipage}[t][1cm][t]{\textwidth}
    \begin{multicols}{2}
        \indent To better understand the information encoded in a graph's HCA, 
        we consider Fig. \ref{fig:features} in conjunction with Figs. 
        \ref{fig:hc_scenario01} through \ref{fig:hc_scenario17}. Recalling that 
        heat content asymptotics are the coefficients of a Maclaurin expansion 
        for a graph's heat content curve, we compare in Figs. 
        \ref{fig:hc_scenario01}, \ref{fig:hc_scenario05}, and 
        \ref{fig:hc_scenario17} plots of a scenario's graph's heat content 
        curves (as a function of heat dissipation time $s$) at 
        different time steps (Figs. \ref{fig:hc01_curves}, 
        \ref{fig:hc05_curves}, \ref{fig:hc17_curves}) to plots of the $L^2$ 
        norms of the corresponding heat curves (Figs. \ref{fig:hc01_l2}, 
        \ref{fig:hc05_l2}, and \ref{fig:hc17_l2}) and $L^2$ norms of the 
        numerical derivatives of corresponding heat curves (Figs. 
        \ref{fig:hc01_derivsl2}, \ref{fig:hc05_derivsl2}, and 
        \ref{fig:hc17_derivsl2}). Figs. \ref{fig:hc01_q0diffs}, 
        \ref{fig:hc05_q0diffs}, and \ref{fig:hc17_q0diffs} show the HCA 
        comparison between a scenario's graph at time step $t$ and a scenario's 
        initial graph ($t=0$). Similarly, Figs. \ref{fig:hc01_qdiffs}, 
        \ref{fig:hc05_qdiffs}, and \ref{fig:hc17_qdiffs} show the HCA comparison 
        between a scenario at time step $t$ and the graph corresponding to the 
        previous time step ($t-1$). Both sets of HCA comparison plots are not 
        only reflective of changes to a scenario's graphs that are observable 
        with by examining specific graph features, but also appear to pick up 
        graph changes not observed by the other methods used in this paper. For 
        example, the first spike in Fig. \ref{fig:hc01_qdiffs} does not appear 
        to correspond to any phenomena in Figs.~\ref{node1},~\ref{edge1}, 
        or~\ref{comp1}, while the second spike in Fig. \ref{fig:hc01_qdiffs} 
        corresponds to the local extrema observed in Fig. 
        \ref{fig:hc01_qdiffs}. On the other hand, the extreme in Fig. 
        \ref{node17} corresponds to an inflection point in Fig. 
        \ref{fig:hc_scenario17}.
    \end{multicols}
\end{minipage}

\vfill
% \\[\intextsep]
\vspace{\intextsep}
\noindent
\begin{minipage}{\textwidth}
    \centering
    \captionsetup[sub]{labelformat=parens}
    \begingroup
        \captionsetup{type=figure}
        \begin{subfigure}[t]{0.32\textwidth}
            \includegraphics[width=\textwidth]{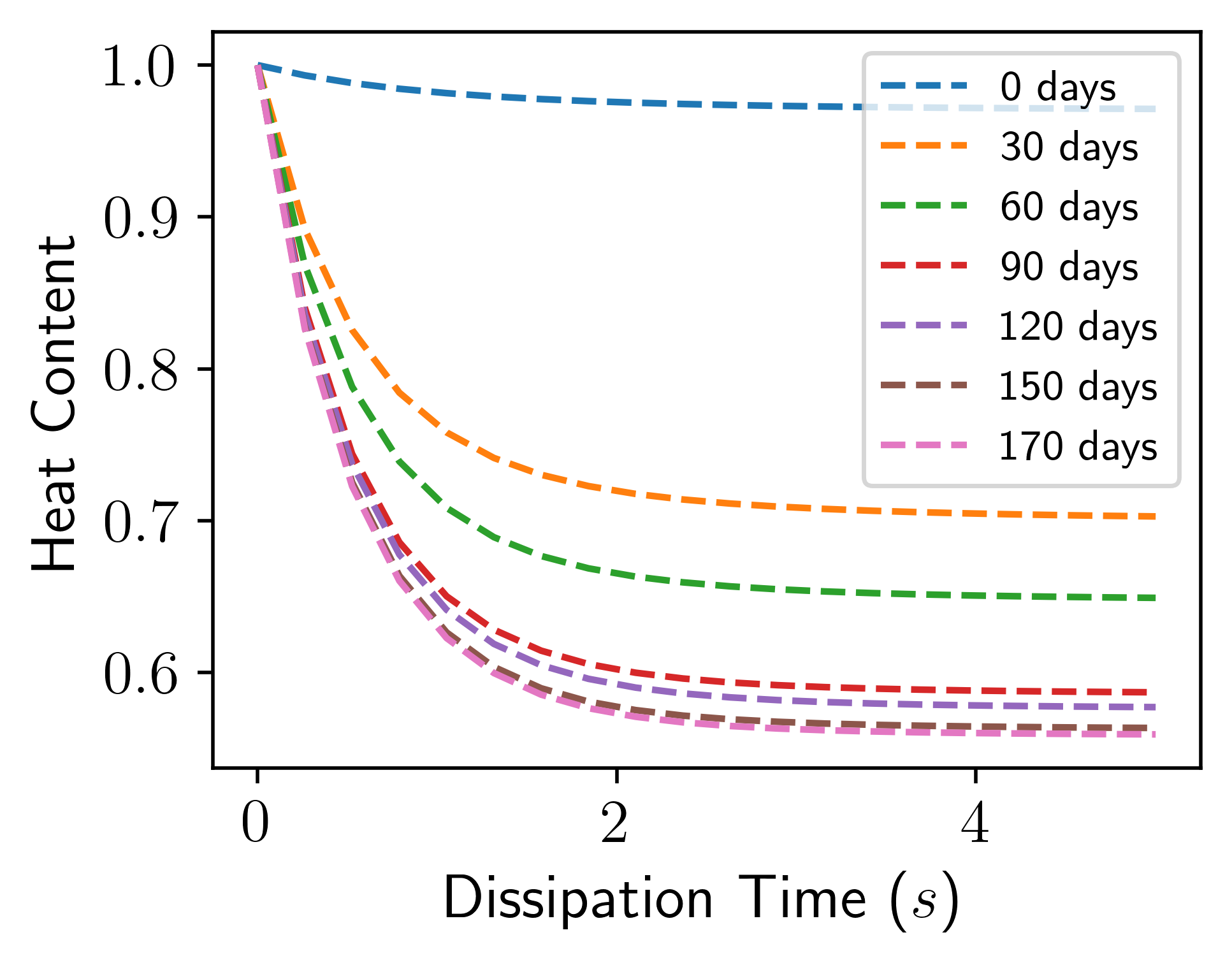}
            \caption{Heat Content Curves (HCCs)}
            \label{fig:hc01_curves}
        \end{subfigure}
        \begin{subfigure}[t]{0.32\textwidth}
            \includegraphics[width=\textwidth]{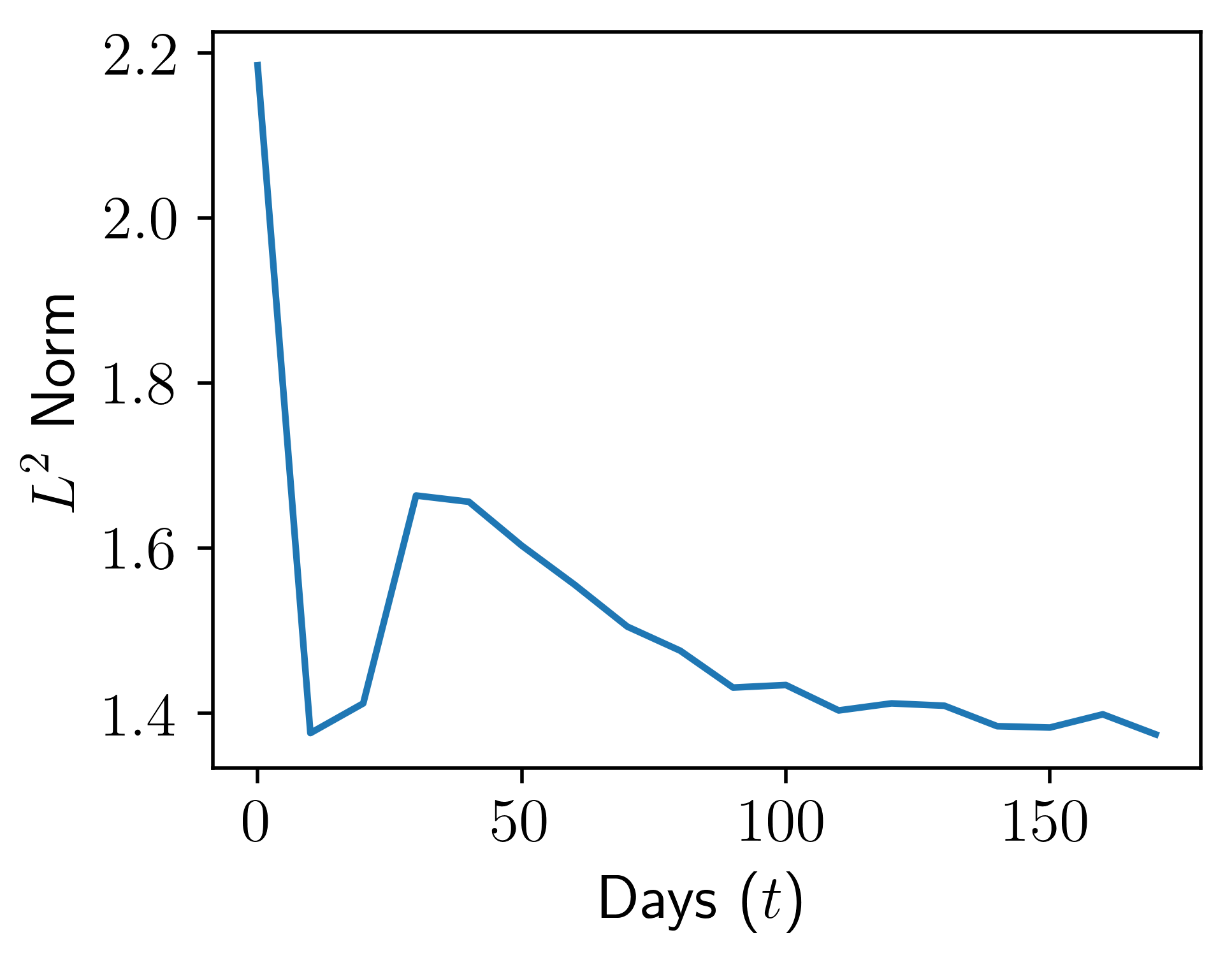}
            \caption{HCC $L^2$ Norms}
            \label{fig:hc01_l2}
        \end{subfigure}
        \begin{subfigure}[t]{0.32\textwidth}
            \includegraphics[width=\textwidth]{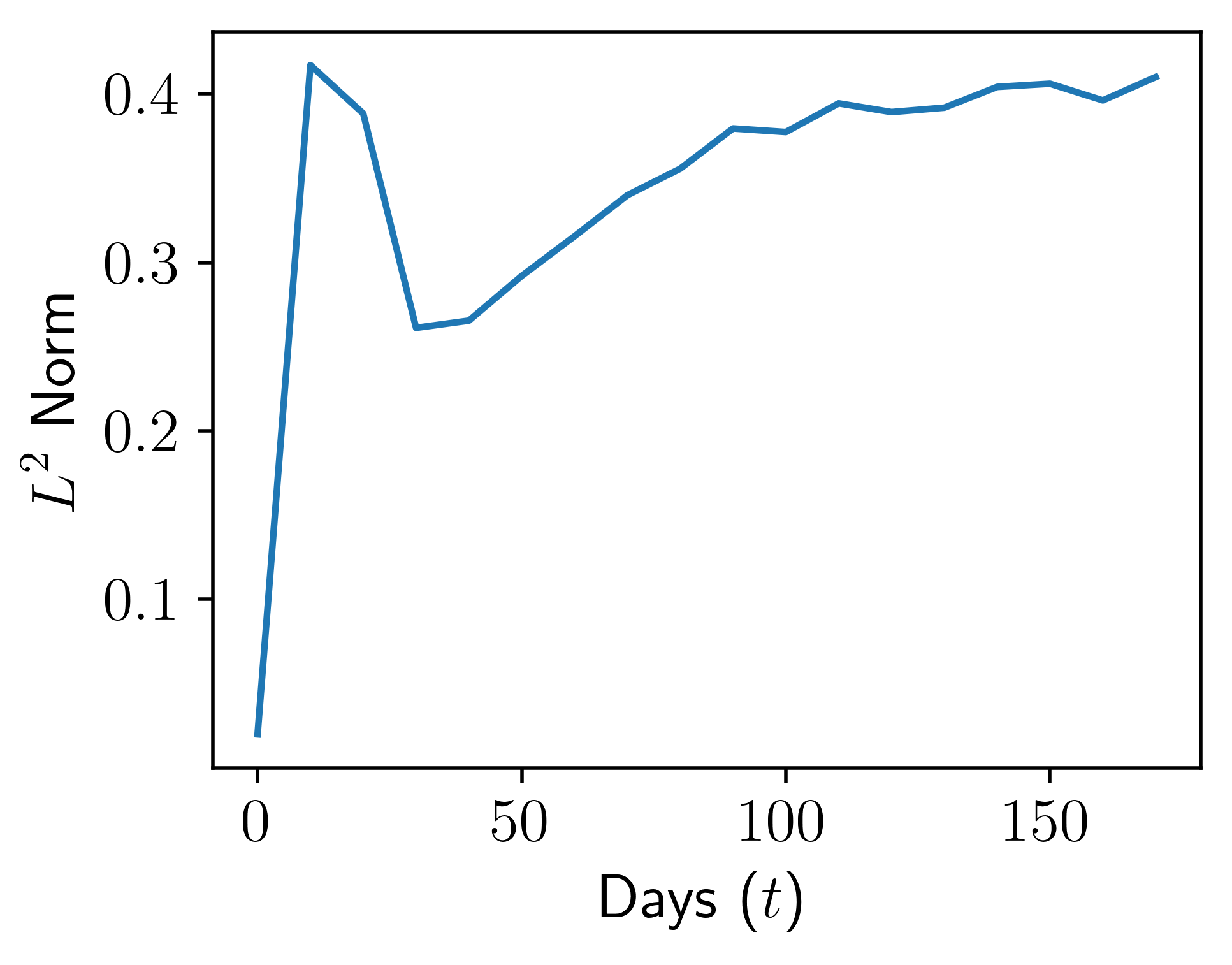}
            \caption{HCC Derivative $L^2$ Norms}
            \label{fig:hc01_derivsl2}
        \end{subfigure}
        \begin{subfigure}[t]{0.32\textwidth}
            \includegraphics[width=\textwidth]{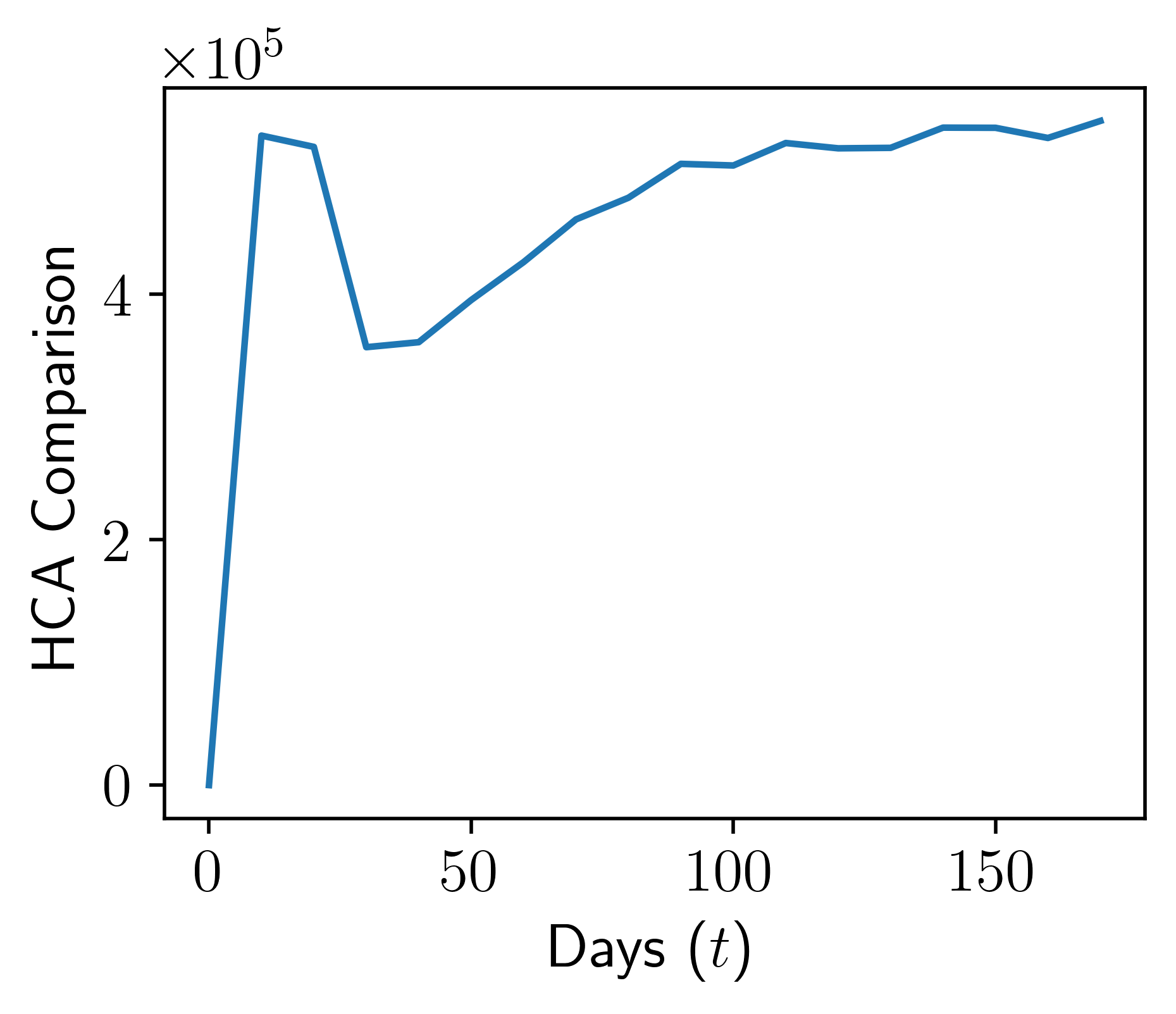}
            \caption{HCA Comparison with Initial Graph}
            \label{fig:hc01_q0diffs}
        \end{subfigure}
        \begin{subfigure}[t]{0.32\textwidth}
            \includegraphics[width=\textwidth]{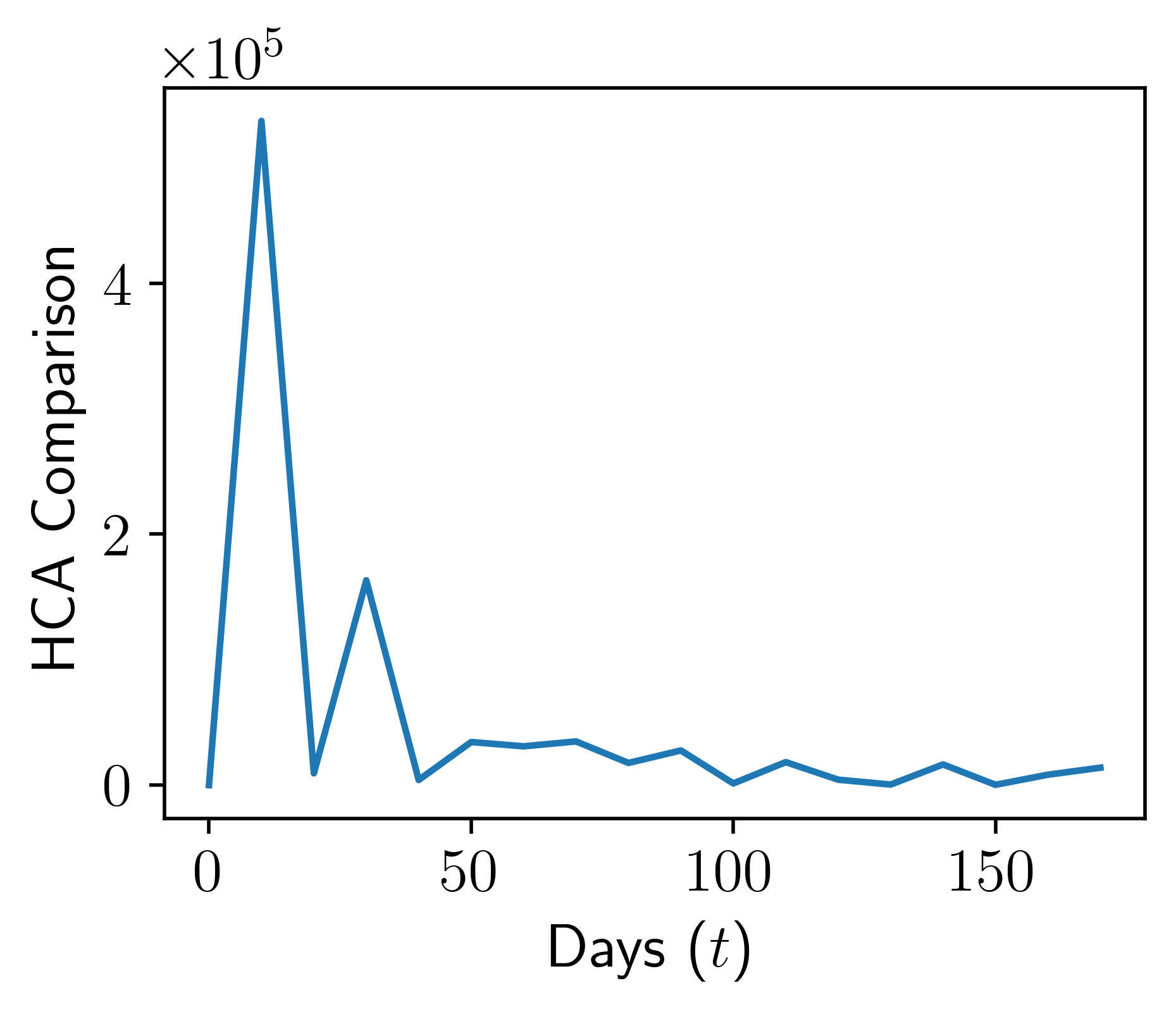}
            \caption{HCA Comparison with Previous Graph}
            \label{fig:hc01_qdiffs}
        \end{subfigure}
    \endgroup
    \addtocounter{figure}{-1}
    \captionof{figure}{\textbf{Heat content curves and HCA metric comparisons 
        (Scenario 1).} Subfigure (\subref{fig:hc01_curves}) shows plots of the 
        heat content curves for the friend networks associated with Scenario 1 
        at selected time steps. We see plots for the $L^2$ norms of these heat 
        content curves and the heat curve derivatives in (\subref{fig:hc01_l2}) 
        and (\subref{fig:hc01_derivsl2}), respectively. Lastly, 
        (\subref{fig:hc01_q0diffs}) and (\subref{fig:hc01_qdiffs}) respectively
        feature plots of the curves $d_{\text{hca}}\big(G(0), \, G(t)\big)$ and
        $d_{\text{hca}}\big(G(t-1), \, G(t)\big)$, where $G(t)$ denotes the 
        associated friend network for this scenario at timestep $t$.
    }
    \label{fig:hc_scenario01}
\end{minipage}

\vfill
\clearpage

\vspace{\intextsep}
\noindent
\begin{minipage}{\textwidth}
    \centering
    \captionsetup[sub]{labelformat=parens}
    \begingroup
        \captionsetup{type=figure}
        \begin{subfigure}[t]{0.32\textwidth}
            \includegraphics[width=\textwidth]{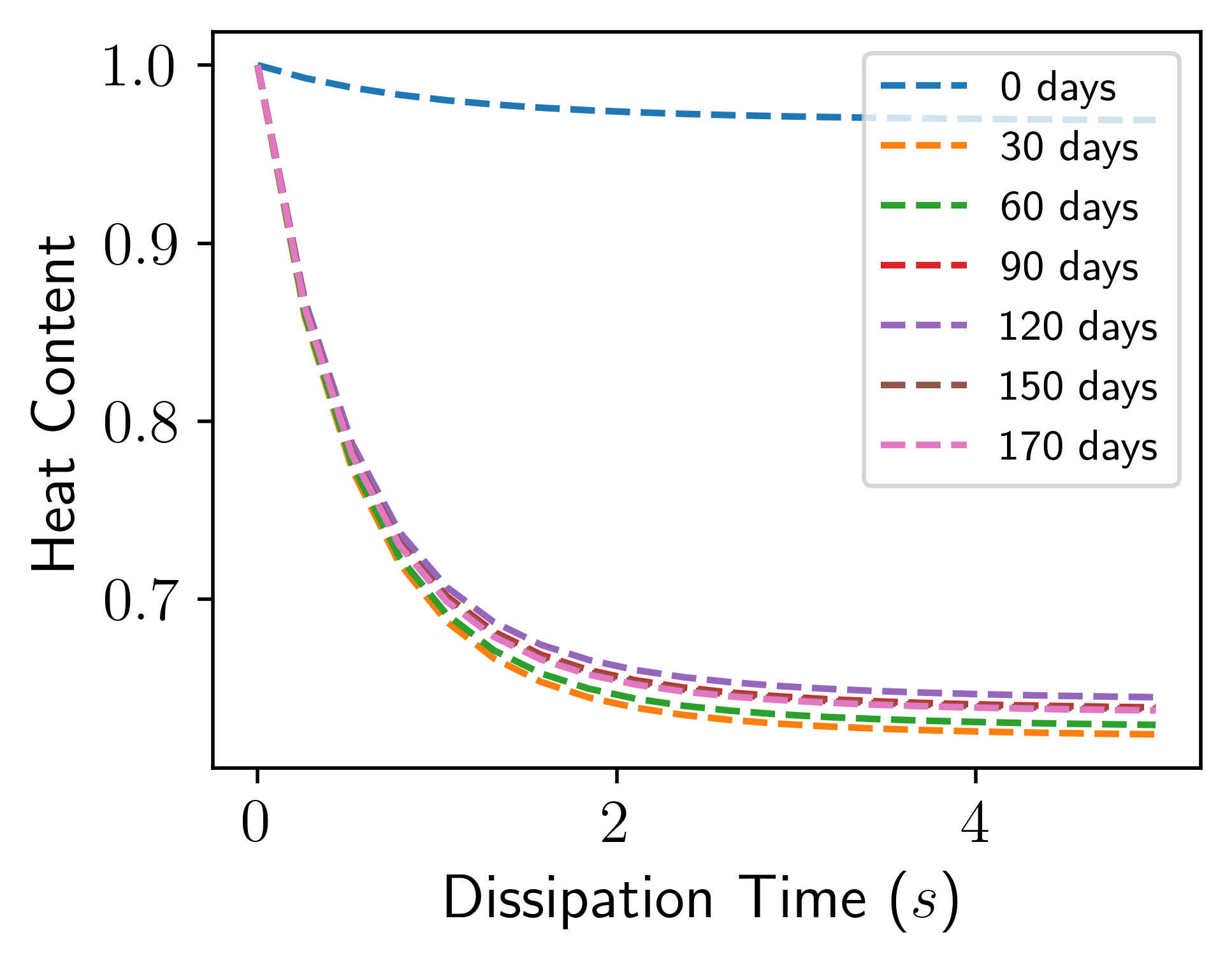}
            \caption{Heat Content Curves (HCCs)}
            \label{fig:hc05_curves}
        \end{subfigure}
        \begin{subfigure}[t]{0.32\textwidth}
            \includegraphics[width=\textwidth]{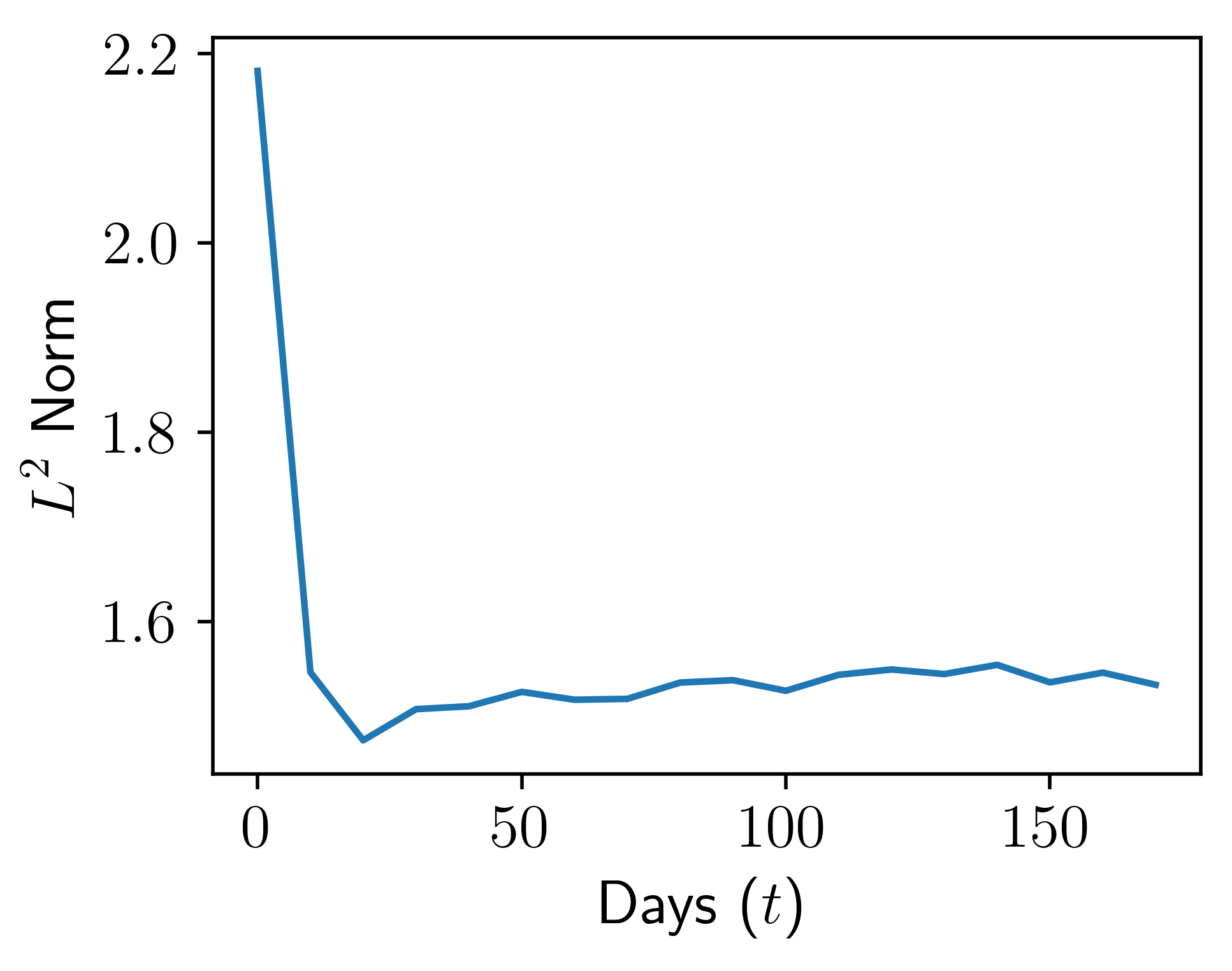}
            \caption{HCC $L^2$ Norms}
            \label{fig:hc05_l2}
        \end{subfigure}
        \begin{subfigure}[t]{0.32\textwidth}
            \includegraphics[width=\textwidth]{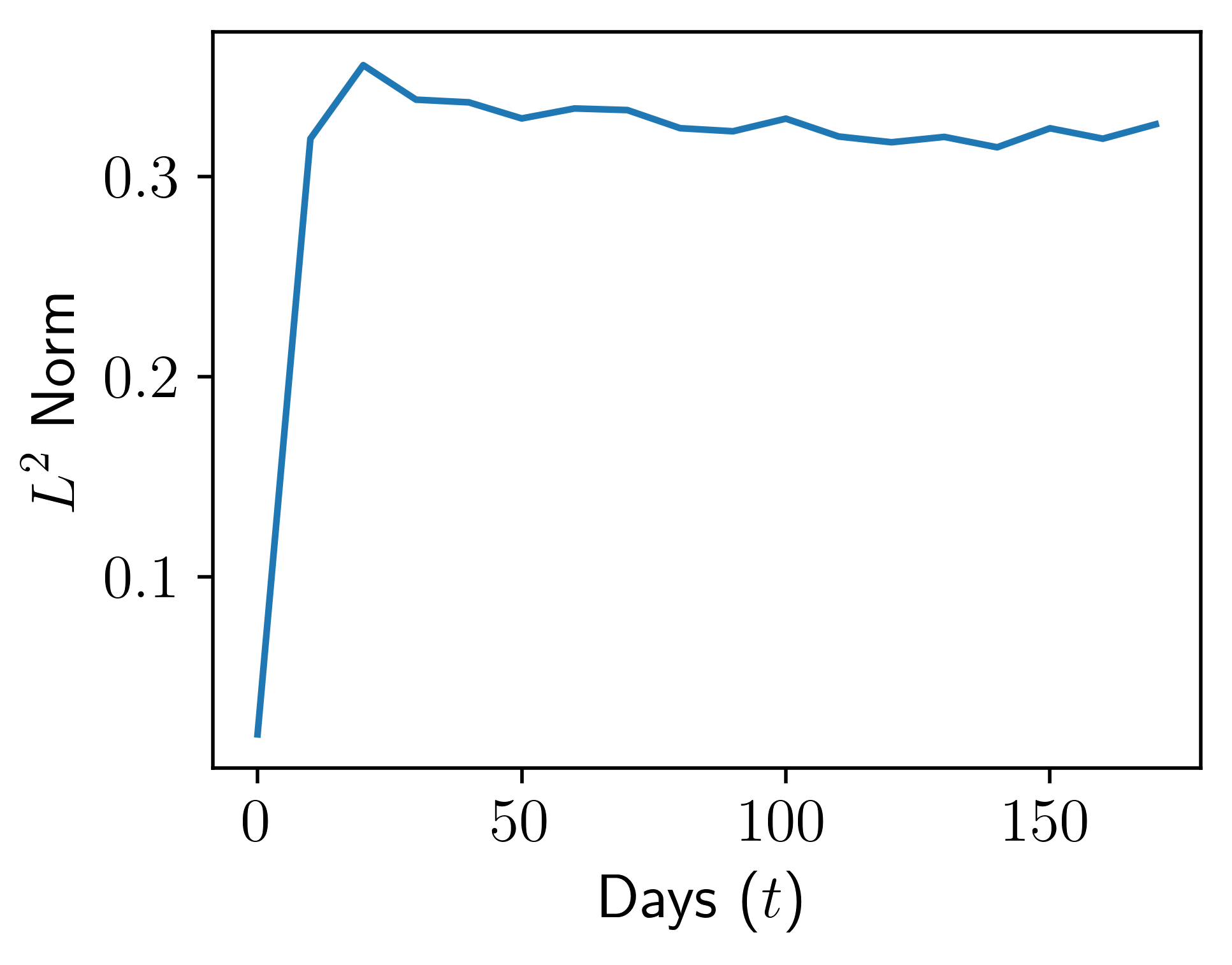}
            \caption{HCC Derivative $L^2$ Norms}
            \label{fig:hc05_derivsl2}
        \end{subfigure}
        \begin{subfigure}[t]{0.4\textwidth}
            \centering
            \includegraphics[width=.8\textwidth]{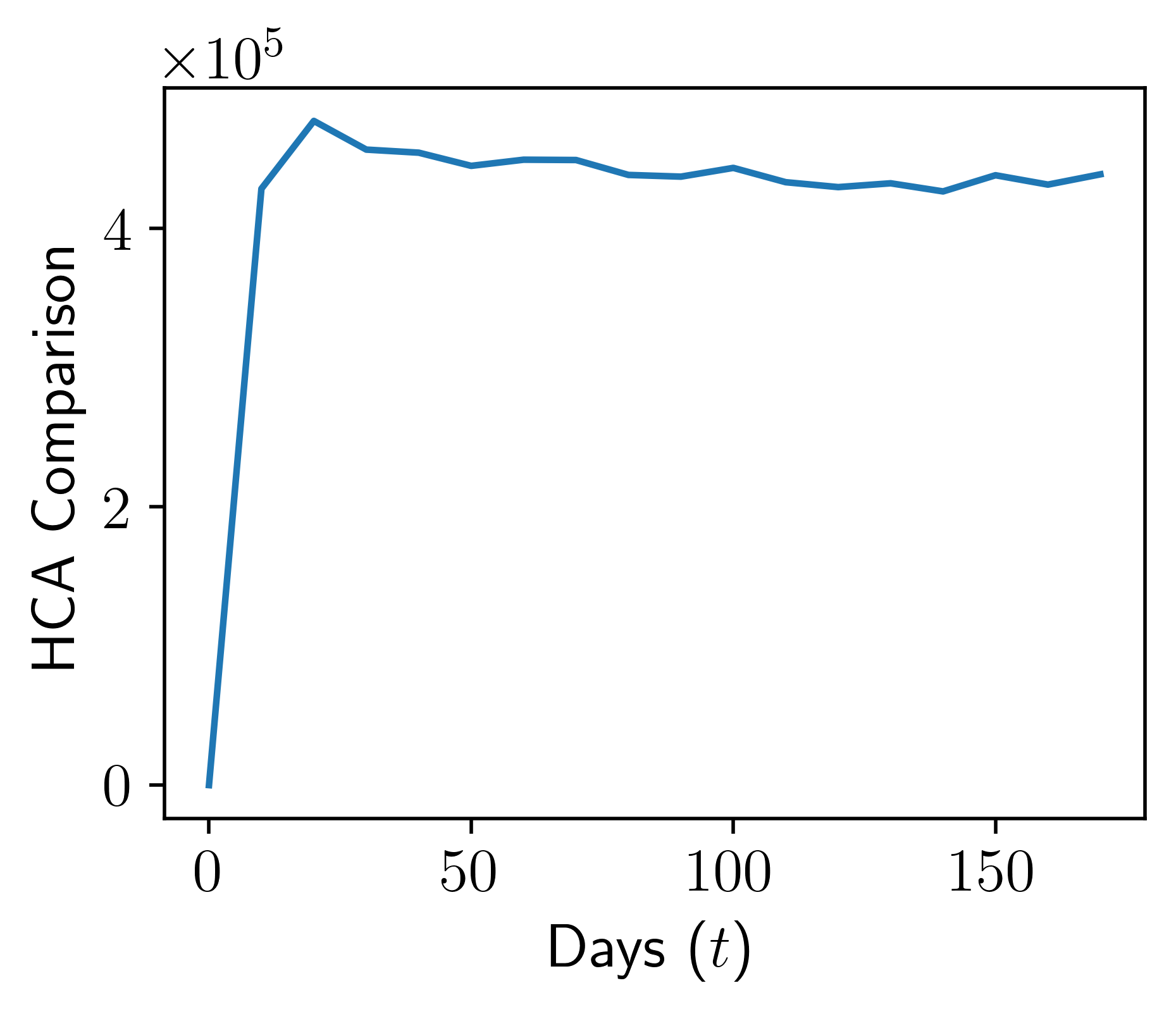}
            \caption{HCA Comparison with Initial Graph}
            \label{fig:hc05_q0diffs}
        \end{subfigure}
        \begin{subfigure}[t]{0.4\textwidth}
            \centering
            \includegraphics[width=.8\textwidth]{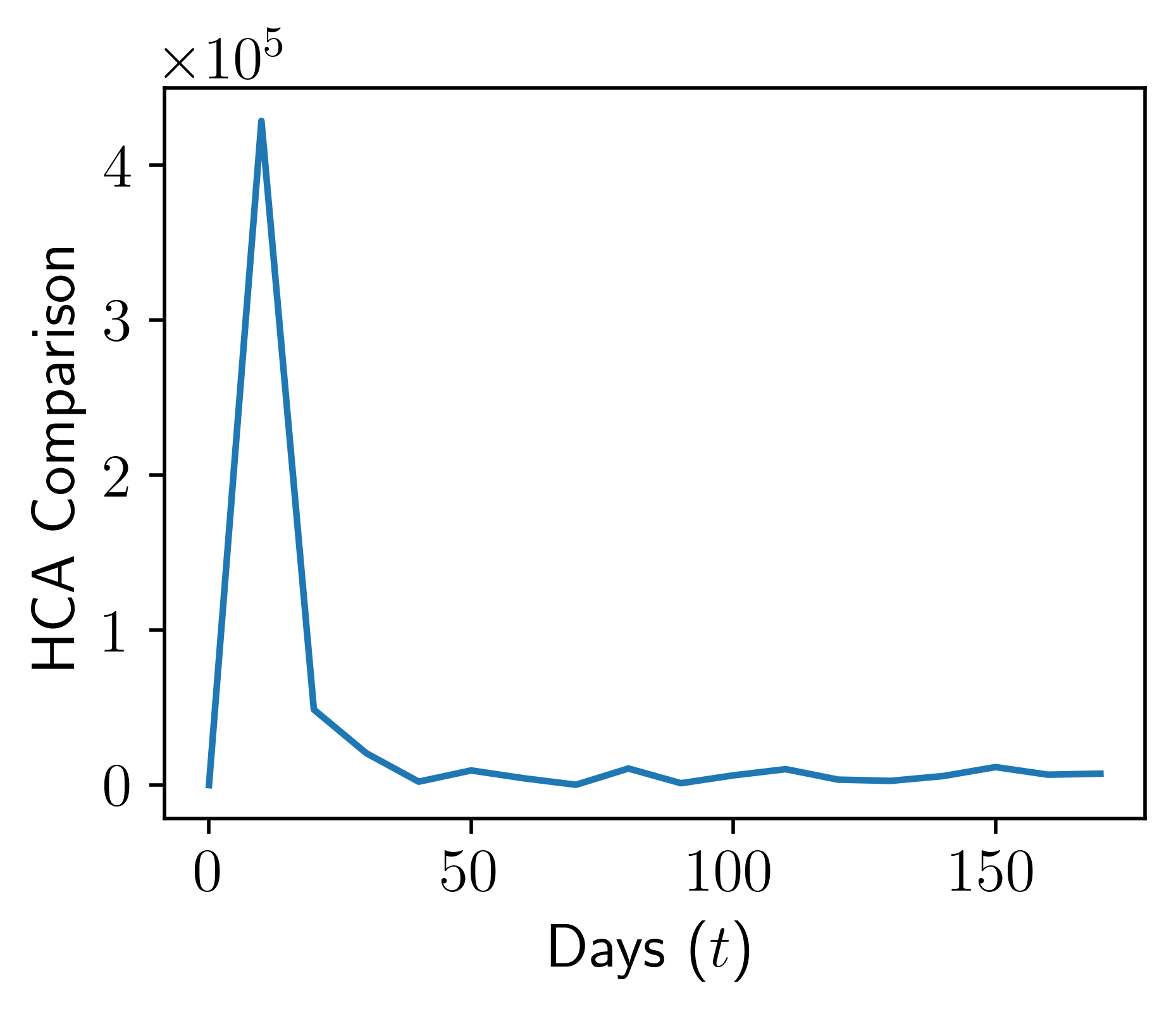}
            \caption{HCA Comparison with Previous Graph}
            \label{fig:hc05_qdiffs}
        \end{subfigure}
    \endgroup
    \addtocounter{figure}{-1}
    \captionof{figure}{\textbf{Heat content curves and HCA metric comparisons (Scenario 5).}}
    \label{fig:hc_scenario05}
\end{minipage}

\vfill
\vspace{\intextsep}
\noindent
\begin{minipage}{\textwidth}
    \centering
    \captionsetup[sub]{labelformat=parens}
    \begingroup
        \captionsetup{type=figure}
        \begin{subfigure}[t]{0.32\textwidth}
            \includegraphics[width=\textwidth]{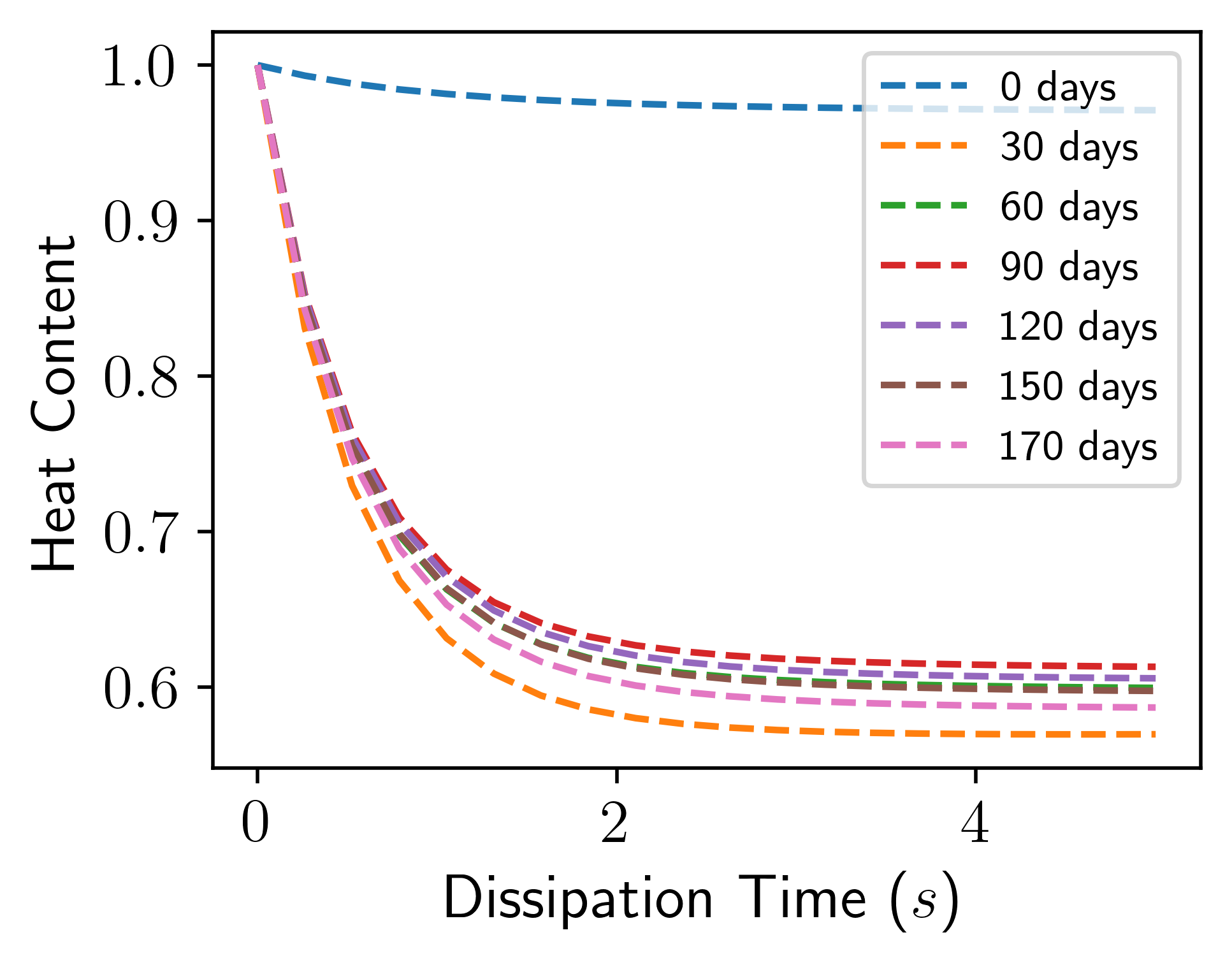}
            \caption{Heat Content Curves (HCCs)}
            \label{fig:hc17_curves}
        \end{subfigure}
        \begin{subfigure}[t]{0.32\textwidth}
            \includegraphics[width=\textwidth]{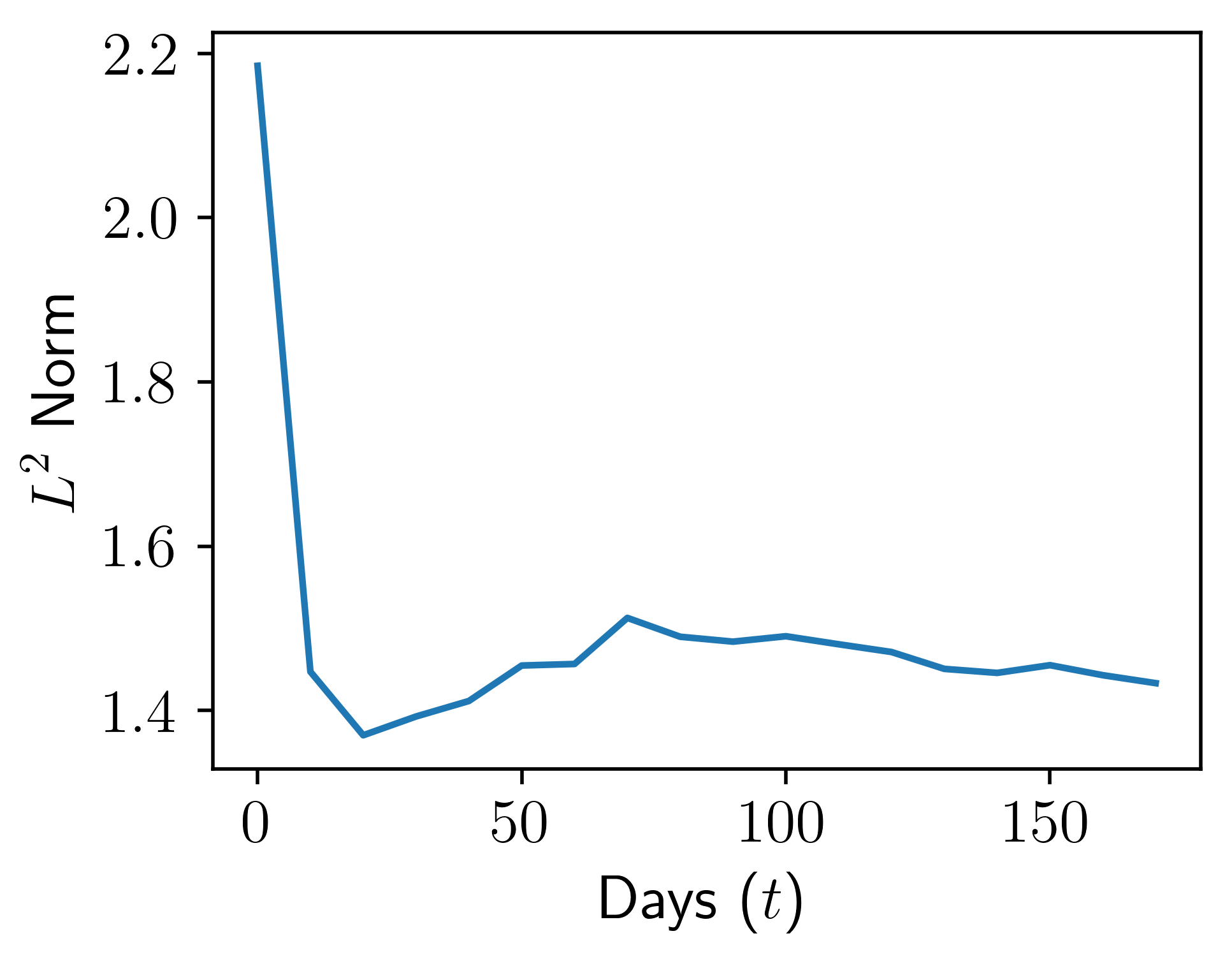}
            \caption{HCC $L^2$ Norms}
            \label{fig:hc17_l2}
        \end{subfigure}
        \begin{subfigure}[t]{0.32\textwidth}
            \includegraphics[width=\textwidth]{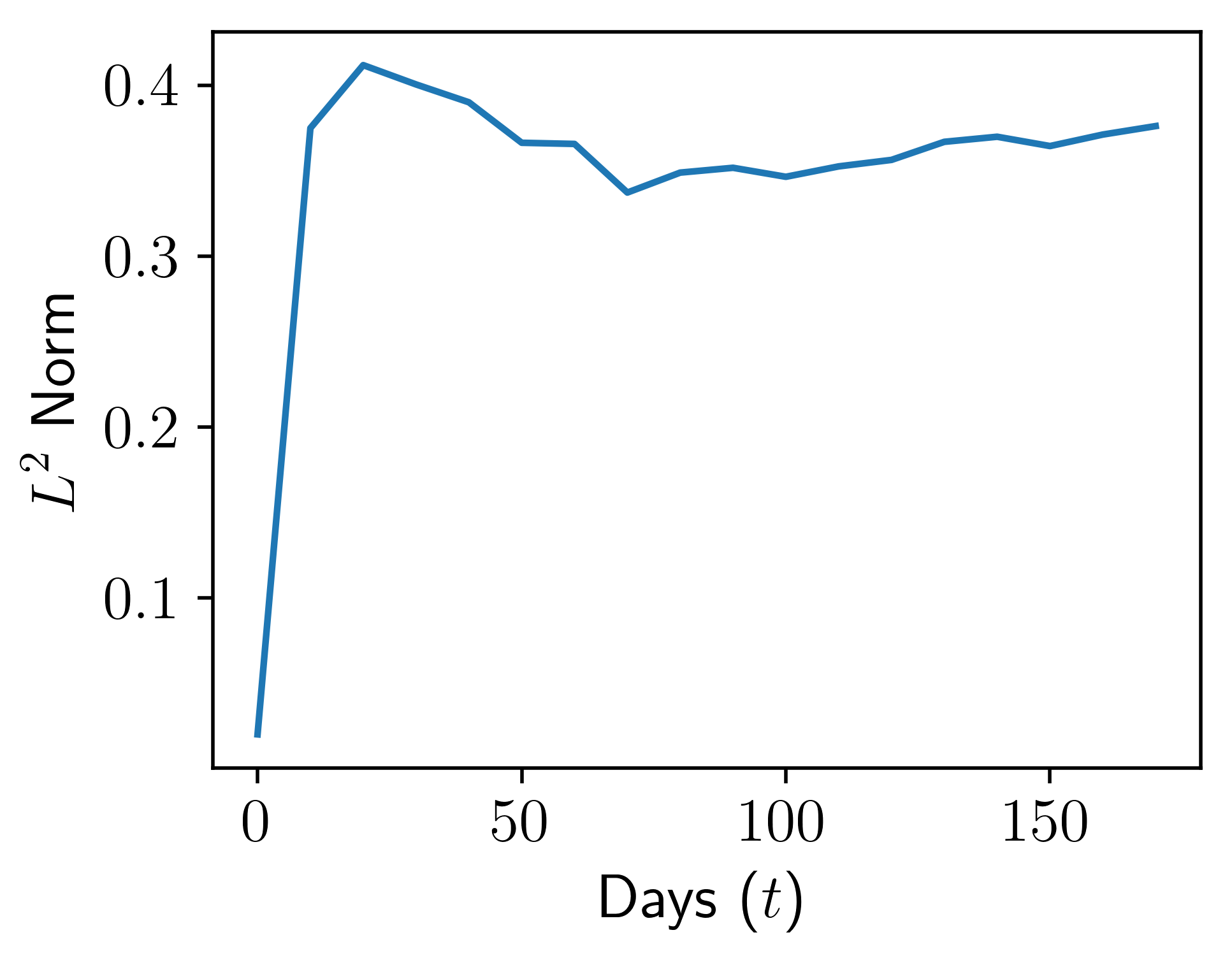}
            \caption{HCC Derivative $L^2$ Norms}
            \label{fig:hc17_derivsl2}
        \end{subfigure}
        \begin{subfigure}[t]{0.4\textwidth}
            \centering
            \includegraphics[width=.8\textwidth]{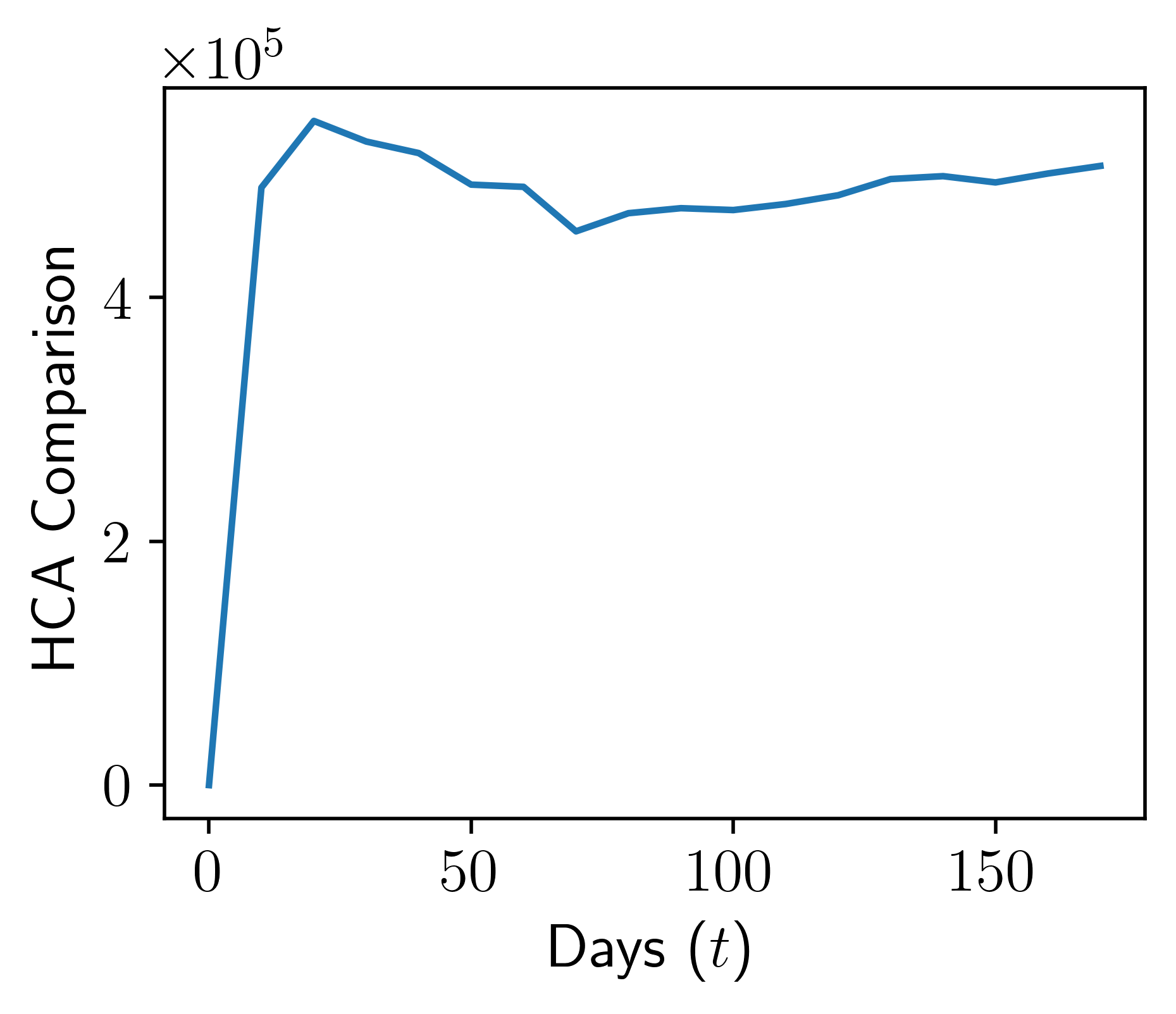}
            \caption{HCA Comparison with Initial Graph}
            \label{fig:hc17_q0diffs}
        \end{subfigure}
        \begin{subfigure}[t]{0.4\textwidth}
            \centering
            \includegraphics[width=.8\textwidth]{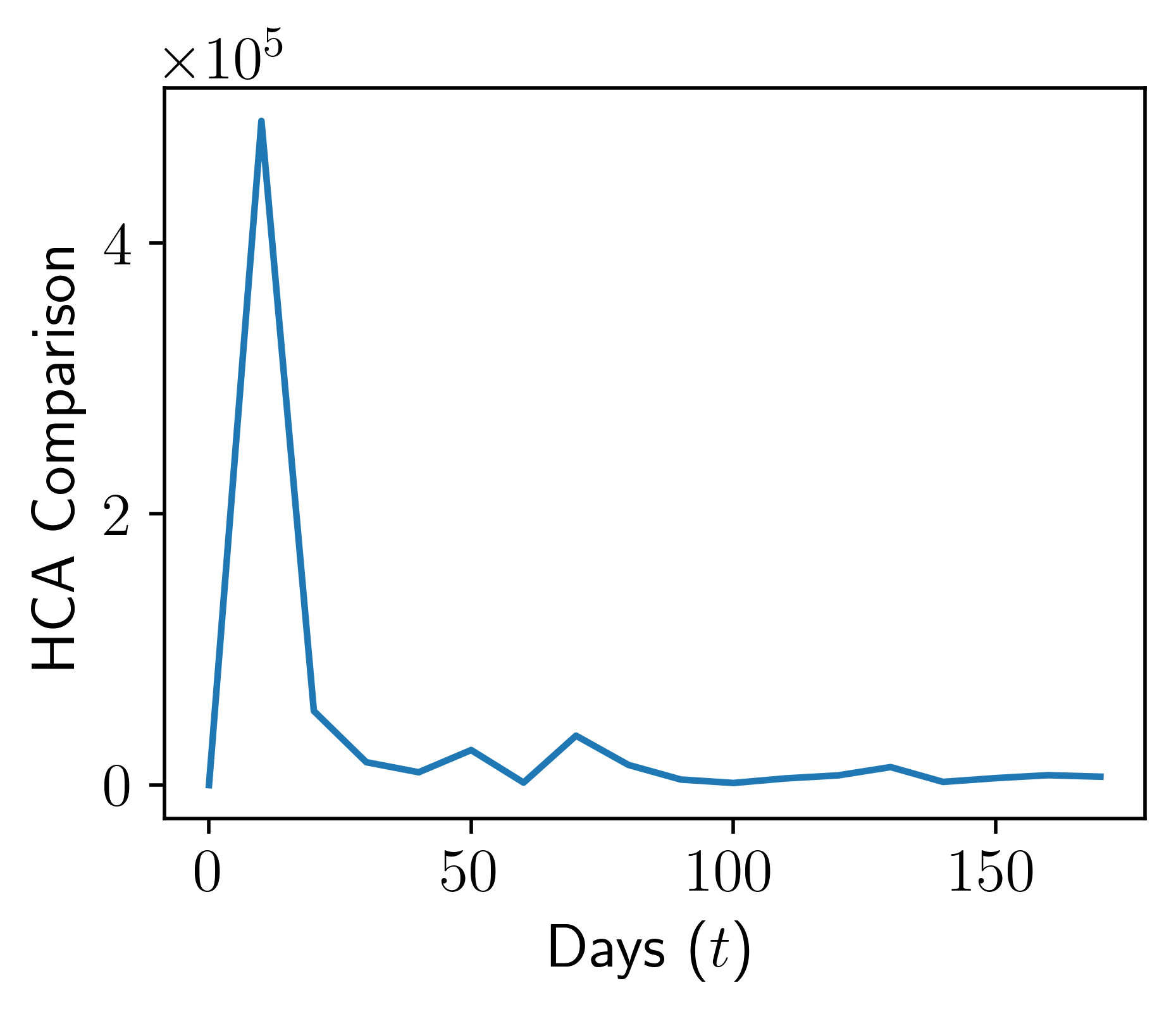}
            \caption{HCA Comparison with Previous Graph}
            \label{fig:hc17_qdiffs}
        \end{subfigure}
    \endgroup
    \addtocounter{figure}{-1}
    \captionof{figure}{\textbf{Heat content curves and HCA metric comparisons (Scenario 17).} }
    \label{fig:hc_scenario17}
% \end{figure*}
\end{minipage}